\documentclass[12pt]{article}
\usepackage{eqsection,latexsym,epsf,cite}

\begin{document}

\title{Mean-field expansion for spin models with medium-range interactions}
\author{
  \\
  {\small Andrea Pelissetto}              \\[-0.2cm]
  {\small\it Dipartimento di Fisica and INFN -- Sezione di Roma I}    \\[-0.2cm]
  {\small\it Universit\`a degli Studi di Roma ``La Sapienza"}        \\[-0.2cm]
  {\small\it I-00185 Roma, ITALIA}          \\[-0.2cm]
  {\small E-mail: {\tt Andrea.Pelissetto@roma1.infn.it}}   \\[-0.2cm]
  \\[-0.1cm]  \and
  {\small Paolo Rossi and Ettore Vicari}              \\[-0.2cm]
  {\small\it Dipartimento di Fisica and INFN -- Sezione di Pisa}    \\[-0.2cm]
  {\small\it Universit\`a degli Studi di Pisa}        \\[-0.2cm]
  {\small\it I-56100 Pisa, ITALIA}          \\[-0.2cm]
  {\small E-mail: {\tt rossi@mailbox.difi.unipi.it}}   \\[-0.2cm]
  {\small E-mail: {\tt vicari@mailbox.difi.unipi.it}}   \\[-0.2cm]
  {\protect\makebox[5in]{\quad}}  
  \\
}
\vspace{0.5cm}
\date{March 25, 1999}

\maketitle
\thispagestyle{empty}   

\vspace{0.2cm}

\begin{abstract}
We study the critical crossover between the Gaussian 
and the Wilson-Fisher fixed point for general $O(N)$-invariant spin
models with medium-range interactions. We perform a systematic 
expansion around the mean-field solution, obtaining the 
universal crossover curves and their leading corrections.
In particular we show that, in three dimensions, the leading correction
scales as $R^{-3}$, $R$ being the range of the interactions.
We compare our results with the existing numerical ones obtained
by Monte Carlo simulations and present a critical discussion of other 
approaches.
\end{abstract}

\clearpage

\newcommand{\be}{\begin{equation}}
\newcommand{\ee}{\end{equation}}
\newcommand{\bea}{\begin{eqnarray}}
\newcommand{\eea}{\end{eqnarray}}
\newcommand{\<}{\langle}
\renewcommand{\>}{\rangle}

\def\spose#1{\hbox to 0pt{#1\hss}}
\def\ltapprox{\mathrel{\spose{\lower 3pt\hbox{$\mathchar"218$}}
 \raise 2.0pt\hbox{$\mathchar"13C$}}}
\def\gtapprox{\mathrel{\spose{\lower 3pt\hbox{$\mathchar"218$}}
 \raise 2.0pt\hbox{$\mathchar"13E$}}}

\def\bsigma{\mbox{\protect\boldmath $\sigma$}}
\def\btau{\mbox{\protect\boldmath $\tau$}}
\def\bphi{\mbox{\protect\boldmath $\phi$}}
\def\bz{\mbox{\protect\boldmath $z$}}
\def\bw{\mbox{\protect\boldmath $w$}}
\def\hatp{\hat p}
\def\hatl{\hat l}
\def\smfrac#1#2{{\textstyle\frac{#1}{#2}}}
\def\case#1#2{{\textstyle\frac{#1}{#2}}}

\def\msbar{ {\overline{\hbox{\scriptsize MS}}} }
\def\normalmsbar{ {\overline{\hbox{\normalsize MS}}} }

\newcommand{\R}{\hbox{{\rm I}\kern-.2em\hbox{\rm R}}}
\newcommand{\N}{\hbox{{\rm I}\kern-.2em\hbox{\rm N}}}

\newcommand{\reff}[1]{(\ref{#1})}

\section{Introduction}

Every physical situation of experimental relevance has at least two scales: 
one scale is intrinsic to the system, 
while the second one is related to experimental conditions.
In statistical mechanics the correlation length $\xi$ is 
related to experimental conditions (it depends on the temperature), 
while the interaction length (Ginzburg parameter) is intrinsic.
The opposite is true in 
quantum field theory: here the correlation length (inverse mass gap) 
is intrinsic, while the interaction scale (inverse momentum) 
depends on the experiment.
Physical predictions are functions of ratios of these two scales and
describe the 
crossover from the correlation-dominated ($\xi/G$ or $p/m$ large) 
to the interaction-dominated ($\xi/G$ or $p/m$ small) regime.
In a properly defined limit they are universal and define the unique flow
between two different fixed points. This universal limit is obtained when
two scales become very large with respect to any other
(microscopic) scale. Their ratio becomes the (universal)
control parameter of the system, whose transition from $0$
to $\infty$ describes the critical crossover.

In this paper we will consider the crossover between the Gaussian fixed
point where mean-field predictions hold (interaction-dominated regime) 
to the standard Wilson-Fisher fixed point (correlation-dominated
regime). In recent years a lot of work has been devoted to understanding
this crossover, either experimentally 
\cite{Corti-Degiorgio_85,Dietler-Cannel_88,Anisimov-etal_95,Anisimov-etal_96,%
Jacob-etal_98}
or theoretically 
\cite{Bagnuls-Bervillier_84,Bagnuls-Bervillier_85,Bagnuls-etal_87,%
Fisher_86,Bagnuls-Bervillier_87,%
Schloms-Dohm_89,Chen-etal_90,Anisimov-etal_92,%
Belyakov-Kiselev_92,Thouless_69,M-B,L-B-B-pre,L-B-B-prl,%
Luijten-Binder_98,PRV_longrange}.
The traditional approach to the crossover between the Gaussian and 
the Wilson-Fisher fixed point starts 
from the standard Landau-Ginzburg Hamiltonian. On a 
$d$-dimensional lattice, it can be written as 
\be
  H\; =\; \sum_{x_1,x_2} \smfrac{1}{2}J({x_1}-{x_2})
\left(\phi_{x_1} - \phi_{x_2}\right)^2
+\, \sum_x \left[ \smfrac{1}{2}  r\phi_x^2 +
{u\over 4!} \phi_x^4 - h_x \cdot \phi_x\right] ,
\label{lham}
\ee
where $\phi_x$ are $N$-dimensional vectors, and
$J(x)$ is the standard nearest-neighbour
coupling. For this model the interaction scale is controlled by the 
coupling $u$ and the relevant parameters are the (thermal) Ginzburg number
$G$ \cite{Ginzburg_60} and its magnetic counterpart $G_h$ 
\cite{L-B-B-pre,PRV_longrange} defined by:
\be
G_{\hphantom{h}} =\, u^{2/(4-d)}, \qquad
G_h =\, u^{(d+2)/[2(4-d)]}.
\ee
Under a renormalization-group (RG) transformation $G$ 
scales like the (reduced) temperature, 
while $G_h$ scales as the magnetic field. For 
$t \equiv r - r_c \ll G$ and $h\ll G_h$ one observes the standard critical 
behaviour, while in the opposite case the behaviour is classical.
The critical crossover limit corresponds to considering 
$t,h,u\to 0$ keeping $\widetilde{t} = t/G$ and $\widetilde{h} = h/G_h$ 
fixed. This limit is universal, i.e. independent of the detailed 
structure of the model: any Hamiltonian of the form \reff{lham} shows
the same universal behaviour as long as the interaction is 
short-ranged, i.e. for any $J(x)$ such that 
$\sum_{x}x^2\, J(x) < + \infty$.
The crossover functions can be related to the RG functions 
of the standard continuum $\phi^4$ theory if one 
expresses them in terms of the zero-momentum four-point renormalized 
coupling $g$ 
\cite{Bagnuls-Bervillier_84,Bagnuls-Bervillier_85,Bagnuls-etal_87,%
Schloms-Dohm_89}.
For the observables that are traditionally studied in statistical mechanics,
for instance the susceptibility or the correlation length, 
the crossover functions can be computed to high precision in
the fixed-dimension expansion in $d=3$ 
\cite{Bagnuls-Bervillier_84,Bagnuls-Bervillier_85,Bagnuls-etal_87}.

Let us now consider the medium-range case. Following 
Refs. \cite{M-B,L-B-B-pre} we assume that $J({x})$
has the following form
\be
J({x})=\, \cases{ J & \qquad for ${x}\in {D}$, \cr
                  0 & \qquad for ${x}\not\in {D}$,
                    }
\label{defJ}
\ee
where $D$ is a lattice domain characterized by some scale $R$.
Explicitly we define $R$ and the corresponding domain volume
$V_R$ by
\be
V_R \, \equiv  \sum_{{x}\in D} 1,
\qquad
R^2 \, \equiv {1\over 2d\,V_R} \sum_{{x}\in D} x^2\; .
\label{defR}
\ee
The shape of ${D}$ is irrelevant for our purposes as 
long as $V_R\sim R^d$ for $R\to\infty$. The constant $J$ defines the 
normalization of the fields. Here we assume $J=1/V_R$, since this 
choice simplifies the discussion of the limit $R\to\infty$.
To understand the connection between the theory with medium-range
interactions and the short-range model let us consider the continuum 
hamiltonian that is obtained replacing in Eq. \reff{lham} the 
lattice sums with the corresponding integrals. Then let us perform a scale
transformation \cite{L-B-B-prl}. We define new (``blocked") coordinates 
$y = x/R$ and rescale the fields according to 
\be
\widehat{\phi}_y = R^{d/2} \phi_{R y}, \qquad
\widehat{h}_{y} = R^{d/2} h_{R y}.
\ee
The rescaled Hamiltonian becomes 
\be
  \widehat{H}\; =\; 
     \int d^d y_1 \, d^dy_2\, \smfrac{1}{2}\widehat{J}(y_1-y_2)
  \left(\widehat{\phi}_{y_1} - \widehat{\phi}_{y_2}\right)^2
 +\int d^dy\, \left[ \smfrac{1}{2}  r\widehat{\phi}_y^2 +
{1\over 4!} {u\over R^d}\, \widehat{\phi}_y^4 - 
    \widehat{h}_y\cdot \widehat{\phi}_y\right] ,
\label{lham1}
\ee
where now the coupling $\widehat{J}(x)$ is of short-range type in the limit 
$R\to\infty$.  Being short-ranged, we can apply the 
previous arguments and define Ginzburg parameters: 
\begin{eqnarray}
&& \hskip -0.7truecm
   G_{\hphantom{h}} =  \left(u R^{-d}\right)^{2/(d-4)} = \;
    u^{2/(d-4)} R^{-2d/(4-d)}, 
\\
&& \hskip -0.7truecm
   G_h = \; R^{-d/2} \left(uR^{-d}\right)^{(d+2)/[2(d-4)]} 
    = \; u^{(d+2)/[2(d-4)]}\, R^{-3d/(4-d)}.
\end{eqnarray}
Therefore, in the medium-range model, the critical crossover limit can 
be defined as $R\to\infty$, $t,h\to 0$, with
$\widetilde{t}\equiv t/G$,
$\widetilde{h}\equiv t/G_h$ fixed. 
The variables that are kept fixed are the same, but a different mechanism 
is responsible for the change of the Ginzburg parameters:
in short-range models we vary $u$ keeping the range $R$ fixed and finite,
while here we keep the interaction strength $u$ fixed and vary the 
range $R$.

In this paper we will study a generalization of the model \reff{lham} 
in the presence of medium-range interactions. We will show explicitly in
perturbation theory the 
equivalence between the crossover functions computed starting from 
the continuum $\phi^4$ model and the results obtained in the 
medium-range model. As a byproduct we will also compute the non-universal
constants relating the two cases so that 
we can compare the field-theory predictions with the numerical 
results of Refs. \cite{L-B-B-pre,L-B-B-prl,Luijten-Binder_98,Luijten-FSS} 
without any free parameters. The calculation will also give us
analytic predictions for the large-$R$ behaviour 
of the critical point. 

In numerical simulations (and also in experiments) the range $R$ is always 
finite. It is therefore important to understand the behaviour of the 
corrections
one should expect. We will show that for $d>2$ the deviations from the 
universal behaviour are of order $R^{-d}$, provided one chooses the 
scale $R$ as in Eq. \reff{defR}. These corrections are non-universal 
and depend on all the details of the microscopic interaction: as a 
consequence they cannot be computed in the continuum field-theory 
framework. In two dimensions the behaviour of the corrections is 
not computable in the perturbative expansion around the mean-field 
solution. Indeed the perturbative limit we consider --- first we expand 
in $1/R$ at $t$ fixed and positive and then we take the limit 
$t\to 0$ --- does not commute with the crossover limit. We conjecture 
that the corrections are of order $\log R^2/R^2$ as already indicated 
by the numerical work of Ref. \cite{L-B-B-prl}. This behaviour has been
explicitly checked in the large-$N$ limit.

The paper is organized as follows. In Sect. \ref{sec2} we review the
computation of the crossover functions in the field-theory framework,
extending the calculations of Refs.
\cite{Bagnuls-Bervillier_84,Bagnuls-Bervillier_85,%
Bagnuls-etal_87}.
In Sect. \ref{sec3} we introduce our model with medium-range interactions and 
in Sect. \ref{sec4} we show that the expansion around the mean-field 
solution is equivalent to the perturbative expansion of the 
$\phi^4$ field theory apart from non-universal computable renormalization
constants. In Sec. \ref{sec5} we discuss the corrections to the universal 
critical behaviour. In Sect. \ref{sec6} we compare our theoretical results 
with the Monte Carlo data of Refs. 
\cite{L-B-B-pre,L-B-B-prl,Luijten-Binder_98,Luijten-FSS} and we present
a critical discussion of the crossover model of Refs.
\cite{Chen-etal_90,Luijten-Maryland}.
A detailed comparison with Monte Carlo results for the 
self-avoiding walk will appear in a separate paper
\cite{nosotros-preparation}.
App. \ref{AppA} contains 
some details about the computation of integrals with medium-range propagators,
while App. \ref{AppB} discusses the medium-range model in the large-$N$
limit verifying explicitly various results presented in the text.

Preliminary results were presented in Ref. 
\cite{CCPRV_Lattice98}.

\section{Critical crossover functions from field theory} \label{sec2}

In this Section we report the computation of the various crossover
functions in the continuum theory. As we described in the introduction,
the idea is the following: consider the
continuum $\phi^4$ theory
\be
H = \int d^d x\left[ {1\over2} (\partial_\mu \phi)^2 +
                     {r\over2} \phi^2 +
                     {u\over 4!} \phi^4 \right],
\ee
where $\phi$ is an $N$-dimensional vector, and consider the limit
$u\to 0$, $t\equiv r - r_c\to 0$, 
with $\widetilde{t}\equiv t/G = t u^{-2/(4-d)}$
fixed.  In this limit we have
\bea
\widetilde{\chi}  &\equiv& \chi\, G \to\, F_\chi (\widetilde{t}), \\
\widetilde{\xi}^2 &\equiv& \xi^2\, G \to\, F_\xi (\widetilde{t}),
\eea
where 
\bea 
 \chi &=& \sum_x \< \phi_0 \cdot \phi_x\>,  \\
 \xi^2 &=& {1\over 2 d \chi} \sum_x x^2 \< \phi_0 \cdot \phi_x\>
\eea
are respectively the susceptibility and the (second-moment) 
correlation length. The functions $F_\chi (\widetilde{t})$ and
$F_\xi (\widetilde{t})$ can be accurately computed by means of perturbative
field-theory calculations.
There are essentially two methods: (a) the
fixed-dimension expansion \cite{Parisi_Cargese,Bagnuls-Bervillier_84,%
Bagnuls-Bervillier_85}, which is at present the most precise
one since seven-loop series are available
\cite{Baker-etal_77_78,Murray-Nickel_91};
(b) the so-called minimal renormalization without $\epsilon$-expansion
\cite{Dohm,Schloms-Dohm_89,K-S-D} which uses five-loop
$\epsilon$-expansion results \cite{C-G-L-T,K-N-S-C-L}. In these two schemes
the crossover functions are expressed in terms of various
RG functions whose perturbative series can be resummed with high accuracy
using standard methods \cite{LeGuillou-ZinnJustin_80,ZinnJustin_book}. 
Here we will consider the first approach
although essentially equivalent results can be obtained using the second
method. Explicitly we have for
$F_\chi(\widetilde{t})$ and $F_\xi(\widetilde{t})$:
\begin{eqnarray}
F_\chi(\widetilde{t}) &=& \chi^* 
   \, \exp\left[ - \int_{y_0}^g dx\, {\gamma(x)\over \nu(x) W(x)}\right], 
\label{Fchi-FT-fixedd}\\
F_\xi(\widetilde{t}) &=& \left(\xi^*\right)^2
   \, \exp\left[ - 2 \int_{y_0}^g dx\, {1\over W(x)}\right], 
\label{Fxi-FT-fixedd}
\end{eqnarray}
where $\widetilde{t}$ is related to the zero-momentum four-point
renormalized coupling $g$ by
\be
\widetilde{t} \, =\, 
  - t_0 \, \int^{g^*}_g dx\, 
 {\gamma(x)\over \nu(x) W(x)} 
  \exp\left[ \int_{y_0}^x dz\, {1\over \nu(z) W(z)} \right],
\label{ttilde-FT-fixedd}
\ee
$\gamma(x)$, $\nu(x)$, and $W(x)$ are the standard RG
functions (see Refs. \cite{Baker-etal_77_78,Murray-Nickel_91} for the 
corresponding perturbative expressions), 
$g^*$ is the critical value\footnote{A review of present estimates
of $g^*$ can be found in Refs. 
\cite{ZinnJustin_book,PV_gstar,Butera-Comi_98,Guida-ZinnJustin_98}.}
of $g$ defined by
$W(g^*) = 0$, and $\chi^*$, $\xi^*$, $t_0$ and $y_0$ are normalization
constants. 

The expressions \reff{Fchi-FT-fixedd}, \reff{Fxi-FT-fixedd} and 
\reff{ttilde-FT-fixedd} are valid for any dimension $d<4$. The first
two equations are always well defined, while Eq. \reff{ttilde-FT-fixedd}
has been obtained with the additional hypothesis that the integral
over $x$ is convergent when the integration is extended up to 
$g^*$. This hypothesis is verified when the system becomes critical 
at a finite value of $\beta$ and shows a standard critical behaviour. 
Therefore Eq. \reff{ttilde-FT-fixedd} is always well defined for 
$d>2$, and, in two dimensions, for $N\le2$. For $N> 2$, one can still
define $\widetilde{t}$ by integrating up to an arbitrary point
$g_0$. For these values of $N$, $\widetilde{t}$ varies between 
$-\infty$ and $+\infty$.

We will normalize the coupling $g$ as in Refs.
\cite{Baker-etal_77_78,LeGuillou-ZinnJustin_80,Murray-Nickel_91} 
so that in the perturbative
limit $g\to 0$, $\widetilde{t} \to \infty$, we have
\be
g \approx \, 
  {1\over 2 (4\pi)^{d/2}} {N+8\over3} \Gamma\left(2 - {d\over2}\right)\,
    \widetilde{t\, }^{(d-4)/2} \, \equiv 
    \lambda_d \widetilde{t\, }^{(d-4)/2}
\ee
This implies that for $y_0\to 0$ we have 
$t_0 \approx (y_0/\lambda_d)^{2/(d-4)}$, and 
$(\xi^*)^2 t_0 \approx \chi^* t_0 \approx 1$. 

For future purposes we will be interested in computing the expansion of 
$F_\chi(\widetilde{t})$ for $\widetilde{t}\to \infty$. 
In two dimensions we have
\be
F_\chi(\widetilde{t}) = \,
 {1\over \widetilde{t}} \left[1 +
    {N+2\over 24 \pi \widetilde{t}} 
      \log\left( {24 \pi \widetilde{t}\over N+8}\right) + 
    {N+8\over 24 \pi \widetilde{t}} + {D_2(N)\over \widetilde{t}} +\,
    O(\widetilde{t}^{-2} \log^2 \widetilde{t})\right],
\label{Fchi-FT-pert-2d}
\ee
where
\bea
D_2(N) &=& {N+8\over 24\pi} 
    \left( - {1\over g^*} + {N+2\over N+8} \log g^*\right) 
\nonumber \\
&& \hskip -0.5truecm 
   - {N+8\over 24\pi} \int_0^{g^*} {dx\over x} 
    \left\{ {\gamma(x)\over \nu(x) W(x)} 
       \exp\left[\int_0^x dz\, \left({1\over \nu(z) W(z)} + {1\over z}\right)
           \right]\, 
      + {1\over x} + {N+2\over N+8}\right\}.
\nonumber \\ 
{} \label{def-D2}
\eea
For $N\ge 2$, $g^*$ should be replaced in $D_2(N)$ with the arbitrary point 
$g_0$ that has been used to define $\widetilde{t}$.

Analogously in three dimensions we have
\bea
F_\chi(\widetilde{t}) &=& 
  {1\over \widetilde{t}} \left[1 +
     {N+2\over 24 \pi \widetilde{t}^{1/2}} - 
     {N+2\over 288 \pi^2 \widetilde{t}} 
           \log\left({48\pi \sqrt{\widetilde{t}}\over N+8}\right)\right.
\nonumber \\
&& \quad \left.
   + {27 N^2 + 52 N - 472\over 20736\pi^2 \widetilde{t}} +
   {D_3(N)\over \widetilde{t}} + O(\widetilde{t}^{-3/2} 
   \log \widetilde{t})\right],
\label{Fchi-FT-pert-3d}
\eea
where
\bea
D_3(N) &=& 
- \left({N+8\over 48\pi}\right)^2\,
    \left[{1\over g^{*\,2}} - {12\over N+8} {1\over g^*} +
     {8(N+2)\over (N+8)^2} \log g^*\right]
\nonumber \\
&& - \left({N+8\over 48\pi}\right)^2\, \int_0^{g^*} {dx\over x^2}
    \left\{ {\gamma(x)\over \nu(x) W(x)}
       \exp\left[\int_0^x dz\, \left({1\over \nu(z) W(z)} + {2\over z}\right)
           \right]\, \right.
\nonumber \\
&& \qquad \left.
      + {2\over x} - {12\over N+8} - {8(N+2)\over (N+8)^2} x \right\}.
\label{D3-def}
\eea
In the large-$N$ limit we have simply
\be
D_3(N) = - {N^2\over (48\pi)^2} + O(N).
\label{D3-largeN}
\ee
The expansions \reff{Fchi-FT-pert-2d} and \reff{Fchi-FT-pert-3d} are 
nothing but the standard perturbative expansions. For generic values of 
$d$ we have
\be
F_\chi(\widetilde{t}) = {1\over \widetilde{t}} 
  \sum_n a_n \widetilde{t}^{-\Delta_{\rm mf}},
\ee
where $\Delta_{\rm mf} = (4-d)/2$. For $d=4-2/n$ additional logarithms 
appear in the expansion. The reason of this phenomenon is well 
known \cite{Symanzik_73,Parisi_79,Bergere-David_82}: the critical 
crossover limit corresponds to the massless limit of the standard 
$\phi^4$ model which is known to have logarithmic singularities for these 
values of the dimension.

The functions $F_\chi(\widetilde{t})$ and $F_\xi(\widetilde{t})$ 
can be computed using the perturbative results of 
Refs. \cite{Baker-etal_77_78,Murray-Nickel_91}, a Borel-Leroy
transform that takes into account the large-order behaviour of the 
perturbative series \cite{Lipatov,B-L-Z}, and a standard resummation technique 
\cite{LeGuillou-ZinnJustin_80}. Explicit expressions can be found 
for $N=1,2,3$ and $d=3$ in Refs. 
\cite{Bagnuls-Bervillier_84,Bagnuls-Bervillier_85}.
Here we compute $F_\chi(\widetilde{t})$ and $F_\xi(\widetilde{t})$ 
for the Ising model in two dimensions using the four-loop
results of Ref. \cite{Baker-etal_77_78}. In order to improve the
precision of the results, by means of appropriate subtractions, 
we have forced the resummed expressions to have the correct 
asymptotic behaviour for $\widetilde{t}\to 0$, i.e.
$F_{\chi}(\widetilde{t})\approx \widetilde{t}^{-\gamma}$, 
$F_{\xi}(\widetilde{t})\approx \widetilde{t}^{-2\nu}$, with 
$\gamma = 7/4$, $\nu = 1$.  The resummed expressions are well fitted by
\bea
F_\chi(\widetilde{t}) &=& 
   {1\over \widetilde{t}}\left[
   1 + {2\log \widetilde{t}\over 3\pi \widetilde{t}} +\, 
   {0.9705\over \widetilde{t}} + {0.3513\over \widetilde{t}^2} +
   {0.01712\over \widetilde{t}^3} +
   {0.001822\over \widetilde{t}^4}\right]^{3/16}, 
\\
F_\xi(\widetilde{t}) &=&
   {1\over \widetilde{t}}\left[
   1 + {\log \widetilde{t}\over 2\pi \widetilde{t}} +\, 
   {0.7377\over \widetilde{t}} + {0.1635\over \widetilde{t}^2} +
   {0.00390\over \widetilde{t}^3} +
   {0.000275\over \widetilde{t}^4}\right]^{1/4}, 
\eea
The resummation errors on $F_\chi(\widetilde{t})$ (resp. 
$F_\xi(\widetilde{t})$) are less than 0.1\% for $\widetilde{t}\gtapprox 0.2$
(resp. $\widetilde{t}\gtapprox 0.1$), at most 3\% (4\%) for smaller values of
$\widetilde{t}$. The fitted expression introduces an additional uncertainty
which is less than 0.3\%.

The constants $D_d(N)$ are non-perturbative constants since they
require the knowledge of the RG functions up to the critical point.
By means of a Borel-Leroy transform, working as before, we obtain in
three dimensions
\bea
D_3(0) &=& 0.002473 (6), \\
D_3(1) &=& 0.002391 (6), \\
D_3(2) &=& 0.002204 (5), \\
D_3(3) &=& 0.001920 (3).
\eea
In two dimensions we have
\be
D_2(1) = - 0.0524 (2).
\ee
The results we have described above apply to the high-temperature phase of the
model. For $N=1$ the critical crossover can also be defined in the 
low-temperature phase \cite{Bagnuls-etal_87} and the crossover 
functions can be computed in terms of (resummed) perturbative 
quantities. Using the perturbative results\footnote{Some perturbative
series contained small errors, see footnote 27 of Ref. 
\cite{PRV_longrange}.} of Ref. \cite{Bagnuls-etal_87}, we can compute 
$F_\chi(\widetilde{t})$ in the low-temperature 
phase of the three-dimensional Ising model. For $|\widetilde{t}\, | > 10^{-3}$ 
the numerical results are well fitted by
\bea
F_\chi(\widetilde{t}) &=&
   {1\over 2|\widetilde{t}|} \left(1 + 
           {0.00019\over |\widetilde{t}|}\right)^{0.2372}
\nonumber \\
&& \hskip -1.5truecm 
    \times \left[
   1 - {0.0140674\over \widetilde{t}^{1/2}} + 
       {0.0021871\over \widetilde{t}} -
       {0.000150048\over \widetilde{t}^{3/2}} + 
       {4.82764 \cdot 10^{-6}\over \widetilde{t}^{2}} - 
       {5.78079 \cdot 10^{-8} \over \widetilde{t}^{5/2}}\right].
\nonumber \\
\label{Fchi-LT}
{}
\eea
The error due to the approximate form 
given above is at most 0.3\%, while the resummation error is 
less than 0.1\%. For $|\widetilde{t}| \ltapprox 10^{-3}$ we can use 
\be 
F_\chi(\widetilde{t}) = |\widetilde{t}|^{-1.2372} 
  \left(0.06125 + 1.3455 \widetilde{t}^{1/2} + 
        1.51525  \widetilde{t} - 124.395 \widetilde{t}^{3/2}\right),
\ee 
with errors of order 0.2\%. The resummation error varies  approximately
from 0.1\% to 3\%.

In the low-temperature phase we can also study the magnetization. 
Using the results of Ref. \cite{Bagnuls-etal_87} 
we have  
\bea
F_{M}(\widetilde{t}) &\equiv& \<\sigma\> u^{-1/2} = \,
    \sqrt{3 |\widetilde{t}|} 
\nonumber \\
&&\times \left(1 + {0.3241\over \widetilde{t}^{1/2} }
    + {0.02751\over \widetilde{t}} + 
      {0.001247\over \widetilde{t}^{3/2}} +
      {0.0000128\over \widetilde{t}^{2}} \right)^{0.0868}\! .
\label{FM-LT}
\eea
For $|\widetilde{t}|> 10^{-4}$, the error on this function is at most of 
order 0.5\%.

We have chosen the expressions \reff{Fchi-LT} and \reff{FM-LT} so that
they reproduce the exact large-$\widetilde{t}$ behaviour of the 
crossover functions:
\bea
F_{\chi}(\widetilde{t}) &=& {1\over 2 |\widetilde{t}|}\, 
   \left(1 - {1\over 16\pi\sqrt{2}} {1\over \widetilde{t}^{1/2}} + 
      O(\widetilde{t}^{-1} \log \widetilde{t}) \right),
\\
F_M (\widetilde{t}) &=& \sqrt{3 |\widetilde{t}|}\,
   \left(1 + {1\over 8\pi\sqrt{2}} {1\over \widetilde{t}^{1/2}} + 
    O(\widetilde{t}^{-1} \log \widetilde{t}) \right).
\eea
Notice that the leading correction to $F_{\chi}(\widetilde{t})$ is 
negative so that $F_{\chi}(\widetilde{t})$ is non-monotonic.
Such a behaviour has also been observed numerically
\cite{L-B-B-prl} and predicted analytically \cite{PRV_longrange} 
in two dimensions.

\section{The models and the critical crossover limit} \label{sec3}

In this Section we will study the critical 
crossover limit in the presence of medium-range interactions 
by performing a systematic expansion around the 
mean-field solution. In particular we will show how to compute the critical
crossover functions as a perturbative expansion in powers 
of $\widetilde{t}^{(d-4)/2}$. 
A general discussion of the mean-field limit
can be found e.g. in Refs. \cite{Brezin-etal_DG6,ZinnJustin_book}.

Generalizing the discussion of the introduction, we assume we have  
a family of couplings $J_\rho(x)$ that are defined on a 
$d$-dimensional cubic lattice and that are parametrized by $\rho$.
For each coupling $J_\rho(x)$, we define,
in analogy with Eq. \reff{defR}, 
the following quantities:
\begin{eqnarray}
V_\rho &\equiv& \sum_x J_\rho(x), \\
R^2 &\equiv& {1\over 2 d V_R} \sum_x x^2 J_\rho(x),
\label{eq3.2}
\end{eqnarray}
and
\be
\overline{\Pi}_\rho(q) \equiv 1 - {1\over V_\rho} J_\rho(q),
\ee
where $J_\rho(q)$ is the Fourier transform of $J_\rho(x)$. In the following
we assume that there is a one-to-one correspondence between
$R$ and $\rho$, so that we will interchangebly think of the various observables
as functions either of $\rho$ or of $R$.

Notice that because of the definition \reff{eq3.2} we have for $q\to 0$
\be
\overline{\Pi}_R(q) \approx R^2 q^2.
\label{Pibar-smallq2}
\ee
We assume that:
\begin{itemize}
\item[(i)] $J_R(x)$ is uniformly bounded in $x$ and $R$,
i.e. $|J_R(x)|<C$, independently of $x$ and $R$;
\item[(ii)] $V_R$ and $R^2$ are finite for finite values of $\rho$. For 
$\rho\to\infty$, $R^2\to+\infty$ and $V_R\sim R^d$;
\item[(iii)] the system is ferromagnetic, i.e. 
$\overline{\Pi}_R(q) > 0$ for all $q\not=0$ in the first Brillouin zone;
\item[(iv)] $\overline{\Pi}_R(q/R)$ has a finite limit for $R\to\infty$
at fixed $q$, i.e. $\overline{\Pi}_R(q/R)\to \Pi(q)$. This assumption is 
equivalent to requiring that $J_R(Rx)$ has a finite limit 
$J_{\infty}(x)$ for $R\to\infty$ at fixed $x$. Furthermore we assume
that $J_{\infty}(x)$ is infinitely differentiable in $x=0$
so that the integral
$\int d^dq\, q^{2n} (1-\Pi(q))$ exists for any $n$. Notice the following 
asymptotic property of $\Pi(q)$: because of Eq. \reff{Pibar-smallq2}
we have $\Pi(q)\approx q^2$ for $q\to 0$.
\end{itemize}
We will investigate the properties of 
a class of models that generalize the theory
defined in Eq. \reff{lham}. We will consider a Hamiltonian of
the form
\be
H = - {N\over2} \sum_{x_1,x_2} J_R(x_1 - x_2) \,
     \varphi_{x_1} \cdot \varphi_{x_2} +
      N \sum_x {h}_x\cdot {\varphi}_x ,
\label{Hsigma}
\ee
where $\varphi_x$ are $N$-dimensional vectors, and a single-site measure 
$d\mu(\varphi_x)$ which we will assume of the form
\be
d\mu(\varphi_x)=\, e^{- V(\varphi_x)} d^N\varphi_x,
\label{dmuexpmV}
\ee
where $V(\varphi_x)$ is an even function (often a polynomial) 
of $\varphi_x$ which is bounded from below and that satisfies
$V(x)\sim |x|^p$, $p > 2$, for $|x|\to\infty$. 
The partition function is defined by
\be
Z[h] \, \equiv \, \int \prod_x d\mu(\varphi_x) e^{-\beta H}.
\label{partition-function}
\ee
The Hamiltonian we defined in the introduction is a particular case 
of the general model we discuss here. Indeed consider a family
of domains $D_R$ and define 
\be
J_R(x)\, \equiv\, \cases{1 & \hskip 1truecm if $x\in D_R$, \cr
                  0 & \hskip 1truecm otherwise,}
\label{Jrhoflat}
\ee
and 
\be
V(\varphi_x) = \, \varphi_x^2 +
	\lambda \left(\varphi_x^2 - 1\right)^2,
\ee
with $\lambda>0$. It is easy to see that $J_R(x)$ satisfies all the 
assumptions, as long as $D_R$ satisfies some simple requirements, 
see App. \ref{AppA.1}.  
To derive the relation between the two models (for simplicity assume
$h_x = 0$) let us rewrite the partition
function as 
\be
Z[h] =\, \int \prod_x d\varphi_x\, 
   \exp\left\{ {\beta N\over 2} \sum_{xy}
   J_R(x-y)\, \varphi_x\cdot \varphi_y - 
   \sum_x \left[\varphi^2_x + \lambda (\varphi^2_x - 1)^2 
       \right]\right\}.
\label{Z-phi4}
\ee
Then we rescale the field 
\be
\varphi_x = \left({\beta N V_R\over 2}\right)^{-1/2}  \phi_x
\ee
and define
\be
r \equiv {4(1 - 2\lambda)\over \beta N V_R} - 2, \qquad
u \equiv {96\lambda\over (\beta N V_R)^2}.
\ee
In terms of $\phi$ we obtain again Eq. \reff{lham} with the coupling
defined by Eq. \reff{defJ}. As we shall see in the following,
the large-$R$ limit is well defined if $\beta$ goes to zero as 
$1/V_R$. Thus $r$ and $u$ remain finite as $R\to\infty$ so that the large-$R$
limit of the Hamiltonian \reff{lham} and of the model defined in this Section 
are identical.

For $\lambda\to\infty$, the model with partition function \reff{Z-phi4}
reduces to the so-called 
$N$-vector model which can be seen as a particular case in which
\be
d\mu(\varphi_x)=\, \delta(\varphi^2_x-1) d^N\varphi_x.
\ee

The coupling $J_R(x)$ defined in Eq. \reff{Hsigma} is ambiguous for $x=0$.
Indeed one can define a new coupling and a new single-site measure 
\begin{eqnarray}
\widehat{J}_R(x) &\equiv& J_R(x) - K \delta_{x,0},    \\
d\widehat{\mu} (\varphi_x) &\equiv& d\mu(\varphi_x) e^{N \beta K \varphi_x^2/2},
\end{eqnarray}
without changing the results. 
This step, which at first sight may appear
trivial, can be interpreted as a mass renormalization needed, as we shall see,
to have a scaling theory. 
It is also needed from a mathematical point of view: in order to make the 
following derivation mathematically rigorous, we will require $K$ to be such
that the Fourier transform $\widehat{J}_R(q)$ satisfies 
$\widehat{J}_R(q)>0$ for all values of $q$.

Let us now discuss the critical limit of the model.
If we consider the critical limit with
$R$ fixed, Eq. \reff{partition-function}
defines a generalized $O(N)$-symmetric
model with short-range interactions. If $d>2$, 
for each value of $R$ there is a critical point\footnote{In two dimensions a 
critical point exists only for $N\le 2$. Theories with $N\ge 3$ are 
asymptotically free and become critical only in the limit $\beta\to \infty$.}
$\beta_{c,R}$; for $\beta\to \beta_{c,R}$ the susceptibility and the 
correlation length have the standard behaviour
\begin{eqnarray}
\chi_R(\beta) &\approx& A_\chi(R) t^{-\gamma} (1 + B_\chi(R) t^\Delta) ,
\label{chiRfisso} \\
\xi_R^2(\beta) &\approx& A_\xi(R) t^{-2\nu} (1 + B_\xi (R) t^\Delta),
\end{eqnarray}
where $t \equiv (\beta_{c,R} - \beta)/\beta_{c,R}$ and we have neglected 
additional subleading corrections.
The exponents $\gamma$, $\nu$ and $\Delta$ do not depend on $R$. 
On the other hand, the 
amplitudes are non-universal\footnote{However some ratios of 
correction-to-scaling
amplitudes are universal. For instance $B_\xi (R)/B_\chi(R)$ is 
universal and therefore independent of $R$.}.
For $R\to\infty$, they behave as \cite{L-B-B-pre,L-B-B-prl}
\begin{eqnarray} 
A_\chi(R) &\approx &   A_\chi^\infty R^{2 d(1-\gamma)/(4 - d)} ,\qquad
A_\xi(R) \approx A_\xi^\infty   R^{4 (2 - d\nu)/(4-d)},  \nonumber \\ 
B_\chi(R) &\approx &   B_\chi^\infty R^{2 d \Delta/(4 - d)} ,\qquad \qquad
B_\xi(R) \approx  B_\xi^\infty  R^{2 d \Delta /(4-d)} .
\label{eq3.4}
\end{eqnarray}
The critical point $\beta_{c,R}$ depends explicitly on $R$. For 
$R\to\infty$ we have $\beta_{c,R}\sim 1/R^d$ with corrections that 
will be computed in the next section.

Let us now define the critical crossover limit. 
In this case we consider the limit \cite{L-B-B-pre,L-B-B-prl}
$R\to\infty$, $t\to0$, with 
$R^{2d/(4-d)}t\equiv \widetilde{t}$ fixed.
We will show perturbatively in the next section that
\begin{eqnarray}
\widetilde{\chi}_R \equiv 
         R^{-2d/(4-d)} \chi_R(\beta) &\to& f_\chi(\widetilde{t}) ,
\label{fchi} \\
\widetilde{\xi}^2_R \equiv 
R^{-8/(4-d)} \xi^2_R(\beta) &\to& f_\xi(\widetilde{t}) ,
\label{fxi}
\end{eqnarray}
where the functions $f_\chi(\widetilde{t})$ and 
$f_\xi(\widetilde{t})$ are universal apart from an overall 
rescaling of $\widetilde{t}$ and a constant factor, in agreement with the
argument presented in the introduction. 

There exists an equivalent way to define the 
crossover limit which is due to Thouless \cite{Thouless_69}. 
Let $\beta^{\rm (exp)}_{c,R}$ be the expansion of $\beta_{c,R}$ for 
$R\to\infty$ up to terms of order $R^{-2d/(4-d)}/V_R$, i.e. such that
\be
\lim_{R\to\infty} R^{2d/(4-d)} \beta_{c,R}^{-1}\, 
     (\beta_{c,R} - \beta^{\rm (exp)}_{c,R}) = 
      b_c,
\label{def-bc}
\ee
with $|b_c|<+\infty$. Then introduce 
\be
\widehat{t} =\, R^{2d/(4-d)} \beta_{c,R}^{{\rm (exp)}\,-1}\, 
              (\beta^{\rm (exp)}_{c,R} - \beta).
\label{def-that}
\ee
It is trivial to see that in the standard crossover limit 
$\widetilde{t} = \widehat{t} + b_c$. Therefore the crossover limit 
can be defined considering the limit $R\to\infty$, 
$\beta\to\beta^{\rm (exp)}_{c,R}$ with $\widehat{t}$ fixed. The crossover 
functions will be identical to the previous ones apart from a 
shift. Thouless' definition of critical crossover has an important 
advantage. It allows the definition of the critical 
crossover limit in models that do not have a critical point for finite values
of $R$: indeed, even if $\beta_{c,R}$ does not exist, one can define 
a quantity $\beta^{\rm (exp)}_{c,R}$ and a variable $\widehat{t}$ such that 
the limit $R\to\infty$ with $\widehat{t}$ fixed exists\footnote{This is the
case of two-dimensional models with $N\ge 3$, see the discussion in 
Sec. \ref{sec4.2}, and of one-dimensional models with $N\ge 1$. 
In the latter case we can take 
$\beta^{\rm (exp)}_{c,R}=1/(\overline{a}_2 V_R)$ where $\overline{a}_2$
is defined in Sec. \ref{sec4.1}.}. 
Moreover, as we shall see,
$\beta^{\rm (exp)}_{c,R}$ is known analytically: therefore, in the 
analysis of Monte Carlo data, no
computation of $\beta_{c,R}$ is needed and a source of errors is eliminated.

\section{Mean-field perturbative expansion} \label{sec4}

\subsection{General framework} \label{sec4.1}

The starting point of our expansion is the identity 
\cite{Baker_62,Hubbard_72}
--- we use matrix notation and drop the subscript $R$ from $J_R(x)$ to 
simplify the notation --- 
\begin{equation}
\exp\left[ {N\beta\over 2} \varphi \widehat{J} \varphi \right]
=\left( {\rm det} N\beta \widehat{J}\right)^{-N/2}
\int {d^N\phi\over (2\pi)^{N/2}}
\exp \left\{ N \left[ -{1\over 2\beta}
\phi \widehat{J}^{-1} \phi + \phi \varphi\right] \right\},
\label{fmf}
\end{equation}
where $\phi$ is another  $N$-dimensional vector field.
The second ingredient is the single-site integral that defines the 
function $A(\phi)$:
\begin{equation}
z e^{N A(\phi)} \equiv 
\int d\widehat{\mu} (\varphi) \;e^{N\phi \varphi} ,
\label{Adef1}
\ee
where $z$ is a normalization factor ensuring $A(0)= 0$. 
If we choose the single-site measure given in
Eq. \reff{dmuexpmV}, we obtain the explicit formula
\be
z e^{N A(\phi)} =\,
2 \pi^{N/2} \int_0^\infty dx\, x^{N-1} e^{-V(x) + N \beta K x^2/2}\, 
\left( {2\over Nx|\phi|}\right)^{N/2-1}
I_{N/2-1}(Nx|\phi|).
\label{Adef}
\end{equation}
Notice that Eq. \reff{Adef} has a regular expansion in 
powers of $\phi^2$, giving finally 
\begin{equation}
A(\phi) = \sum_{k=1}^\infty {a_{2k}\over (2k)!} (\phi^{2})^k.
\label{Aexp}
\ee
The first few coefficients are given by
\begin{eqnarray}
&& \hskip -1truecm
a_2 =\, f_2, \\[2mm]
&& \hskip -1truecm
a_4 =\, {3 N\over N+2} \left[N f_4 - (N+2) f_2^2\right], \\
&& \hskip -1truecm
a_6 =\, {15 N^2\over (N+2)(N+4)}
      \left[ N^2 f_6 - 3 N (N+4) f_4 f_2 + 2 (N+2)(N+4) f^3_2\right],
\end{eqnarray}
where
\be
f_k =\, {\int_0^\infty dx \, x^{N+k-1} e^{-V(x) + N\beta K x^2/2} \over 
         \int_0^\infty dx \, x^{N-1} e^{-V(x) + N\beta K x^2/2}}.
\ee
The results for the $N$-vector model are obtained setting $f_k=1$ in the 
previous formulae.

The coefficients $a_{2k}$ depend  on the various parameters
that appear in the single-site measure, and on $\beta K$. 
In the following we will assume $V(x)$ to be independent of $R$, although
considering $R$-dependent potentials does not introduce any significant
change in the discussion as long as $\lim_{R\to\infty} V(x)$ exists and 
is finite. Under this assumption $a_{2k}$ depends on $R$ only through
the combination $\beta K$.
In the following we will always consider 
the limit $R\to\infty$ with $\beta K\to 0$, and therefore we will 
introduce
\be
\overline{a}_{2k} = \lim_{\beta K\to 0} a_{2k}.
\ee
We will discuss the generic case\footnote{Formally the discussion we will
present requires only $\overline{a}_4\not=0$. However for 
$\overline{a}_4>0$ the expansion we obtain would correspond to a 
$\phi^4$ model with negative coupling. We therefore expect that 
potentials $V(x)$ such that $\overline{a}_4>0$ correspond to 
non-critical models. This can be checked explicitly in the 
large-$N$ limit for a potential containing a $\phi^4$ and a 
$\phi^6$ coupling. In the large-$N$ limit one can check explicitly that 
$\overline{a}_4<0$ is a necessary condition in order 
to obtain the critical crossover limit, see
Sec. \ref{AppB.2}.}
in which $\overline{a}_4<0$. If one tunes the parameters appropriately 
one can obtain $\overline{a}_4=0$ and a different critical limit. We will not 
consider these cases here. For 
a discussion in the large-$N$ limit, see App. \ref{AppB.2}.
In our expansion around the mean-field solution we will need 
the expansion of $a_2$ in powers of $\beta K$. Explicitly we have:
\bea
a_2 &=& \overline{a}_2 + 
    \beta K\left[ {N+2\over 6N} \overline{a}_4 + \overline{a}_2^2\right]
\nonumber \\
    && + \beta^2 K^2 \left[ {(N+2)(N+4)\over 120 N^2} \overline{a}_6
   + {N+2\over 2N}\overline{a}_4 \overline{a}_2 +
     \overline{a}_2^3\right] + O(\beta^3 K^3).
\label{a2_expansion}
\eea
Using Eqs. \reff{fmf} and \reff{Adef1} we have
\begin{equation}
Z[h] \propto
\int \prod_x d\phi_x\,  \exp \left\{{N}  \left[-{1\over 2\beta}
\phi \widehat{J}^{-1}\phi + \sum_x A(\phi_x+h_x)\right] \right\}.
\label{pf}
\end{equation}
The correlation functions of the $\varphi$-fields are obtained by
taking derivatives with respect to $h_x$. Using the equations of motion,
it is easy to relate them to correlations of the $\phi$-fields. For
instance, for $h_x=0$, we have
\be
\<\varphi_x\cdot\varphi_y\>\, =\, 
 - {1\over\beta}\left( \widehat{J}^{-1}\right)_{xy} + 
   {1\over\beta^2} \sum_{wz} 
   \left( \widehat{J}^{-1}\right)_{xw} 
   \left( \widehat{J}^{-1}\right)_{yz} 
   \< \phi_w \cdot \phi_z \>,
\ee
where the expectation value $\<\varphi_x\cdot\varphi_y\>$ 
is obtained using the 
Hamiltonian \reff{Hsigma}, while $\< \phi_w \cdot \phi_z \>$ 
is computed in the model \reff{pf}. 

Now let us consider the 
(formal) perturbative expansion of the theory \reff{pf} with 
$h_x = 0$. It corresponds 
to a scalar model with propagator given by
\be
\widehat{\Delta}(q) = {1\over N}
    {\beta \widehat{J}(q) \over 1 - a_2 \beta \widehat{J}(q)}
\ee
and vertices $\phi^4$, $\phi^6$, ..., that can be read from the 
expansion of the function $A(\phi)$, see Eq. \reff{Aexp}. 
Let us now define
\be
\overline{t} \equiv {1\over a_2 V_R \beta} - 1 + {K\over V_R},
\label{tbar-def}
\ee
and consider the limit
$\beta\to 0$, $R\to\infty$ with $K$ and $\overline{t}$ fixed.
If $\overline{t}>0$ this formal perturbative expansion defines
an expansion in powers of $R^{-d}$. 
To prove this result let us first rewrite the propagator as
\be
\widehat{\Delta}(q) \, =\, 
   {1\over N} \left[
   - {\beta K\over 1 + a_2 \beta K} + \, 
     {1\over a_2} {1\over 1 + a_2 \beta K} 
     {1 - \overline{\Pi}(q)\over \overline{\Pi}(q) + \overline{t}}
     \right]\, =\, \widehat{\Delta}_1 + \widehat{\Delta}_2(q).
\label{sommawidehatDelta}
\ee
Now let us consider a generic $l$-loop graph. It has the generic form
\be
\left(\prod_i a_{2k_i}\right) Q(N) \int \prod_{i<j} {dq_{ij}\over (2\pi)^d}\,
    \prod_{i<j} \widehat{\Delta}(q_{ij}) \, \prod_i (2\pi)^d
    \delta\left(\sum_j q_{ij}\right),
\label{genericgraph}
\ee
where $Q(N)$ is an $N$-dependent constant. Let us now expand 
$\widehat{\Delta}(q)$ in the graph using Eq. \reff{sommawidehatDelta}. 
We obtain 
a sum of terms that can be represented as graphs in which each line is 
associated to a propagator $\widehat{\Delta}_2(q)$; these graphs are 
obtained from the original one contracting the lines corresponding to
$\widehat{\Delta}_1$. A generic term has the form 
\be
(\widehat{\Delta}_1)^n\, \int \prod_{i<j} {dq_{ij}\over (2\pi)^d}\,
    \prod_{i<j} \widehat{\Delta}_2(q_{ij}) \, \prod_i (2\pi)^d
    \delta\left(\sum_j q_{ij}\right),
\label{eq:C2x}
\ee
corresponding to an $m$-loop subgraph.
Now notice that, because
of the assumptions we have made at the beginning, 
$\widehat{\Delta}_2(q/R)$ converges, 
for $R\to\infty$ at fixed $q$, to a function ${\Delta}_2(q)$ given by
\be
\Delta_2 (q) = {1\over a_2 N} {1-\Pi(q)\over \Pi(q) + \overline{t}},
\ee
that is integrable for all positive $\overline{t}$. 
Then change variables in Eq. \reff{eq:C2x}, 
setting $q_{ij} = p_{ij} R$, and then
take the limit $R\to\infty$ in the integrand, keeping 
$\overline{t}$ fixed. We obtain
\be
{(\widehat{\Delta}_1)^n\over R^{dm}}\, 
\int \prod_{i<j} {dq_{ij}\over (2\pi)^d}\, 
    \prod_{i<j} {\Delta}_2(q_{ij}) \, \prod_i (2\pi)^d 
    \delta\left(\sum_j q_{ij}\right),
\label{genericgraph-hat}
\ee
where the integration is extended over $\R^{dm}$. The integral is 
$R$-independent, and,
as long as $\overline{t}$ is positive, it is finite.
Since $\widehat{\Delta}_1\sim R^{-d}$, 
the leading contribution of this graph behaves as $R^{-d (m + n)}$. 
Obviously $m+n\ge l$ [the equality corresponds to those cases in which 
all lines with $\widehat{\Delta}_1$ belong to one-loop tadpoles], 
thus proving that 
the leading contribution of an $l$-loop graph behaves as $R^{-d l}$.

It is immediate to see that the perturbative expansion is not uniform as 
$\overline{t}\to 0$. Indeed for $d\le 4$ infrared divergences appear,
so that the coefficients of the expansion diverge as $\overline{t}\to 0$.
If one expands these coefficients in the limit $\overline{t}\to 0$ 
and chooses $K$ appropriately --- this corresponds to a mass renormalization
and fixes the expression of $\beta^{\rm (exp)}_{c,R}$ --- 
one obtains a new series which is a perturbative expansion in 
powers of $\widetilde{t}^{(d-4)/2}$ with additional logarithms if 
$d = 4 - 2/n$, $n\in \N$. The resulting expression can then be interpreted 
as the expansion of the critical crossover functions for 
$\widetilde{t}\to\infty$, {\em provided} that the limit 
$R\to\infty$ at $\overline{t}$ fixed 
followed by $\overline{t}\to 0$ is identical to the 
crossover limit. It is important to stress that this commutativity 
is not an obvious fact and indeed it is not true at the level of the 
corrections to the universal behaviour. In the following we will also show
that graphs containing 6-leg (or higher-order) vertices can be neglected 
in the critical crossover limit. In other words one can simply consider 
the $\phi^4$ theory obtained from Eq. \reff{pf}. This will explicitly
give the relation between the crossover functions computed in the medium-range
model and those obtained in the field-theory framework of Sect. \ref{sec2}.

We will first discuss the two-dimensional case, in which all these features 
can be understood easily, then we will present the general case.

\subsection{Two-dimensional crossover limit} \label{sec4.2}

We wish now to discuss the critical crossover limit in two dimensions 
using the perturbative expansion
of the model \reff{pf}. Let us consider the zero-momentum correlation function 
\be
G^{(m,n)} \equiv \sum_{x_2,\ldots,x_m,y_1,\ldots,y_n}
\< (\phi^2)_{x_1} (\phi^2)_{x_2}\cdots (\phi^2)_{x_m} 
    \phi_{y_1} \cdots \phi_{y_n}\>,
\label{defGmn}
\ee
which contains $m$ insertions of $\phi^2$ and $n$ fields $\phi$, and its 
one-particle irreducible counterpart $\Gamma^{(m,n)}$.

Let us consider a generic $l$-loop graph contributing to 
$\Gamma^{(m,n)}$, and, to begin with, let us suppose that it does  
not contain tadpoles.
We will be interested in the crossover limit $\overline{t}\to 0$,
$R\to \infty$ with $R^2 \overline{t}$ fixed.
After expanding
$\widehat{\Delta}(q) = \widehat{\Delta}_1 + \widehat{\Delta}_2(q)$, 
we will obtain, apart from 
numerical factors, an expression which is a sum of terms of the form
\reff{eq:C2x}. We wish now to show that, if $K$ increases with $R$ 
at most logarithmically,  we can disregard all terms containing 
$\widehat{\Delta}_1$. In other words, we can simply substitute 
$\widehat{\Delta}(q)$ with $\widehat{\Delta}_2(q)$ in the 
original graph. To prove this result,
we begin by taking the limit $R\to\infty$ keeping $\overline{t}$ 
fixed and rewriting 
the integral of Eq. \reff{eq:C2x} in the form of Eq. 
\reff{genericgraph-hat}. We should now consider the behaviour of this integral
for $\overline{t}\to 0$. Simple power counting indicates 
that the integral is infrared divergent in this limit.
Let us suppose that the graph associated 
to Eq. \reff{genericgraph-hat} does not contain tadpoles. In this case, 
in order to compute the leading infrared contribution, we can 
substitute $\Delta_2(q)$ with its small-$q$ expansion 
$1/(a_2 N (q^2 + \overline{t}))$ and then
rescale $q^2 = \overline{t} p^2$, obtaining
\be
{(\widehat{\Delta}_1)^n (R^2 \overline{t})^{m-M_{int}}\over 
 R^{2(2m-M_{int})} }
\int \prod_{i<j} {dp_{ij}\over (2\pi)^d}\,
    \prod_{i<j} {1\over p_{ij}^2 + 1} \, \prod_i (2\pi)^d
    \delta\left(\sum_j p_{ij}\right),
\label{eq:C.2.2.x}
\ee
where $m$ is the number of loops and $M_{int}$ the number of internal lines.
Since, by hypothesis, the graph associated to Eq. \reff{eq:C.2.2.x} does not 
contain tadpoles, the integral is finite
and thus the prefactor gives its behaviour in the crossover limit.
If the integral in Eq. \reff{genericgraph-hat} is associated to a 
graph with tadpoles, it can be written as the product of an integral 
associated to a graph without tadpoles, and therefore behaving 
as in Eq. \reff{eq:C.2.2.x}, and a power of the one-loop tadpole integral.
Now, using Eq. \reff{I1R-2d}, we have
\be
N\int {d^2q\over (2\pi)^2} \Delta_2(q) \approx
{R^2\over a_2 (1 + a_2 \beta K)} I_{1,R}(\overline{t}) \approx\, 
- {1\over 4\pi a_2}\log \overline{t} + {C_2\over a_2}.
\label{tadpolo-int1}
\ee
Therefore neglecting logarithms all contributions behave as
$R^{-2(2 m - M_{int})} \widehat{\Delta}_1^n$ in the crossover limit. Now,
if $K$ increases at most logarithmically, $\widehat{\Delta}_1$ behaves as 
$1/R^2$ modulo logarithms. Therefore all terms of the form 
\reff{genericgraph-hat} behave as
\be
  {(\log R^2)^p \over 
  R^{2(2m -M_{int}+n)} },
\ee
for some $p$.
Now, if $N_{int}$ is the number of internal lines of the original $l$-loop
graph, $N_{int} = M_{int} + n$. Moreover if the original graph
does not contain tadpoles $m+n>l$ for $n\ge 1$. Thus, for $n\ge 1$,
\be
2 m + n - M_{int} = 2(m+n) - N_{int} > 2l - N_{int}.
\ee
Therefore contributions with $\widehat{\Delta}_1$ decrease faster and 
can be neglected. Thus the original $l$-loop graph we started from 
can be computed replacing $\widehat{\Delta}(q)$ with 
$\widehat{\Delta}_2(q)$ and scales as
\be
{1\over R^{2l}} \overline{t}^{l-N_{int}} = 
  {\left(R^2 \overline{t}\right)^{l-N_{int}}\over 
   R^{2(2l -N_{int})} },
\label{generic-scaling-2d}
\ee
without any logarithm.

We should now consider graphs with tadpoles. As we already discussed these
contributions can be written as the product of an integral associated to 
a graph without
tadpoles, and therefore behaving according to Eq. \reff{generic-scaling-2d},
and a power of the one-loop tadpole diagram.

Now, using Eq. \reff{tadpolo-int1}, we have
\be
N\int {d^2k\over (2\pi)^2} \widehat{\Delta}(k) =\, 
- {\beta K\over 1 + a_2 \beta K} + {1\over R^2} \left[
- {1\over 4\pi a_2}\log \overline{t} + {C_2\over a_2}\right] + 
 o(R^{-2}).
\ee
The parameter $K$ is free and we can take it $R$-dependent at our will. 
We should also define $\beta_{c,R}^{\rm (exp)}$, 
i.e. the scaling behaviour of the 
temperature $\beta$. To fix these two variables we require that 
the tadpole scales as $R^{-2}$ for $R\to\infty$ without logarithms 
and that $\overline{t}\sim R^{-2}$. This can be achieved by taking
\begin{eqnarray}
K &=& {V_R\over 4 \pi R^2} \log R^2 + {c_0\over R^2} V_R, \\
\widehat{t} &=& R^2\left[1 - 
    {N+2\over 24 \pi N } {\overline{a}_4\over\overline{a}_2^2}
     {1\over R^2}\log R^2 - 
     \overline{a}_2 \beta V_R\right],
\end{eqnarray}
where $c_0$ is an arbitrary constant. In the derivation 
we have used the expansion of $a_2$ in powers of $\beta K$, see 
Eq. \reff{a2_expansion}.
With this choice we have
\begin{eqnarray}
\overline{t} &=& 
    {1\over R^2} (\widehat{t} + \widehat{c}_0) + o(R^{-2}), \\
N \int {d^2 k\over (2\pi)^2}
   \widehat{\Delta}(k) &=&
  {1\over \overline{a}_2 R^2}
   \left[-{1\over 4\pi} \log(\widehat{t} + \widehat{c}_0) + 
    C_2 - c_0\right] +\,
   o(R^{-2}),
\end{eqnarray}
where
\be
\widehat{c}_0 =\, - {N+2\over 6 N} {\overline{a}_4\over\overline{a}_2^2} c_0.
\label{widehatc0-def}
\ee
Therefore, with this choice of $K$, in the limit $R\to\infty$, 
$\beta\to 0$, with $\widehat{t}$ fixed, the tadpole scales as 
$R^{-2}$, without logarithms of $R^2$.
It follows that all graphs 
scale as in Eq. \reff{generic-scaling-2d} with additional logarithms 
of $(\overline{t}R^2)$.

Now, a simple topological argument gives 
\be
2 l - N_{int} = 2 - m - {n\over 2} + {1\over 2} \sum_k (k-4) V_k,
\label{topologicalconstraint}
\ee
where $V_k$ is the number of $k$-leg vertices. Therefore, the graph scales as
\be
{\left(R^2 \overline{t}\right)^{l-N_{int}}\over R^{4 - 2m -n}} 
   R^{-{1/2} \sum_k (k-4) V_k},
\ee
with additional powers of $\log(R^2 \overline{t})$.
In the limit $R\to\infty$, $\overline{t}\to 0$, with 
$\widehat{t}$ fixed, the previous formula shows that graphs
that have one or more vertices with six or more legs vanish faster for 
$R\to\infty$ than graphs containing only 4-leg vertices. This means
that in this limit we can simply ignore the $\phi^6$, $\phi^8$, 
$\ldots$, terms in Eq. \reff{pf}. In conclusion we obtain
\be
\widetilde{\Gamma}^{(m,n)} \equiv 
   \Gamma^{(m,n)} R^{4 - 2 m - n} = \,
     (\widehat{t} + \widehat{c}_0)^{2 - m - n/2}\,
      \sum_{l} \widetilde{\Gamma}^{(m,n)}_l,
\ee
where $\widetilde{\Gamma}^{(m,n)}_l$ behaves as 
$\widehat{t}^{-l}$ times logarithms of $\widehat{t}$. 
Therefore the loop expansion provides the expansion of the critical
crossover functions in the limit $\widehat{t}\to\infty$. For instance,
if one considers the susceptibility, one obtains at two loops,
\begin{eqnarray}
&& \hskip -1truecm f_{\chi}(\widehat{t}) \equiv \lim_{R\to\infty}\chi R^{-2} = 
    {\overline{a}_2\over \widehat{t} + \widehat{c}_0} -
    {N+2\over 6 N} {\overline{a}_4\over\overline{a}_2}
    {1\over (\widehat{t} + \widehat{c}_0)^2} 
    \left[{1\over 4\pi} \log(\widehat{t} + \widehat{c}_0) + c_0 - C_2\right] 
\nonumber \\
&& 
   + {(N+2)^2\over 36 N^2} {\overline{a}_4^2\over\overline{a}_2^3}
    {1\over (\widehat{t} + \widehat{c}_0)^3}
    \left[\left({1\over 4\pi} \log(\widehat{t} + \widehat{c}_0) + c_0 - 
          C_2\right)^2 \right.
\nonumber \\
&& \qquad\left.
- {1\over 4\pi} 
  \left({1\over 4\pi} \log(\widehat{t} + \widehat{c}_0) + c_0 - C_2\right) +
              {2 H\over N+2}\right] +\, 
   O\left((\widehat{t} + \widehat{c}_0)^{-4}\right),
\label{fchi-2d-twoloops}
\end{eqnarray}
where $H = {1\over 24\pi^2} \psi'({1\over3}) - {1\over36}$.

This result should not depend on $c_0$. Expanding
in $\widehat{t}$ for $\widehat{t}\to\infty$ it is easy to check that no 
dependence remains and we can simply set $c_0 = 0$.

From the discussion we have presented, it is clear that the crossover 
functions can be obtained directly in the standard $\phi^4$ theory
with hamiltonian
\be
\int d^dx\, \left[
\smfrac{1}{2} \sum_\mu \partial_\mu \phi \cdot \partial_\mu \phi + 
      \smfrac{1}{2} r  \phi^2 + 
      \smfrac{1}{4!} u \phi^4\right].
\label{sec4:lambdaphi4}
\ee
Indeed all graphs, except the tadpole, have been computed using 
the propagator of this theory. The tadpole has been dealt with 
differently as it should be expected: indeed this is the only
ultraviolet divergent diagram. Therefore we have proved that 
we can rewrite
\be
f_\chi(\widehat{t}) =\, \mu_\chi F_\chi\left[s(\widehat{t} + \tau)\right],
\ee
where $F_\chi$ is the crossover curve computed in the short-range theory
and that was explicitly expressed in terms of RG functions in 
Sec. \ref{sec2}. In field-theoretic terms $\mu_\chi$ is the 
field renormalization, $s$ is related to the difference in the 
normalization of the fields and of the coupling constant and $\tau$ is the 
additive shift due to the mass renormalization. Comparing the 
expansion \reff{Fchi-FT-pert-2d} 
with Eq. \reff{fchi-2d-twoloops}
we obtain
\bea
\hskip -1truecm 
\mu_\chi &=& - {N\overline{a}_2^3\over \overline{a}_4} \\
\hskip -1truecm 
s       &=& - {N\overline{a}_2^2\over \overline{a}_4} \\
\hskip -1truecm 
\tau &=& - {\overline{a}_4\over N\overline{a}_2^2}
    \left[ {N+2\over 24 \pi} 
      \log\left( {24\pi N\overline{a}_2^2\over (N+8)|\overline{a}_4|}
          \right) + 
       D_2(N) + {N+8\over 24\pi} + {N+2\over 6} C_2\right],
\eea
where $D_2(N)$ is defined in Eq. \reff{def-D2}. 

For the $N$-vector model
we obtain 
\bea
\mu_\chi &=& {N+2 \over 6} ,
\label{chirisc-sigma-2d} \\
s       &=& {N+2 \over 6} ,
\label{tSR-that-sigma-2d}
\\
\tau &=& {1\over 4\pi} \log\left({4\pi(N+2)\over N+8}\right) +
          C_2 + {6 D_2(N)\over N+2} + {1\over 4\pi} {N+8\over N+2}.
\eea
As a final remark we wish to notice that we have followed here 
Thouless' approach to the critical crossover. For $N\ge 3$ this is the 
only possibility since no critical point exists. For $N < 2$ the crossover 
functions defined in this way differ by a simple shift given by
the constant $\tau$ computed above. 

For $N< 2$ our results give also the large-$R$ behaviour of the 
critical point. Since the critical theory corresponds to $\widetilde{t} = 0$
where $\widetilde{t}$ is the parameter appearing in the field-theory 
crossover functions, we have
\be
\beta_{c,R}=\, {1\over \overline{a}_2 V_R} 
    \left[ 1 -
    {N+2\over 24 \pi N} {\overline{a}_4\over\overline{a}_2^2}\,
     {1\over R^2}\log R^2 + {\tau\over R^2} 
    + o(R^{-2})\right].
\ee
Notice that this result depends explicitly on the single-site measure 
through $\overline{a}_2$ and $\overline{a}_4$, while it does not depend 
on the hopping coupling $J(x)$. Indeed all the dependence on $J(x)$ is 
encoded in the expansion variable $R^2$.
For the $N$-vector model we obtain the simpler and $N$-independent 
expression
\be
\beta_{c,R}=\, {1\over  V_R} 
    \left[ 1 + {1\over 4 \pi R^2} \log R^2 + {\tau\over R^2} +
    o(R^{-2})\right].
\ee
The presence of a logarithmic factor in $\beta_{c,R}$ 
was already predicted by Thouless \cite{Thouless_69} in a modified Ising model.

\subsection{Three-dimensional crossover limit} \label{sec4.3}

The ideas of the previous paragraphs can be generalized to any
dimension $d<4$. For generic values of $d$ one works as follows:
first one considers the graphs with $l$ loops contributing to the 
one-particle irreducible two-point function, where $l$ satisfies
\be
d l \le {2 d\over 4 - d},
\ee
computing their contribution in the limit 
$R\to\infty$, $\beta\to 0$ with $\overline{t} R^{2d/(4-d)}$ fixed. Then one
fixes the scaling behaviour of $K$ in order to cancel all the terms that scale
faster than $R^{-2d/(4-d)}$. The expression of $\beta_{c,R}^{\rm (exp)}$ 
is obtained requiring
$\overline{t} R^{2d/(4-d)}$ to be constant for $\beta\to 0$ and 
$R\to\infty$. The perturbative expansion becomes an expansion in powers 
of $\widehat{t}^{(4-d)/2}$ with additional logarithms when 
$2/(4-d)$ is an integer, i.e. for 
$d=4-2/n$. The reason for the appearance of these singular terms is well known
\cite{Symanzik_73,Parisi_79,Bergere-David_82}: the critical crossover limit
corresponds to the massless limit of the standard $\phi^4$ theory
which is known to have logarithmic singularities for these values of $d$.

Let us now discuss in more detail the three-dimensional case. 
For $d=3$ we should consider the one-loop and two-loop graphs 
in the expansion of the two-point function $\<\phi^a \phi^b\>$.
If $\Gamma^{(0,2)}$ is the irreducible two-point function at zero 
external momentum, we have at two loops
\begin{eqnarray}
\Gamma^{(0,2)} &\approx&
   {\overline{t}\over 1 - K/V_R}  - {N+2\over 6N} a_4 T_1 - 
   {(N+2)^2\over 36 N^2} a_4^2 T_1 T_2 
\nonumber \\
&& - {N+2\over 18 N^2} a_4^2 T_3 - 
     {(N+2)(N+4)\over 120 N^2} a_6 T_1^2,
\label{chim1-3d}
\end{eqnarray}
where
\bea
   T_1 & = &  N\int {d^3q\over (2\pi)^3} \widehat{\Delta} (q),
\\
   T_2 & = &  N^2 \int {d^3q\over (2\pi)^3} \widehat{\Delta}(q)^2,
\\
T_3 & = & N^3 \int {d^3q\over (2\pi)^3} {d^3k\over (2\pi)^3}\,
        \widehat{\Delta}(q) \widehat{\Delta}(k) \widehat{\Delta}(q+k).
\eea
Now, using Eq. \reff{I1R-dgt2}, for $\beta K \sim O(R^{-3})$ --- we will show
in the following this is
the correct asymptotic behaviour --- we have,
\bea
   T_1 & =&  - {\beta K\over 1 + a_2 \beta K} + 
      {1\over a_2 (1 + a_2 \beta K)} I_{1,R}(\overline{t})
\nonumber \\
&=& {1\over a_2}(1 - a_2 \beta K)\left(\overline{I}_{1,R} - a_2 \beta K\right)
  - {1\over 4 \pi a_2} {1\over R^6} 
       \left(\overline{t} R^6\right)^{1/2} + o(R^{-6}).
\label{T1-expansion}
\eea
The estimate of $T_2$ is easy and we find, cf. Eq. \reff{Jdgt2},
\be
T_2 = {1\over 8 \pi a_2^2} \left(\overline{t} R^6\right)^{-1/2} +
      o(R^0).
\label{T2asyn}
\ee
Finally, using Eqs. \reff{I1R-dgt2}  and \reff{I2R-d3} we obtain for 
$T_3$ --- we will call the associated graph ``two-loop watermelon" ---
\be
T_3 =\, {1\over a_2^3 R^6}\left[ 
 - {1\over 32 \pi^2} \log  \overline{t} + C_3\right] + o(R^{-6}),
\label{T3-expansion}
\ee
where $C_3$ is a constant given in Eq. \reff{defC3}.

Now let us consider the terms that scale slower than $R^{-6}$ for 
$R\to\infty$ at $\overline{t} R^6$ fixed. Using the previous results we have
\be
-{N+2\over 6N} {a_4\over a_2} (1 - a_2 \beta K) 
     (\overline{I}_{1,R} - a_2 \beta K) - 
    {N+2\over 192 \pi^2 N^2} {a_4^2\over a_2^3 R^6} \log R^2 -
    {(N+2)^2\over 36 N^2} a_4^2 T_1 T_2.
\label{eq:4.3.x}
\ee
At first we neglect the last term proportional to $T_1 T_2$.
If $K$ has a finite limit for $R\to\infty$ and $\overline{t}\sim R^{-6}$
in the same limit, using Eq. \reff{tbar-def}, we obtain
$\beta = 1/(\overline{a}_2 V_R) (1 + O(R^{-3}))$.
We can then determine
$K$ by requiring the expression \reff{eq:4.3.x} to be of order 
$R^{-6}$. This gives
\be
K=\, V_R\left[\overline{I}_{1,R} + {1\over 32 \pi^2 N} 
   {\overline{a}_4\over \overline{a}_2^2} {1\over R^6} \log R^2 + 
   {c_0\over R^6}\right],
\label{K-scaling-3d}
\ee
where $c_0$ is arbitrary. The scaling behaviour of $\beta$ is obtained 
requiring $\overline{t}\sim O(R^{-6})$. If we introduce a variable
$\widehat{t}$ related to $\beta$ by
\be
\beta\, =\, {1\over \overline{a}_2 V_R}\left\{ 
    1 - {N+2\over 6N} {\overline{a}_4\over\overline{a}_2^2}
     \left[ \overline{I}_{1,R} + 
        {1\over 32\pi^2 N} {\overline{a}_4\over\overline{a}_2^2}
        {1\over R^6}\log R^2\right]- {\widehat{t}\over R^6}\right\},
\label{beta-scaling-3d}
\ee
and define the critical crossover limit as the limit $\beta\to0$, 
$R\to\infty$ with $\widehat{t}$ fixed, we find that indeed 
$\overline{t}\sim R^{-6}$. More precisely, using
Eqs. \reff{tbar-def} and \reff{a2_expansion}, we have
\be
\overline{t} R^6=\, 
\widehat{t} + \widehat{c}_0 + c_1 + o(R^0),
\ee
where
\be
c_1 =\, - \sigma^2\left[
    {(N+2)(N+4)\over 120 N^2} {\overline{a}_6\over\overline{a}^3_2} -
    {(N+2)^2\over 18 N^2} {\overline{a}_4^2\over\overline{a}^4_2} +
    {N+2\over 6N} {\overline{a}_4\over\overline{a}^2_2}\right],
\ee
$\sigma$ is defined in Eq. \reff{sigma-def}, 
and $\widehat{c}_0$ is related to $c_0$ by Eq. \reff{widehatc0-def}. 
Notice that for $R\to\infty$, $K$ converges to a constant, cf. Eq.
\reff{Ibar1R-asintotico}, and $\beta\approx 1/(\overline{a}_2 V_R)$ as 
it was claimed before. 

Using Eqs. \reff{beta-scaling-3d} and \reff{K-scaling-3d} we can rewrite
$T_1$ as 
\be
T_1 = \, - {1\over 32\pi^2 N} {\overline{a}_4\over \overline{a}_2^3}
    {1\over R^6} \log R^2 + O(R^{-6}) = 
    \, - {\overline{a}_4\over 3N} T_3 + O(R^{-6}).
\label{T1-espressione2}
\ee
Let us now go back to Eq. \reff{eq:4.3.x}. We should now deal with the 
term proportional to $T_1 T_2$ that was neglected in the previous 
treatment. We have
\be
\Gamma^{(0,2)} = \, 
    {(N+2)^2\over 9216 \pi^3 N^3} {\overline{a}_4^3\over\overline{a}_2^5}
    \left(\overline{t} R^6\right)^{-1/2} {1\over R^6} \log R^2 +
    O(R^6).
\ee
Because of 
the presence of the logarithm, this term does not scale correctly.
However it is of order $1/\widehat{t}^{\,1/2}$, and terms of this order 
appear also at three loops. We will now show that this contribution is canceled 
exactly by the contribution of the 
three-loop graph in which the tadpole has been replaced by 
the two-loop watermelon. This graph is associated to the integral
\be
T_4 = \, N^5 \int {d^3 p\over (2\pi)^3}{d^3 q\over (2\pi)^3} 
                  {d^3 r\over (2\pi)^3} 
  \widehat{\Delta}(p)^2 \widehat{\Delta}(q) \widehat{\Delta}(r)
  \widehat{\Delta}(p+q+r).
\ee
It can be rewritten as
\bea
T_4 &=& \, T_3 T_2 +\, 
\nonumber \\
&& \hskip -0.5truecm + N^5 \int {d^3 p\over (2\pi)^3}{d^3 q\over (2\pi)^3} 
                  {d^3 r\over (2\pi)^3} 
  \widehat{\Delta}(p)^2 \widehat{\Delta}(q) \widehat{\Delta}(r)
  [\widehat{\Delta}(p+q+r)-\widehat{\Delta}(q+r)].
\eea
In the crossover limit the last term can be easily computed using
the technique presented in App. \ref{AppA.3}. It is easy to 
see that we can neglect the contributions due to $\widehat{\Delta}_1$,
and that we can replace
\be
\widehat{\Delta}_2(k) \to \, {1\over \overline{a}_2 N} 
   {1\over k^2 R^2 + \overline{t}}.
\ee
Thus, neglecting terms $o(R^{-6})$, we have
\bea
T_4 &=& \, T_3 T_2 +\, 
 {(\overline{t} R^6)^{-1/2}\over \overline{a}_2^5 R^6}\, 
  \int {d^3 p\over (2\pi)^3}{d^3 q\over (2\pi)^3} 
                  {d^3 r\over (2\pi)^3} 
\nonumber \\
&& \quad \times
  {1\over (p^2 + 1)^2 (q^2 + 1) (r^2 + 1)}
  \left[ {1\over (p+q+r)^2 + 1} - 
         {1\over (q+r)^2 + 1}\right].
\label{T4-asintotico}
\eea
Thus, if we keep only terms of order $R^{-6}\log R^2$, we have simply
$T_4 \approx T_3 T_2$. Then including the combinatorial and 
group factors we find the total contribution
\be
- {(N+2)^2\over 108 N^3} \overline{a}_4^3 T_4 \approx 
- {(N+2)^2\over 108 N^3} \overline{a}_4^3 T_3 T_2 + O(R^{-6}) \approx
  {(N+2)^2\over 36 N^2} \overline{a}_4^2  T_1 T_2 + O(R^{-6}),
\label{T4-asintotico2}
\ee
where we have used Eq. \reff{T1-espressione2}. Comparing 
with Eq. \reff{eq:4.3.x} we see that the logarithms
cancel, as claimed at the beginning. 

This calculation illustrates the 
general mechanism in three dimensions. Consider a graph with only 
$\phi^4$ vertices that does not
have tadpoles or two-loop watermelons as subgraphs. In the crossover 
limit the contribution of this  graph
 can be computed neglecting $\widehat{\Delta}_1$ and substituting 
$\widehat{\Delta}_2(q)$ with its small-$q$ expression. Following the argument 
presented in two dimensions it is easy to see that an $l$-loop 
contribution scales exactly as $\widehat{t}^{\, 1-l/2}$. The same argument
we have presented in two dimensions proves also that the contributions 
of graphs with $\phi^6$, $\phi^8$, $\ldots$ vertices are suppressed and 
can be neglected in the critical crossover limit. Finally consider a
graph with only $\phi^4$ vertices that 
has tadpoles or two-loop watermelons as subgraphs. 
This graph generates  non-scaling terms involving powers of $\log R^2$. 
However the logarithmic contribution associated to each tadpole 
is exactly canceled, using Eq. \reff{T1-espressione2}, by the 
analogous term which is associated to the graph in which the 
tadpole is replaced by the two-loop watermelon. 
The sum of all contributions scales as $R^{-6}$ without logarithms of $R^2$.
Our discussion proves therefore that the perturbative expansion corresponds
to an expansion in powers of $1/\widehat{t}^{\, 1/2}$ of the critical
crossover functions with additional logarithms of 
$\widehat{t}$. Explicitly for the susceptibility we obtain
at two loops
\begin{eqnarray}
&& (f_{\chi}(\widehat{t}))^{-1}\equiv \lim_{R\to\infty} \chi^{-1} R^6 =\,
   \overline{a}_2 \widehat{t} +
   {N+2\over 24\pi N} {\overline{a}_4\over \overline{a}_2}
  \widehat{t}^{\, 1/2} + \overline{a}_2 c_1
  + {N+2\over 6N} {\overline{a}_4\over \overline{a}_2}
  \sigma^2
\nonumber \\
&& \qquad
- {N+2\over 18 N^2} {\overline{a}_4^2\over \overline{a}_2^3}
 \left[ - {1\over 32 \pi^2} \log\widehat{t} + C_3\right] +
  {(N+2)^2\over 1152\pi^2 N^2} {\overline{a}_4^2\over \overline{a}_2^3}
  +O(\widehat{t}^{-1/2} \log \widehat{t}).
\label{fchi-2loop-3d}
\end{eqnarray}
As expected, the contribution proportional to
$a_6$ at two-loops disappears in the crossover limit: again only
the $\phi^4$-vertex is relevant. The result is also independent on
$c_0$.

The discussion we have presented shows also that the critical crossover 
functions can be computed in the standard continuum $\phi^4$ theory. 
Therefore, as we did in two dimensions, we can write
\be
f_\chi(\widehat{t}) =\, \mu_\chi F_\chi\left[s(\widehat{t} + \tau)\right],
\ee
where $F_\chi$ is the crossover curve computed in the short-range theory
and that was explicitly expressed in terms of RG functions in
Sec. \ref{sec2}.
Comparing the expansion \reff{Fchi-FT-pert-3d} with Eq. 
\reff{fchi-2loop-3d}, we obtain
\bea
\mu_{\chi} &=& {N^2 \overline{a}_2^3\over \overline{a}_4^2},
\label{chi-chiSR-3d} \\
s &=& {N^2 \overline{a}_2^4\over \overline{a}_4^2},
\label{tSR-that-3d} \\
\tau &=& - {N+2\over 288\pi^2 N^2} {\overline{a}_4^2\over \overline{a}_2^4}
\log\left({48\pi N\overline{a}_2^2\over (N+8)|\overline{a}_4|}\right) +\, 
  {\overline{a}_4^2\over \overline{a}_2^4} {D_3(N)\over N^2} +
  {\overline{a}_4^2\over \overline{a}_2^4} 
  {9N^2 - 20 N - 544\over 20736\pi^2 N^2}
\nonumber \\
&& +\, c_1 + {N+2\over 6N} {\overline{a}_4\over\overline{a}_2^2} \sigma^2 - 
   {N+2\over 18 N^2} {\overline{a}_4^2\over \overline{a}_2^4} C_3,
\eea
where $D_3(N)$ is defined in Eq. \reff{D3-def}.
For the $N$-vector model these equations become 
\bea
\mu_{\chi} &=& {(N+2)^2\over 36},
\label{chi-chiSR-3d-sigma}
\\
s &=& {(N+2)^2\over 36}, 
\label{tSR-that-3d-sigma}
\\
\tau &=& - {1\over 8\pi^2(N+2)} \log\left({8\pi (N+2)\over (N+8)}\right) 
  + {36  D_3(N) \over (N+2)^2} + {9N^2 - 20 N - 544\over 576\pi^2 (N+2)^2} -
   {2 C_3\over N+2}.
\nonumber \\
{}
\eea
As remarked in the previous Section, Eq. \reff{beta-scaling-3d} gives us 
the behaviour of the critical point $\beta_{c,R}$ up to terms of order 
$R^{-6}$.
We have
\be
\beta_{c,R} = \, {1\over \overline{a}_2 V_R}\left\{
    1 - {N+2\over 6N} {\overline{a}_4\over\overline{a}_2^2}
     \left[ \overline{I}_{1,R} +
        {1\over 32\pi^2 N} {\overline{a}_4\over\overline{a}_2^2}
        {1\over R^6}\log R^2\right] + {\tau\over R^6} + o(R^{-6})\right\}.
\label{betacR-3d}
\ee
In the $N$-vector case, this expression simplifies becoming
\be
\beta_{c,R} = \,{1\over V_R} \left[
           1 + \overline{I}_{1,R} - {3\over 16\pi^2} {1\over N+2} 
           {1\over R^6} \log R^2 + {\tau\over R^6} + o(R^{-6})\right].
\label{betacR-3d-Nvector}
\ee
The first correction to $\beta_{c,R}$ was derived in Refs.
\cite{Brout_60,Dalton-Domb_66,Vaks-Larkin-Pikin_66}, while the 
presence of the logarithmic correction was predicted in a modified Ising 
model with long-range interactions by Thouless \cite{Thouless_69}.
Notice that for $N\to\infty$ the logarithmic term disappears and 
that, using Eq. \reff{D3-largeN},
$\tau\sim O(1/N)$ so that the 
resulting formula gives the exact large-$N$ prediction for $\beta_{c,R}$.

\section{Corrections to the critical crossover functions} \label{sec5}

In this Section we will study the corrections to the critical 
crossover functions. First we will present the results for the 
three-dimensional case, which is the easiest one, then
we will discuss the corrections in two dimensions. To derive the
behaviour of these corrections we must introduce some additional hypothesis 
on the function $J_R(x)$. We will thus assume:
\begin{itemize}
\item[(v)] $\sum_x (x^2)^2 J_R(x)$ is finite. It follows 
 $\Pi(q)\approx q^2 + O(q^4)$.
\item[(vi)] In the limit $R\to\infty$ at fixed $q$, 
$\overline{\Pi}_R(q/R)$ has an expansion in powers of $1/R^2$. Explicitly
\be
\overline{\Pi}_R(q/R) = \Pi(q) + \sum_{n=1}^\infty 
   {1\over R^{2n}} \Pi_n(q).
\ee
Notice that, because of property
(v), $\Pi_n(q)\sim q^4$ for $q\to 0$.
\end{itemize}

\subsection{Corrections in three dimensions} \label{sec5.1}

We begin by discussing the three-dimensional case. We will show at 
two loops --- but we conjecture that this is true to all orders of 
perturbation theory --- that the leading correction to the scaling 
behaviour is of order $R^{-3}$, {\em provided} one appropriately
defines $\widehat{t}$. In other words we will show that 
\be
\widetilde{\chi} = f_\chi(\widehat{t}) + 
      {1\over R^d} g_{\chi}(\widehat{t}) + 
\ldots
\label{chitilde-expansion-correzioni}
\ee
in the crossover limit for a suitable definition of $\widehat{t}$. 
This type of behaviour should be true for any dimension $2<d<4$. 
It is obvious that an expansion of the form 
\reff{chitilde-expansion-correzioni} cannot be valid generically.
Indeed if $\beta_{c,R}^{\rm (exp)}$ is such that 
Eq. \reff{chitilde-expansion-correzioni} holds --- 
$\widehat{t}$ is defined in Eq. \reff{def-that} --- 
consider $\beta_{{\rm New},c,R}^{\rm (exp)}$ defined by
\be
\beta_{{\rm New},c,R}^{\rm (exp)} = \beta_{c,R}^{\rm (exp)} \left(
  1 + A R^{-6-\alpha}\right),
\ee
with $0<\alpha<3$. If $\widehat{t}_{\rm New}$ is the corresponding scaling 
variable we have $\widehat{t}_{\rm New} = \widehat{t} + A R^{-\alpha}$. 
Therefore the two definitions are identical in the critical 
crossover limit. However in the variable $\widehat{t}_{\rm New}$ 
the corrections are of order $R^{-\alpha}$. 

Let us now go back to Eq. \reff{chim1-3d}, and again let us neglect at first 
the contribution proportional to $T_1 T_2$. The expression for $T_1$
appearing in Eq. \reff{T1-expansion} is valid up to terms of order $R^{-9}$
as it can be seen from the results of App. \ref{AppA.2}. Using the 
expressions for $K$ and $\beta$, cf. Eqs. \reff{K-scaling-3d} and 
\reff{beta-scaling-3d}, one finds that the leading correction in $T_1$ is 
of order $\log R^2/R^9$. Let us now consider $T_3$. Using the results 
of App. \ref{AppA.3} we find 
\be
T_3 = {1\over a_2^3} I_{2,R}(\overline{t}) + O(R^{-9}) = \,
    {1\over a_2^3 R^6} \left[ - {1\over 32\pi^2} \log \overline{t} + C_3
    + {F_3\over R^2}\right] + O(R^{-9}).
\ee
Therefore with the definition of $\beta_{c,R}^{\rm (exp)}$ appearing in Eq.
\reff{beta-scaling-3d} we would obtain corrections of order $R^{-2}$ and 
$R^{-3}\log R^2$. We will now show that these corrections can be 
eliminated with a proper redefinition of the expressions of $K$ and 
$\beta_{c,R}^{\rm (exp)}$. Considering for simplicity the $N$-vector 
case [we have $a_2 = 1$, and $a_4=-6N/(N+2)$] we assume
\bea
K &=& V_R \left[ \overline{I}_{1,R} - 
    {3\over 16 \pi^2 (N+2)} {1\over R^6} \log R^2 + 
    {K_1\over R^8} + {K_2\over R^9} \log R^2\right], \\
\beta &=& {1\over V_R} \left[
    1 + \overline{I}_{1,R} - {3\over 16\pi^2(N+2)} {1\over R^6} \log R^2 -
    {\widehat{t} \over R^6} + {b_1\over R^8} + 
    {b_2\over R^9} \log R^2\right],
\eea
where $K_1$, $K_2$, $b_1$, and $b_2$ are constants to be determined.
Then the corrections of order $1/R^8$ and $\log R^2/R^9$ in
$T_1$ and $T_2$ are the following:
\bea
T_1 &=& \ldots + \left[- K_1 + 
   {b_1 - K_1\over 8\pi} (\widehat{t} + \sigma^2)^{-1/2}\right] 
   {1\over R^8} 
\nonumber \\
&& \hskip -1truecm + \left[- K_2 + {3\over 16\pi^2} {\sigma\over N+2} + 
   {1\over 8\pi} (\widehat{t} + \sigma^2)^{-1/2}
   \left(b_2 - K_2 + {3\over 8\pi^2} {\sigma\over N+2}\right)
   \right] {\log R^2\over R^9} + O(R^{-9}),
\nonumber \\[-2mm]
{} 
\\
T_3 &=& \ldots + \left[F_3 + 
   {b_1 - K_1\over 32\pi^2} (\widehat{t} + \sigma^2)^{-1}\right] 
   {1\over R^8} 
\nonumber \\
&& + \left[- {9\sigma\over 32\pi^2}  + 
   {1\over 32\pi^2} (\widehat{t} + \sigma^2)^{-1}
   \left(b_2 - K_2 + {3\over 8\pi^2} {\sigma\over N+2}\right)
   \right] {\log R^2\over R^9} + O(R^{-9}),
\eea
where the dots indicate terms that scale as $R^{-6}$ and $R^{-6}\log R^2$.
To cancel the unwanted corrections we must require that the 
combination $T_1 - 2 T_3/(N+2)$ is free of terms that scale 
as $R^{-8}$ and $R^{-9} \log R^2$. In this way we determine the 
constants $K_1$, $K_2$, $b_1$, and $b_2$. Explicitly
\be
K_1=\, b_1 = - {2 F_3\over N+2}, \qquad\qquad
K_2=\, 2 b_2 = {3\over 4\pi^2} {\sigma\over N+2}.
\ee
We must now consider the terms proportional to $T_1 T_2$. As we already 
discussed before we must consider at the same time the diagram associated
to $T_4$. A simple analysis shows that Eq. \reff{T4-asintotico} has 
corrections of order $O(R^{-9})$ which are therefore negligible in the 
present discussion. Therefore, including the combinatorial and group
factors we must show that 
\be
T_1 T_2 - {2 T_2 T_3\over N+2} = 
   T_2\left ( T_1 - {2 T_3\over N+2}\right)
\ee
is free of terms that scale as $R^{-8}$ and $R^{-9}\log R^2$. We have 
already shown that the term in parenthesis has this property. 
For $T_2$, using Eq. \reff{Jdgt2}, we can show that 
Eq. \reff{T2asyn} is valid up to terms of order $O(R^{-3})$. Therefore
the previous expression is free of the unwanted corrections.
In conclusion we have proved at two loops Eq.
\reff{chitilde-expansion-correzioni}. We conjecture this is true to all 
orders: graphs without tadpoles or insertions of the two-loop watermelon
should have corrections of order $R^{-9}$, while terms of order 
$R^{-8}$ and $R^{-9}\log R^2$ that appear in graphs with tadpoles
or two-loop watermelon insertions should cancel with the mechanism we 
presented above. At the order we are considering graphs containing
vertices with more than four legs should still be negligible: at two 
loops the contribution proportional to $a_6$ in Eq. 
\reff{chim1-3d} scales as $\log^2 R^2/R^{12}$.

Using the perturbative expansion
we can compute the function $g_\chi(\widehat{t})$
in the limit $\widehat{t}\to\infty$. We have 
\be
g_\chi(\widehat{t}\, ) = -{E_3\over \widehat{t}}
     + O(\widehat{t}^{-3/2}),
\ee
where $E_3$ is defined in Eq. \reff{Ed-def}. Notice that this behaviour 
cannot be changed by modifying the definition of $\beta_{c,R}^{\rm (exp)}$. 

Let us now show that if one uses $\widetilde{t}$ defined using 
the exact $\beta_{c,R}$ one directly obtains the expansion
\reff{chitilde-expansion-correzioni}.
To prove this fact, assume the opposite and 
write
\be
\widetilde{\chi} = f_\chi (\widetilde{t}) + 
            {1\over R^\alpha} h_\chi (\widetilde{t}) + o(R^{-\alpha}),
\label{eq:5.12}
\ee
with $\alpha < 3$. For $\widetilde{t}\to 0$ and any value of $R$, 
$\widetilde{\chi} \sim \widetilde{t}^{-\gamma}$. Therefore
$f_\chi (\widetilde{t})\sim \widetilde{t}^{-\gamma}$ and 
$h_\chi (\widetilde{t})\sim \widetilde{t}^{-\gamma}$ in this limit.
Now, it follows from our discussion that the term of order $R^{-\alpha}$
can be eliminated if we introduce a new variable 
$\widehat{t} = \widetilde{t} + A R^{-\alpha}$. Substituting in 
Eq. \reff{eq:5.12} and expanding in $R^{-\alpha}$ we have
\be
\widetilde{\chi} = f_\chi (\widehat{t}) + {A\over R^\alpha} 
         {f'}_\chi (\widehat{t}) + 
            {1\over R^\alpha} h_\chi (\widehat{t}) + o(R^{-\alpha}).
\ee
Cancellation of the terms of order $R^{-\alpha}$ requires 
$A {f'}_\chi (\widehat{t}) + h_\chi (\widehat{t}) = 0$. However this 
relation cannot be true since ${f'}_\chi (\widehat{t})\sim 
\widehat{t}^{-\gamma-1}$ for $\widehat{t}\to 0$. Therefore 
$h_\chi (\widetilde{t}) = 0$. We have therefore showed that the 
variable $\widetilde{t}$ is a particularly good one, since it 
automatically eliminates a whole class of corrections to the leading 
behaviour. Another consequence of these results is that we can now estimate
the order of the neglected terms in Eqs. \reff{betacR-3d}, 
\reff{betacR-3d-Nvector}:
the terms $o(R^{-6})$ are of order $R^{-8}$.

In App. \ref{AppB.2} we compute the function $g_\chi(\widetilde{t})$ 
for our general model in the large-$N$ limit. The graph of 
$g_\chi(\widetilde{t})/f_\chi(\widetilde{t})$
for some particular cases is reported in Fig. \ref{correzioni_largeN}. 
We consider: (1) the $N$-vector model, (2) the potential 
$V(\varphi) = N(\varphi^4 - \varphi^2)$, and (3) the potential
$V(\varphi) = N(\varphi^6 + \varphi^4 - \varphi^2) $.
Although the function is
not universal since it depends explicitly on various constants 
whose value is specific of the model one uses, the qualitative features
are similar in all cases: $g_\chi(\widetilde{t})/f_\chi(\widetilde{t})$
interpolates smoothly between the values for $\widetilde{t}=0$ and 
$\widetilde{t}=\infty$. Notice however that the function is decreasing 
in the $N$-vector model, while it is increasing in the other two cases.

Finally let us notice that the result \reff{chitilde-expansion-correzioni}
depends crucially on our use of $R$ as scale and on the specific 
field normalizations used in the definition of our model. In general with 
an arbitrary scale $\rho$ and arbitrarily-normalized fields we have
\be
\widetilde{\chi} =\, A(\rho) f_\chi(B(\rho) \widetilde{t})
   +\, {1\over \rho^d} g_{\chi} (\widetilde{t}) +\ldots,
\ee
where $A(\infty)$ and $B(\infty)$ are non-vanishing constants.

\subsection{Corrections in two dimensions} \label{sec5.2}

We wish now to discuss the corrections to the universal crossover curves in two
dimensions. If we repeat the perturbative analysis we have performed in the 
previous Section we face immediately a difficulty. Working at one loop
and using the results of App. \ref{AppA.2} we find corrections to the crossover
functions of order $\log^2 R^2/R^2$ and $\log R^2/R^2$. However, at 
variance with the three-dimensional case, only the former terms can be 
canceled with a redefinition of $K$ and $\widehat{t}$: the terms 
proportional to $\log R^2/R^2$ cannot be canceled. Indeed let us suppose
that 
\bea
K &=& {V_R\over R^2}\left[{1\over 4\pi} \log R^2 + 
       {K_0\over R^2} \log^2 R^2 +
       {K_1\over R^2} \log R^2\right],
\\
\beta = &=& {1\over V_R} \left[1 + {1\over 4\pi R^2} \log R^2 +
       {b_0\over R^4}\log^2 R^2 +
       {b_1\over R^4}\log R^2 - {\widehat{t}\over R^2}\right].
\eea
At one loop we obtain
\be
\Gamma^{(0,2)} =\, \ldots + 
    \Gamma_2(\widehat{t}) {\log^2 R^2\over R^4} + 
    \Gamma_1(\widehat{t}) {\log^2 R\over R^4} + 
    \Gamma_0(\widehat{t}) {1\over R^4} + o(R^{-4}),
\ee
where the dots indicate terms that scale as $R^{-2}$. 
The coefficient $\Gamma_2(\widehat{t})$ is given by
\be
\Gamma_2(\widehat{t}) =\, -b_0
  - {1\over 4 \pi \widehat{t}}\left[K_0 - b_0 + {1\over 16\pi^2}\right].
\ee
This term can be canceled setting
\be
b_0 =\, 0, \qquad\qquad K_0 =\, - {1\over 16 \pi^2}.
\ee
Let us now consider $\Gamma_1(\widehat{t})$. We have
\be
\Gamma_1(\widehat{t}) = 
   {1\over 8\pi} (4\alpha_1 + 3\alpha_2 + 2) \widehat{t} + 
   O(\log \widehat{t}),
\label{eq:6.6}
\ee
where $\alpha_1$ and $\alpha_2$ are determined by the 
low-momentum expansion of $\Pi(q)$, cf. Eq. \reff{Pi-smallv-expansion}.
This term does not depend on $K_1$ or $b_1$ and it is therefore impossible
to eliminate it. Therefore at one loop we obtain correction terms of 
order $\log R^2/R^4$. 
At two loops terms proportional to 
$\log R^2/R^2$ pop in and in general we have
\be
\widetilde{\chi} f_\chi(\widehat{t})^{-1} \approx 
  1 + {1\over R^2} \sum_{n=0}^\infty \log^n R^2 g_n(\widehat{t}).
\ee
The presence of this infinite series of logarithms may indicate that 
perturbation theory does not provide us with the correct corrections
and that a resummation of the perturbative series is needed. In other 
words the perturbative limit, $R\to\infty$ at $\overline{t}$ fixed
followed by $\overline{t}\to 0$ may not commute with the crossover 
limit $R\to\infty$, $\overline{t}\to 0$ at $\overline{t} R^2$ fixed at the 
level of the corrections to the universal behaviour. This phenomenon is not 
new in two-dimensional models. Indeed a similar non-commutativity
appears in the corrections to the finite-size scaling functions 
\cite{Caracciolo-Pelissetto_98,Caracciolo-etal_98}. 

In the large-$N$ limit, cf. App. \ref{AppB.1}, the corrections can be 
computed exactly and in this case one finds
\be
\widetilde{\chi}=\, f_\chi(\widehat{t})
   \left[1 + A(\widehat{t}) {\log R^2\over R^2} +
             B(\widehat{t}) {1\over R^2}\right] + o(R^{-2}).
\label{correzioni-2d-largeN}
\ee
However this simple behaviour may be due to the large-$N$ limit. 
In general, as long as $N\ge 3$, we do not expect a change in the exponent,
but a more complicated behaviour of the logarithmic corrections would
not be surprising. By analogy with what has been found for the 
finite-size scaling corrections in Refs.
\cite{Caracciolo-Pelissetto_98,Caracciolo-etal_98}, 
we could have a behaviour of the form
\be
\widetilde{\chi} f_\chi(\widehat{t})^{-1} \approx 
  1 + {\log R^2\over R^2} \sum_{n=0}^\infty 
  {g_n(\widehat{t})\over (\log R^2)^n}.
\ee
Also for the Ising model it is unlikely that a new exponent appears. 
The numerical work of Refs. \cite{L-B-B-pre,L-B-B-prl} confirms 
this expectation: indeed they find that the corrections to scaling
are well described in terms of a behaviour of the form
\reff{correzioni-2d-largeN}. In these works $A(\widehat{t})$ and 
$B(\widehat{t})$ are assumed independent of $\widehat{t}$. Of course 
this is an approximation, but it is not surprising it works well, since 
these two functions should be slowly varying, as indicated by the 
large-$N$ solution.

One may wonder if the non-commutativity we have discussed above 
is peculiar of two-dimensional models. A simple analysis indicates 
that a similar problem should also appear in three dimensions 
if one considers the corrections of order $R^{-6}$.
Indeed at this order $T_3$ gives rise to terms $\log R^2$ that cannot
be eliminated by changing the scaling of $K$ and $\beta$.

\section{Discussion} \label{sec6}

In this Section we wish to compare the analytic results obtained in the 
previous Sections with the numerical ones presented in Refs.
\cite{L-B-B-pre,L-B-B-prl,Luijten-Binder_98,Luijten-FSS} and discuss
other approaches to the crossover problem.

Let us first compare our results for $\beta_{c,R}$ with the numerical
determinations of Refs. \cite{L-B-B-pre,Luijten-FSS}. These simulations
are performed in the Ising model with coupling given in
Eq. \reff{Jrhoflat} and domain family
\be
D_\rho =\, \left\{x: \sum_{i=1}^d x^2_i \le \rho^2\right\}.
\ee
Using the numerical results of App. \ref{AppA} for the constants 
$C_2$ and $C_3$ and the numerical estimates of Sec. \ref{sec2}
for the non-perturbative constants $D_2(N)$ and $D_3(N)$, we obtain
in two dimensions the asymptotic expression
\be
\beta_{c,R} V_R\approx 1 + {1\over 4\pi R^2} \log R^2 + 
     {0.1975(5)\over R^2} \approx 
   1 + {1\over R^2} (0.0796 \log R^2 + 0.1975),
\label{betacR-sec6-2d}
\ee
while in three dimensions 
\be
\beta_{c,R} V_R\approx 1 + \overline{I}_{1,R} - 
   {1\over 16 \pi^2 R^6} \log R^2 - {0.0017(1)\over R^6},
\ee
where $\overline{I}_{1,R}$ is defined in Eq. \reff{Ibar1R-def}.
Numerical estimates for selected values of $\rho$ are reported in 
Table \ref{table_sigma}. If one is interested in the expression
of $\beta_{c,R} V_R$ up to terms of order $o(R^{-3})$, one 
can replace $\overline{I}_{1,R}$ with the asymptotic expression 
\reff{Ibar1R-asintotico}.
In two dimensions Eq. \reff{betacR-sec6-2d} agrees approximately
with the fit of the numerical data of Ref. \cite{L-B-B-prl}. 
They quote\footnote{Notice the different normalization of $R^2$:
$R^2$ in Ref. \cite{L-B-B-prl} is four times our definition of $R^2$.}
\be
\beta_{c,R} V_R\approx 1 + {1\over R^2} (
    0.076 (3)\, \log R^2 + 0.172(7)\, ).
\ee
To understand better the discrepancies we have considered 
\be
\Delta(R) = \left({\beta_{c,R,{\rm approx}}\over \beta_{c,R,{\rm exact}} } - 1
            \right) R^2,
\ee
where $\beta_{c,R,{\rm approx}}$ is the asymptotic form \reff{betacR-sec6-2d},
while $\beta_{c,R,{\rm exact}}$ is the exact value determined in the 
Monte Carlo simulation. Asymptotically we should observe
$\Delta(R)\to 0$. However for the values of $\rho$ used in the 
simulation $\Delta(R)$ shows a somewhat erratic behaviour. For 
$\rho^2 = 32,50,72,100,140$, we have 
$\Delta(R) = -0.0812$, $0.1649$, $0.1660$, $-0.1943$, $-0.1124$ with an error
of approximately $5\cdot 10^{-4}$ due mainly to the uncertainty 
in Eq. \reff{betacR-sec6-2d} (the error on $\beta_{c,R,{\rm exact}}$ is 
much smaller). Clearly these values of $\rho$ are too small for the 
asymptotic expansion to be valid. Similar discrepancies are observed in 
three dimensions. The non-monotonic behaviour of the corrections 
appears to be a general phenomenon for the family of domains used 
in the simulations, and it is probably connected with the fact that 
the shape is not natural on a cubic lattice.
Similar oscillation with $R$ are observed in lattice integrals.
For instance, from the results of Table \ref{table_sigma} in App. \ref{AppA.2},
one can see that the integral
$\overline{I}_{1,R}$ does not have a monotonic behaviour even for 
$\rho^2\approx 10^3$. 

Let us now compare the results for the crossover curves. 
In Figs. \ref{gammaeff2d}, \ref{nueff2d}, \ref{gammaeff}, and 
\ref{betaeff}, we report the graph of the effective exponents 
$\gamma_{\rm eff}$, $\nu_{\rm eff}$ and $\beta_{\rm eff}$ defined by
\bea
\gamma_{\rm eff} (\widetilde{t}) &=& 
- {\widetilde{t}\over f_\chi(\widetilde{t})}
   {df_\chi(\widetilde{t})\over d\widetilde{t}},
\\
\nu_{\rm eff} (\widetilde{t}) &=&
- {\widetilde{t}\over 2 f_\xi(\widetilde{t})}
   {df_\xi(\widetilde{t})\over d\widetilde{t}},
\\
\beta_{\rm eff} (\widetilde{t}) &=&
   {\widetilde{t}\over f_M(\widetilde{t})}
   {df_M(\widetilde{t})\over d\widetilde{t}},
\eea
for the Ising model
in two and three dimensions, using the field-theory results presented 
in Sec. \ref{sec2} and the rescalings 
\reff{chirisc-sigma-2d}, \reff{tSR-that-sigma-2d},
\reff{chi-chiSR-3d-sigma}, and \reff{tSR-that-3d-sigma}.
In two dimensions we can compare our results for 
$\gamma_{\rm eff}(\widetilde{t})$ with the numerical ones of 
Ref. \cite{L-B-B-prl}.
In Fig. \ref{gammaeff2d} we report also the curve
\be
\gamma_{\rm eff}(\widetilde{t}) = 1 + 
   {3\over 4} { 1 + 0.339\, \widetilde{t}^{\, 1/2}\over 
                1 - 0.115\, \widetilde{t}^{\, 1/2} + 4.027\, \widetilde{t}}\, ,
\ee
whis is a rough interpolation of the numerical data. 
The agreement is very good,
showing nicely the equivalence of medium-range and field-theory
calculations.

In Figs. \ref{gammaeff} and \ref{betaeff} we report the results for 
$\gamma_{\rm eff}(\widetilde{t})$ and 
$\beta_{\rm eff}(\widetilde{t})$ in three dimensions. 
As already discussed in Ref. \cite{Luijten-Binder_98}, in the 
high-temperature phase, $\gamma_{\rm eff}(\widetilde{t})$ agrees 
nicely with the Monte Carlo data in the mean-field region 
while discrepancies appear in the 
neighbourhood of the Wilson-Fisher point. However,
for $\widetilde{t}\to 0$, only data with small values of $\rho$
are present, so that the differences that are observed should be 
due to the corrections to the universal behaviour. 
The low-temperature phase 
shows a similar behaviour: good agreement in the mean-field region,
and a difference near the Wilson-Fisher point where again only
point with small $\rho$ are available \cite{Luijten-private}.
We can also compare the results for the magnetization. 
In Fig. \ref{scalingrel} we report the combination
$2 - \gamma^{-}_{\rm eff} - 2 \beta_{\rm eff}$ which should be 
compared with the analogous figure appearing in 
Ref. \cite{Luijten-Binder_98}: the behaviour of the two curves is 
completely analogous.

We wish now to discuss a different approach to the
crossover that has been developed in Refs.
\cite{Chen-etal_90,Chen-etal_90b,Anisimov-etal_92,Anisimov-etal_96}
following the so-called RG matching
\cite{Nicoll-Bhattacharjee_81,Nicoll-Albright_85} and that has been
applied successfully to many different experimental situations
\cite{Chen-etal_90,Chen-etal_90b,Anisimov-etal_95,Anisimov-etal_96,%
Melnichenko-etal_97,Jacob-etal_98}. These papers consider phenomenological
parametrizations  which are able to describe the crossover even outside 
the universal critical regime. Let us now introduce this model 
in the formulation of Ref. \cite{Luijten-Maryland}, which is intended to 
apply directly to our class of Hamiltonians. If $t$ is the reduced 
temperature, one introduces two functions $\kappa(t)$ and $Y(t)$
defined by the set of equations
\bea
\kappa(t)^2 &=& c_t\, t\, Y(t)^{(2\nu - 1)/\Delta},
\\
1 - (1 - \overline{u}) Y(t) &=& \overline{u} 
  \left[1 + \left( {\Lambda\over \kappa(t)}\right)^2\right]^{1/2} 
  Y(t)^{\nu/\Delta}.
\label{defY-Sengers}
\eea
Notice that, although three non-universal constants
$c_t$, $\Lambda$ and $\overline{u}$ appear in these equations,
$Y(t)$ and $\kappa^2(t)/c_t$ depend only 
on $\overline{u}$ and on the combination $\sqrt{c_t}/\Lambda$. 
The susceptibility is given by
\be
\chi^{-1} = c_\rho^2 c_t {t\over t+1} Y(t)^{(\gamma-1)/\Delta} (1 + y),
\ee
where
\be
y = {u^* \nu\over 2 \Delta} 
  \left\{ 2 \left({\kappa(t)\over \Lambda}\right)^2 
      \left[1 + \left( {\Lambda\over \kappa(t)}\right)^2\right]
      \left( {\nu\over \Delta} + 
       {(1 - \overline{u}) Y(t)\over 1 - (1 - \overline{u}) Y(t)}\right)
       - {2 \nu - 1\over \Delta}\right\}^{-1},
\label{def-y-Sengers}
\ee
where $u^*$ is a numerical constant, $u^* = 0.472$, and $c_\rho$ is another
normalization non-universal parameter. Notice that $\chi^{-1}/(c_\rho^2 c_t)$
depends only on $\overline{u}$ and $\sqrt{c_t}/\Lambda$, so that 
$\Lambda$ or $c_t$ could be fixed to any value without loss of generality.
In order to interpret the 
Monte Carlo results of Ref. \cite{Luijten-Binder_98}, 
Ref. \cite{Luijten-Maryland} further 
assumes that $c_t$ and $\overline{u}$ scale as 
\be
c_t = {c_{t0}\over R^2}, \qquad\qquad 
\overline{u} =\, {\overline{u}_0\over R^4}.
\label{scaling-costanti-Luijten}
\ee
In order to have the correct scaling of $\chi$, one should also set 
$c_\rho = c_{\rho 0} R$. Then in the critical crossover limit 
$t\to0$, $R\to\infty$, with $\widetilde{t}\equiv tR^6$ fixed, we obtain
\be
\widetilde{\chi}^{-1}=\, 
 c_{\rho 0}^2 c_{t 0}\, \widetilde{t}\,
   Y_0(\widetilde{t})^{(\gamma - 1)/\Delta}\, 
   \left[1 + {u^*\nu\over 2} 
   {1 - Y_0(\widetilde{t})\over 1 + (2\Delta - 1)Y_0(\widetilde{t})}\right],
\label{fchi-Sengers}
\ee
where $Y_0(\widetilde{t})$ satisfies the equation 
\be
1 - Y_0(\widetilde{t}) =\, 
   {1\over \sqrt{\alpha^2 \widetilde{t}}} \,
  Y_0(\widetilde{t})^{1/2\Delta},
\ee
with 
\be
\alpha \equiv\, {\sqrt{c_{t0}}\over \overline{u}_0 \Lambda}.
\ee
Eq. \reff{fchi-Sengers} defines the universal crossover function 
in this approach,
the model-dependence being included in the constants $\alpha$ and 
$c_{\rho 0}^2 c_{t 0}$. In order to understand the accuracy of this approach
we can compare $\widetilde{\chi}$ obtained from Eq. \reff{fchi-Sengers} 
with the very precise results of Bagnuls and Bervillier
\cite{Bagnuls-Bervillier_84,Bagnuls-Bervillier_85}. 
First of all let us compare the 
asymptotic behaviour for $\widetilde{t}\to0$ and $\widetilde{t}\to\infty$.
In the mean-field limit $\widetilde{t}\to\infty$ we have 
\be
\widetilde{\chi}^{-1}=\, c_{\rho 0}^2 c_{t 0}\, \widetilde{t}\, 
  \left[1 - {g_2\over \sqrt{\alpha^2\widetilde{t}}} + 
   O(\widetilde{t}^{-1})\right] ,
\ee
where $g_2\approx 0.311$. Using the results\footnote{Notice 
that we use a different normalization for $R^2$: our $R^2$ is 
$1/6$ of $R^2$ used in Ref. \cite{Luijten-Binder_98}.}
of the fit of 
Ref. \cite{Luijten-Maryland}, $\overline{u}_0 = 1.22/36$,
$c_{t0} = 1.72/6$, $\Lambda = \pi$, we have
\be
\widetilde{\chi}^{-1}=\, c_{\rho 0}^2 c_{t 0}\, \widetilde{t}\,
  \left[1 - {a\over \sqrt{\widetilde{t}}} +
   O(\widetilde{t}^{-1})\right],
\ee
with $a \approx 0.062$, to be compared with the exact result,
cf. Eq. \reff{fchi-2loop-3d}, $a = 1/(4 \pi) \approx 0.0796$.
Notice that if we wish to reproduce the correct behaviour for 
$\widetilde{t}\to\infty$, we should also require 
$c_{\rho 0}^2 c_{t 0} = 1$.
Analogously for $\widetilde{t}\to 0$ we have
\be
\widetilde{\chi}^{-1}=\, c_{\rho 0}^2 c_{t 0} 
 \left(1 + {u^*\nu\over2}\right) \alpha^{2(\gamma - 1)}
\widetilde{t}^{\gamma}\, 
  \left(1 - g_1 \alpha^{2\Delta} \widetilde{t}^\Delta 
        + O(\widetilde{t}^{2\Delta}) \right),
\ee
where $g_1\approx 0.618$. Using $c_{\rho 0}^2 c_{t 0} = 1$, we obtain 
numerically
\be
\widetilde{\chi}^{-1}= 2.49\, \widetilde{t}^{\gamma}
  \left(1 - 3.42\, \widetilde{t}^\Delta
        + O(\widetilde{t}^{2\Delta}) \right),
\label{chi-smallt-Sengers}
\ee
to be compared with 
\be
\widetilde{\chi}^{-1}= (2.70\pm 0.04)\, \widetilde{t}^{\gamma}
  \left(1 - (4.0\pm 0.1)\, \widetilde{t}^\Delta
        + O(\widetilde{t}^{2\Delta}) \right),
\label{chi-smallt-sec6}
\ee
obtained using the results of Ref. \cite{Bagnuls-Bervillier_85} and 
Eqs. \reff{chi-chiSR-3d-sigma}, \reff{tSR-that-3d-sigma}. 
Finally we report in Fig. \ref{confrontoSengersBB}
\be
\Delta_\chi(\widetilde{t}) = 
   {f_{\chi,{\rm phen}}(\widetilde{t})\over 
    f_{\chi,{\rm BB}} (\widetilde{t})}\; ,
\ee
where $f_{\chi,{\rm phen}}(\widetilde{t})$ is given by Eq. 
\reff{fchi-Sengers} with the numerical values of 
Ref. \cite{Luijten-Maryland}, and $f_{\chi,{\rm BB}} (\widetilde{t})$ 
is obtained using the expressions of 
Ref. \cite{Bagnuls-Bervillier_85} and fixing the non-universal constants 
with the help of Eqs. \reff{chi-chiSR-3d-sigma}, \reff{tSR-that-3d-sigma} 
(therefore $f_{\chi,{\rm BB}} (\widetilde{t})$  does not 
have any free parameter). The agreement is overall good --- the difference
is less than 1.5\% --- except in a small neighbourhood of the 
Wilson-Fisher point where the difference increases to 8\% 
as it can be seen comparing Eqs. \reff{chi-smallt-Sengers} and 
\reff{chi-smallt-sec6}. Notice however that the region where the 
discrepancies are large is outside the domain investigated in the 
Monte Carlo simulation of Ref. \cite{Luijten-Binder_98}.
Similar discrepancies 
were already observed in Ref. \cite{Anisimov-etal_92}.

Let us now consider the corrections to the leading behaviour.
If we use the expressions \reff{scaling-costanti-Luijten} we find corrections
of order $R^{-4}$, in contrast with the theoretical analysis we have presented.
However there is a simple modification  
that gives the correct corrections and that 
does not change the leading behaviour we have discussed before. 
It is enough to assume that, for $R\to\infty$,
\be
\overline{u} \to {\overline{u}_0\over R^3},\qquad 
c_{t}\to c_{t0}, \qquad c_{\rho} \to c_{\rho 0}.
\ee
Notice that this scaling of $\overline{u}$ is more natural, since,
as we discussed in the introduction and in Ref. \cite{PRV_longrange}, 
any coupling constant $u$ should 
scale as $R^{-3}$ in the crossover limit. 
Observe also that if one wishes to keep the interpretation of 
$\kappa(t)$ as an inverse correlation length, then 
$\kappa(t) \sim (R/\xi)$, i.e. $1/\kappa(t)$ is a correlation measured in units
of the interaction range.
Using these rescalings, we can write
\be
\widetilde{\chi} = 
  f_\chi(\widetilde{t}) + 
  {\overline{u}_0\over R^3} g_\chi(\widetilde{t}) + O(R^{-6}),
\ee
where $g_\chi(\widetilde{t})$ depends only on $\alpha^2 \widetilde{t}$
apart from a multiplicative constant. By means of an explicit computation
we obtain
\bea
\hskip -1truecm
{g_\chi(\widetilde{t})\over f_\chi(\widetilde{t})} &= &
\, - {2 (\gamma - 1) Y_0(\widetilde{t})\over 
      (2 \Delta - 1) Y_0(\widetilde{t})+1} 
\nonumber \\
&& \hskip -1truecm
- \left[(2 \Delta - 1) Y_0(\widetilde{t}) + 1 
   + {u^*\nu\over2} (1 - Y_0(\widetilde{t}))\right]^{-1}
    {\Delta u^* \nu\, Y_0(\widetilde{t}) (1 - Y_0(\widetilde{t}))\over 
      [(2 \Delta - 1) Y_0(\widetilde{t})+1]^2}.
\label{gchi-Sengers}
\eea
For $\widetilde{t}\to \infty$ we have
\be
{g_\chi(\widetilde{t})\over f_\chi(\widetilde{t})} \to
   - {\gamma - 1\over \Delta} \approx - 0.45,
\label{gchiSlarget}
\ee
while, for $\widetilde{t}\to 0$, we have
\be
{g_\chi(\widetilde{t})\over f_\chi(\widetilde{t})} \approx 
  -g_1 \alpha^{2\Delta} \widetilde{t}^\Delta.
\ee
The behaviour ${g_\chi(\widetilde{t})/ f_\chi(\widetilde{t})} \sim 
\widetilde{t}^\Delta$ for $\widetilde{t}\to 0$ is not what
one should expect in general, see Fig. \ref{correzioni_largeN} 
for an example in 
the large-$N$ limit, and it is related to our assumptions on
$c_\rho$ and $c_t$. If we include $1/R^3$ corrections, i.e. assume
\be
c_\rho = c_{\rho 0} + {c_{\rho 1}\over R^3}, \qquad
c_t = c_{t 0} + {c_{t 1}\over R^3}, 
\ee
then ${g_\chi(\widetilde{t})/f_\chi(\widetilde{t})}$ would tend to a 
non-vanishing constant for $\widetilde{t}\to 0$. For $\widetilde{t}\to\infty$,
we should compare Eq. \reff{gchiSlarget} with the exact result 
${g_\chi(\widetilde{t})/f_\chi(\widetilde{t})} = -E_3/\overline{u}_0$
derived in Sec. \ref{sec5.1}.

A graph of ${g_\chi(\widetilde{t})/ f_\chi(\widetilde{t})}$ as a
function of $\alpha^2 \widetilde{t}$ is reported in Fig. 
\ref{ratiofg_Sengers}.
It shows a behaviour analogous to that found in the large-$N$ limit,
and also the numerical size of the corrections is similar.

It should be emphasized
that the function $g_\chi(\widetilde{t})$ is non-universal and that it 
cannot be determined in continuum field theory. 
Therefore 
the expression \reff{gchi-Sengers} cannot be justified and represents 
some natural --- but nonetheless totally arbitrary --- generalization of the
field-theory results. For the model at hand it provides a reasonable 
qualitative
approximation, but this is not true for any model one can consider. For 
instance, in the large-$N$ limit, the ratio 
${g_\chi(\widetilde{t})/f_\chi(\widetilde{t})}$ can be either 
decreasing or increasing, see Fig. \ref{correzioni_largeN}, depending 
on the Hamiltonian. On the other hand, 
the function in Eq. \reff{gchi-Sengers} is decreasing for any choice of the
parameters. Therefore this
phenomenological extension is not even guaranteed to be qualitatively 
correct. 
If one is interested in phenomenological interpolations that can describe the 
crossover even outside the universal regime, one could proceed in a 
more straightforward way, distinguishing clearly what can be predicted 
using the field-theory approach (the limiting universal curve) and what
is introduced phenomenologically (the corrections to the universal
behaviour). For instance one could 
use the essentially exact $f_\chi(\widetilde{t})$ derived from 
perturbative field theory
and any arbitrary reasonable definition for the corrections depending
on some parameters that could be fitted to obtain the best agreement between
data and model. In this way one could also obtain good phenomenological 
interpolations of the numerical (or experimental) data.

Finally we wish to comment on the role played by the terms 
$[1 + \Lambda^2/\kappa(t)^2]$ appearing in Eqs. 
\reff{defY-Sengers}, \reff{def-y-Sengers}. In our large-$R$ expansion
they can be simply replaced by $\Lambda^2/\kappa(t)^2$ with 
corrections of order $R^{-6}$ (of order $R^{-8}$ with the original scalings).
Therefore, these terms that were introduced in 
Refs. \cite{Chen-etal_90,Chen-etal_90b} in order to improve 
the behaviour in the mean-field region, represents a way 
to introduce additional corrections of order $R^{-6}$. In the analysis of the 
numerical results of Ref. \cite{Luijten-Binder_98} they play little
role, since
\be
   {\kappa(t)^2\over \Lambda^2} = {c_t\over \Lambda^2} 
     t Y(t)^{(2\nu-1)/\Delta}\approx 0.17 {t\over 6 R^2} 
     Y(t)^{(2\nu-1)/\Delta}
\ee
and $t\ltapprox 0.05$, $Y(t)\ltapprox 1$, $6 R^2 \gtapprox 1$. In practice 
the leading term and the first correction $g_\chi(\widetilde{t})$ already 
provide a good interpolation of the data of Ref. \cite{Luijten-Binder_98}.

\section*{Acknowledgments}
We thank Erik Luijten for useful correspondence.

\appendix

\section{Integrals with medium-range propagators} \label{AppA}

\subsection{Lattice propagators}   \label{AppA.1}

In this appendix we will compute the quantity
\be
\sum_{x\in D} e^{ik\cdot x},
\label{eq:A.1}
\ee
for two choices of interaction domain $D\subset Z^d$. For integer $\rho$ 
we define
\begin{eqnarray}
D^{(1)}_\rho
   &\equiv& \left\{x\in Z^d\, :\, |x_i|\le \rho\;\;\;
       {\rm for}\;\;\; i:1,\ldots,d\right\},
\label{domainD1}
\\
D^{(2)}_\rho
   &\equiv& \left\{x\in Z^d\, :\, \sum_{i=1}^d |x_i| \le \rho\right\}.
\end{eqnarray}
In order to compare with the numerical results of Refs. 
\cite{L-B-B-pre,L-B-B-prl,Luijten-Binder_98,Luijten-FSS} 
we will be also interested in the following family of domains
\be
D^{(3)}_\rho
   \equiv\, \left\{x\in Z^d\, :\, \sum_{i=1}^d x_i^2 \le \rho^2\right\}.
\ee
We will not be able to compute \reff{eq:A.1} for this class of domains. 
However we will obtain some numerical results that will be used in the main
text.

Let us compute \reff{eq:A.1} for $D^{(1)}_\rho$. 
The computation is trivial and we obtain
\be
\Omega^{(1)}_{\rho,d} \; \equiv \;
\sum_{x\in D^{(1)}} e^{ik\cdot x} \; =\;
\prod_{i=1}^d {\sin k_i L\over \sin (k_i/2)}\; ,
\ee
where $L = \rho + 1/2$. Correspondingly we find
\be
V_\rho\, =\, (2\rho+1)^d = (2 L)^d, \qquad \qquad
R^2 \,=\, {1\over6} \rho(\rho+1) = {4L^2 -1\over 24}.
\ee

Let us now consider the second case. The computation is now much more
involved. The result can be expressed in terms of the
determinant of two $d$-dimensional matrices. Define
\begin{eqnarray}
A_{ij} &\equiv& \cases{
       (\cos k_j)^{i-1} & \quad for\ $i=1,\ldots,d-1$,\ $j=1,\ldots, d$; \cr
       f_d(k_j)         & \quad for\ $i=d$, $j=1,\ldots, d$;
       } 
\\[2mm]
B_{ij} &\equiv& (\cos k_j)^{i-1}.
\end{eqnarray}
Then 
\be
\Omega^{(2)}_{\rho,d} \; \equiv \;
\sum_{x\in D^{(2)}} e^{ik\cdot x} \; =\;
  {{\rm det}\, A \over {\rm det}\, B}\; .
\label{eqA8}
\ee
The result depends on the function of a single variable $f_d(k)$ given by
\be
f_d(k)\, =\; - 2 \cos{k\over2}\; \left(\sin k\right)^{d-2} \, 
                 \cos\left( k L + {d \pi\over2}\right)
\ee
where, as before, $L = \rho + 1/2$. Explicitly in two and three
dimensions we have
\begin{eqnarray}
f_2(k) &=& 2 \cos{k\over2}\, \cos kL, \\
f_3(k) &=& - 2 \cos{k\over2}\, \sin k\, \sin kL.
\end{eqnarray}

Expanding in powers of $k$ it is possible to compute $V_\rho$ and 
$R$. In two dimensions we obtain
\begin{eqnarray}
V_\rho &=& 2 \rho^2 + 2\rho + 1\, =\, {1\over2} (4 L^2 + 1), \\
V_\rho R^2 &=& {1\over6}\rho(1+\rho)(1+\rho+\rho^2) 
       \, =\, {1\over 96} (4 L^2 + 3) (4 L^2 - 1) .
\end{eqnarray}
In three dimensions we have
\begin{eqnarray}
V_\rho &=& {1\over3} (2\rho + 1) (2 \rho^2 + 2\rho + 3) \, =\, 
   {L\over3} (4 L^2 + 5), \\
V_\rho R^2 &=& {1\over30} \rho (\rho + 1) (2\rho + 1) (\rho^2 + \rho + 3)
\, =\, {1\over 240} L (4 L^2 - 1) (4 L^2 + 11).
\end{eqnarray}
For large values of $\rho$ one finds 
\be
V_\rho \to {(2 L)^d\over d!}, \qquad \qquad
R^2 \to {L^2\over (d+1) (d+2)}.
\ee
To prove Eq. \reff{eqA8}, let us suppose that the result has the form
\be
\Omega^{(2)}_{\rho,d} (k) =\; 
i^d \sum_{i=1}^d \alpha_{i,d} \left(e^{ik_i L} + \beta_{i,d} e^{-ik_i L}\right),
\label{Ansatz}
\ee
where $\alpha_{i,d}$ and $\beta_{i,d}$
depend on $k$ but not on $L$. This Ansatz is a 
natural generalization of the result that can be obtained in two and three
dimensions by direct computation. Using the fact that
\be
\Omega^{(2)}_{\rho,d} (k_1,\ldots,k_d) =\; 
\sum_{n=-\rho}^\rho e^{ik_d n} 
\Omega^{(2)}_{{\rho-|n|},d-1} (k_1,\ldots,k_{d-1}),  
\label{ricorrenza}
\ee
we obtain $\beta_{i,d} = (-1)^d$ and the following recursion relations:
\begin{eqnarray}
&& \hskip -1truecm
\alpha_{i,d} \; =\; \alpha_{i,d-1} {\sin k_i\over \cos k_i - \cos k_d} ,
\label{alphaidrec}
\\
&& \hskip -1truecm
\alpha_{d,d} \; =\; \sum_{i=1}^{d-1} \alpha_{i,d-1} 
       {1\over \cos k_i - \cos k_d} 
\left[\sin\left( {k_i - k_d\over 2}\right) -
     (-1)^d \sin\left( {k_i + k_d\over2}\right)\right],
\label{alphaddrec}
\end{eqnarray}
where in Eq. \reff{alphaidrec} $i=1,\ldots,d-1$. Using 
$\alpha_{1,1} = (2 \sin (k/2))^{-1}$, Eq. \reff{alphaidrec}, and 
the obvious symmetry under permutation of the labels of the coordinates,
we obtain, for $1\le i\le d$,
\be
\alpha_{i,d}\; =\; 
    {(\sin k_i)^{d-2}\over \prod_{j=1,j\not=i}^d (\cos k_i - \cos k_j)}\,
    \cos{k_i\over2}. 
\label{alphaid}
\ee
We should now prove that this expression solves Eq. \reff{alphaddrec} which
is the consistency condition of the Ansatz \reff{Ansatz}.
Assuming $d$ even, we can rewrite Eq. \reff{alphaddrec} as
\be
\sum_{i=1}^d {(\sin k_i)^{d-2} \over 
     \prod_{j=1,j\not=i}^d(\cos k_i - \cos k_j)} =\, 0\; .
\label{codizionedd}
\ee
Let us now use the following result:
given $x_1,\ldots,x_n$, consider the $n$-dimensional
matrix $M_{ij} = x_i^{j-1}$. Then it is easy to see that 
(in the mathematical literature this determinant is known as 
Vandermonde determinant)
\be
\det M\; =\; (-1)^{n(n-1)/2} \prod_{i=1}^{n-1} \prod_{j=i+1}^n (x_i - x_j).
\label{detToeplitz}
\ee
Now define the matrix
\be
C_{ij} \equiv\, \cases{(\cos k_j)^{i-1} &\qquad for \ $i=1,\ldots,d-1$; \cr
                  (\sin k_j)^{d-2} &\qquad for \ $i=d$.}
\ee
Using Eq. \reff{detToeplitz},
it is easy to convince oneself that Eq. \reff{codizionedd} can be written as 
\be
{\det C\over \det B} = 0.
\ee
Since $d$ is even, 
one can express $(\sin k_j)^{d-2}$ as a sum of $\cos^{2i}k_j$, with 
$0\le i\le d-1$. Thus the last row of $C$ is a linear combination of the
previous rows and therefore $\det C = 0$. When $d$ is odd 
the discussion is analogous.
We have thus proved that the consistency condition 
\reff{alphaddrec} is satisfied. Therefore the Ansatz 
\reff{Ansatz} with $\alpha_{id}$ given by Eq. \reff{alphaid} 
is the solution of the recurrence relation \reff{ricorrenza} that
uniquely defines $\Omega^{(2)}_{\rho,d}(k)$.
Using again Eq. \reff{detToeplitz} we obtain the result
\reff{eqA8}.

If $J_\rho(x)$ defined in Sec. \ref{sec3} is given by  Eq. \reff{Jrhoflat},
then $J_\rho(q) = \Omega_{\rho,d}(q)$. We wish now to prove that
$\Omega_{\rho,d}(q)$ satisfies the properties mentioned at the 
beginning of Sect. \ref{sec3}. Property (i) is obvious, while properties
(ii) and (iii) depend on $D_\rho$: they are satisfied if
$V_\rho\sim R^d$ and if, for 
any $x\in Z^d$, there exist\footnote{Notice that it is not sufficient
that $V_\rho\sim R^d$ to ensure property (iii). For instance consider in 
one dimension the set $D_\rho = \{x = 2 n, \, n\in Z,\,  |n|\le \rho\}$.}
$x_1,\ldots, x_d\in D_\rho$ such that $x = \sum_i \alpha_i x_i$, 
$\alpha_i\in Z$. To check the 
fourth property, define $v\equiv k R$ and consider the limit of 
$\Omega_{\rho,d}(v/R)$ at fixed $v$. An easy computation for the 
domains $D^{(1)}$ and $D^{(2)}$ gives
\be
{\Omega_{\rho,d}(v/R)\over V_R} \to \Omega_0(v)
\label{defOmega0}
\ee
with
\begin{eqnarray}
\Omega_0(v)  &=&
\prod_{i=1}^d {\sin u_i\over u_i},
\\
\Omega_0(v) &=&
- d! \sum_{i=1}^d {u_i^{d-2} \cos (u_i + d \pi/2) \over
                   \prod_{j\not=i} (u^2_j - u^2_i) },
\end{eqnarray}
where
\be
u = v\; \lim_{R\to\infty} {L\over R}.
\ee
Notice that in two dimensions there is a simple relation for $\Omega_0(v)$ 
for the two domains $D^{(1)}$ and $D^{(2)}$. Indeed
\be
\Omega_0^{(2)}(u_x,u_y) = \Omega_0^{(1)}
\left( {u_x+u_y\over2}, {u_x-u_y\over2} \right).
\label{relation-Omega0-D1-D2}
\ee
For the family of domains $D^{(3)}$, $\Omega_0(v)$ was computed in App. A
of Ref. \cite{L-B-B-pre} finding
\be
\Omega_0(v) =\, \Gamma\left({d\over2} + 1\right)
                \left({|u|\over2}\right)^{-d/2} J_{d/2}(|u|),
\ee
where
\be
u_i = \sqrt{2(d+2)}\, v_i.
\ee
From these expressions
it is easy to see that $\Pi(q) = 1 - \Omega_0(q)$ satisfies condition (iv).

For $D^{(1)}$ and $D^{(2)}$ it is also easy to show explicitly 
properties (v) and (vi) of Sec. \ref{sec5}. Indeed in the limit
$R\to\infty$ at $v$ fixed we have
\be
{1\over V_\rho}\, \Omega_{\rho,d}(k) \; =\;
   \sum_{n=0}^\infty {1\over R^{2n}} \Omega_n(v),
\label{expanOmega}
\ee
without odd powers of $1/R$. Indeed, from the explicit results
we immediately see that $\Omega_{\rho,d}(k)$ is even under
the transformations $k\to -k$ and $L\to -L$. Moreover $L^2$ is
an analytic function of $R^2$. Therefore for $v$ fixed,
$\Omega_{\rho,d}(k)$ is even in $R$, proving
Eq. \reff{expanOmega}.
Since we used the explicit expressions we computed before this proof
applies only to the two cases we have studied. However we
conjecture this is a general property of every family of domains
that is cubic invariant.

\subsection{One-loop integrals} \label{AppA.2}

Let us now consider, for $d<4$, the following class of one-loop integrals:
\bea
I_{1,R}(m^2) &\equiv&\, \int {d^dk\over (2\pi)^d} 
   {1-\overline{\Pi}_R(k)\over \overline{\Pi}_R(k) + m^2}, 
\label{I1R-def} \\
J_{1,R}(m^2) &\equiv&\, \int {d^dk\over (2\pi)^d} 
   {1-\overline{\Pi}_R(k)\over (\overline{\Pi}_R(k) + m^2)^2}, 
\label{J1R-def}
\eea
where $\overline{\Pi}_R(q)$ is a function with the properties 
mentioned at the beginning of Sect. \ref{sec3} and \ref{sec5}. 
We will be interested in computing these
integrals in the crossover limit that corresponds to $R\to\infty$,
$m\to 0$, with $m^2 R^{2d/(4-d)}$ fixed. The integral \reff{I1R-def} 
exists for $m=0$ only for $d>2$, while the second one is always infrared 
divergent. In order to analyze the asymptotic behaviour of these integrals
we will distinguish three cases: (a) $d>2$; (b) $d<2$; (c) $d=2$.

\subsubsection{Case (a): $2<d<4$} \label{AppA.2.1}

For $d>2$ the integral \reff{I1R-def} is well
defined for $m^2\to 0$ and thus we begin by studying 
\be
\overline{I}_{1,R}\equiv\lim_{m^2\to 0} I_{1,R}(m^2).
\label{Ibar1R-def}
\ee
We wish to compute its asymptotic behaviour for $R\to\infty$. Defining 
$p=k R$, we rewrite
\be
\overline{I}_{1,R} = 
   {1\over R^d} \int {d^dp\over (2\pi)^d}
   {1-\overline{\Pi}_R(p/R)\over \overline{\Pi}_R(p/R) + m^2}
     \to {1\over R^d} \int {d^dp\over (2\pi)^d} 
    {1 - \Pi(p)\over \Pi(p)},
\ee
using property (iv). The last integral can be extended over all $\R^d$.
Thus we obtain
\be
\overline{I}_{1,R} \approx\, {\sigma\over R^d} 
\label{Ibar1R-asintotico}
\ee
with
\be
\sigma\, \equiv\, \int {d^dp\over (2\pi)^d} 
    {1 - \Pi(p)\over \Pi(p)}.
\label{sigma-def}
\ee
If additionally we assume that $\overline{\Pi}_R(q)$ satisfies properties
(v) and (vi) of Sec. \ref{sec5},
we can easily prove that $\overline{I}_{1,R}$ admits an 
expansion of the form
\be
\overline{I}_{1,R} =\, 
   {\sigma\over R^d} + {1\over R^d} \sum_{n=1}^\infty {\sigma_n\over R^{2n}}.
\ee
In three dimensions, for 
$D^{(1)}$, $D^{(2)}$, $D^{(3)}$, we have respectively\footnote{
Luijten \cite{Luijten-FSS} has noticed that an approximate expression 
of $\sigma$ can be obtained using the numerical results of 
Ref. \cite{Dalton-Domb_66}. He writes 
$\sigma\approx 4.46 \lim_{R\to\infty} R^3/V_R$.
For the three domains one obtains $0.0379$, $0.0374$, $0.0337$ respectively;
these estimates are not very far from the exact values.}
\begin{eqnarray}
\overline{I}_{1,R} &=& {0.0435562069\over R^3} + O(R^{-5}), \nonumber \\
\overline{I}_{1,R} &=& {0.04336529\over R^3} + O(R^{-5}), \nonumber \\
\overline{I}_{1,R} &=& {0.04139\over R^3} + o(R^{-3}).
\end{eqnarray}
Estimates of $\overline{I}_{1,R}$ for various values of $R$ are reported 
in Table \ref{table_sigma}.
\begin{table}
\begin{center}
\begin{tabular}{|l|rrr|}
\hline
$\rho$ & $D^{(1)}$ & $D^{(2)}$ & $D^{(3)}$ \\
\hline
3  & 0.042971778 & 0.043960387 & 0.041600702 \\
4  & 0.043202728 & 0.043921767 & 0.041279504 \\
5  & 0.043319601 & 0.043713672 & 0.041423800 \\
6  & 0.043386811 & 0.043664053 & 0.041387988 \\
7  & 0.043428975 & 0.043574469 & 0.041394901 \\
8  & 0.043457153 & 0.043547206 & 0.041384933 \\
10 & 0.043491295 & 0.043486698 & 0.041398502 \\
12 & 0.043510406 & 0.043451767 & 0.041386965 \\
14 & 0.043522170 & 0.043429899 & 0.041392394 \\
16 & 0.043529921 & 0.043415345 & 0.041389669 \\
18 & 0.043535297 & 0.043405187 & 0.041392740 \\
20 & 0.043539178 & 0.043397824 & 0.041391612 \\
\hline
\end{tabular}
\end{center}
\caption{Estimates of $R^3\overline{I}_{1,R}$ for various values of $\rho$
for the three domains introduced in the text.
 }
\label{table_sigma}
\end{table}

Let us now go back to $I_{1,R}(m^2)$ and let us compute the leading correction
depending on $m^2$ in the crossover limit. We rewrite
\be
I_{1,R}(m^2) = \overline{I}_{1,R} - 
   m^2 \int {d^dq\over (2\pi)^d} 
   {1 - \overline{\Pi}_R(q)\over 
   \overline{\Pi}_R(q)(\overline{\Pi}_R(q) + m^2)}.
\ee
Setting $q = v/R$ and using property (iv) of $\overline{\Pi}_R(q)$ we find
the leading term as $R\to\infty$
\be
I_{1,R}(m^2) \approx \overline{I}_{1,R} - 
   {m^2\over R^d} \int {d^dv\over (2\pi)^d}
   {1 - \Pi(v)\over \Pi(v) (\Pi(v) + m^2)},
\ee
where the integration is extended over $\R^d$. If $\overline{\Pi}_R(q)$ 
satisfies properties (v) and (vi) the neglected terms are of order
$m^2/R^{d+2}$ [in three dimensions they are of order $R^{-11}$].
Now, for $d<4$, the last
integral is infrared divergent for $m\to 0$. Thus the leading contribution
in the limit $m\to 0$ is obtained replacing $\Pi(v)$ with $v^2$ in the 
denominator and with $1$ in the numerator. We have therefore
\bea
I_{1,R}(m^2) &\approx&
   \overline{I}_{1,R} -{m^2\over R^d} \int {d^dv\over (2\pi)^d}
   {1\over v^2(v^2 + m^2)} 
\nonumber \\
   && -{m^2\over R^d} \int {d^dv\over (2\pi)^d} 
   \left[ {1 - \Pi(v)\over \Pi(v) (\Pi(v) + m^2)} - 
          {1\over v^2(v^2 + m^2)} \right].
\eea
The first integral can be computed exactly, while in the second one 
we can simply take the limit $m\to 0$.
We have finally
\be
I_{1,R}(m^2) \approx \overline{I}_{1,R} 
    + (4\pi)^{-d/2} \Gamma\left(1 - {d\over2}\right) {m^{d-2}\over R^d} + 
    E_d {m^2\over R^d} + 
    o(R^{-d(6-d)/(4-d)}),
\label{I1R-dgt2}
\ee
where
\be
E_d \equiv - \int {d^dv\over (2\pi)^d} \left[ {1 - \Pi(v)\over \Pi(v)^2} - 
  {1\over (v^2)^2}\right].
\label{Ed-def}
\ee
Numerically, in three dimensions, we have for the interaction 
\reff{Jrhoflat} and the domains $D^{(1)}$, $D^{(2)}$ and $D^{(3)}$
\bea
E_3 &=&   0.058391 \qquad \hbox{\rm for $D^{(1)}$},
\nonumber \\
E_3 &=&   0.058545 \qquad \hbox{\rm for $D^{(2)}$},
\nonumber \\
E_3 &=&   0.0635\hphantom{45} \qquad \hbox{\rm for $D^{(3)}$}. 
\eea
The computation of $J_{1,R}(m^2)$ is analogous. We obtain
\be
J_{1,R}(m^2) \approx 
   (4\pi)^{-d/2} \Gamma\left(2 - {d\over2}\right) {m^{d-4}\over R^d}
   - {E_d \over R^d} + o(R^{-d}).
\label{Jdgt2}
\ee

\subsubsection{Case (b): $d<2$} \label{AppA.2.2}

Let us now consider $I_{1,R}(m^2)$ for $d<2$. In this case the integral
is infrared divergent.
Following the previous steps, we have
\bea
I_{1,R}(m^2) &\approx&
   {1\over R^d} \int {d^dv\over (2\pi)^d} 
   {1 - \Pi(v) \over \Pi(v) + m^2} + O(R^{-d-2}) 
\nonumber \\
   &=& {1\over R^d} \int {d^dv\over (2\pi)^d} {1\over v^2 + m^2} +
       {1\over R^d} \int {d^dv\over (2\pi)^d} 
   \left[ {1 - \Pi(v) \over \Pi(v)+m^2} - {1\over v^2+m^2} \right] 
\nonumber \\
&\approx &
   (4\pi)^{-d/2} \Gamma\left(1 - {d\over2}\right) {m^{d-2}\over R^d} 
   + {1\over R^d} \int {d^dv\over (2\pi)^d}
   \left[ {1 - \Pi(v) \over \Pi(v)} - {1\over v^2} \right] + o(R^{-d}).
\nonumber \\[-2mm]
{}
\label{I1R-dlt2}
\eea
Analogously 
\be
J_{1,R}(m^2) \approx \,
(4\pi)^{-d/2} \Gamma\left(2 - {d\over2}\right) {m^{d-4}\over R^d} + 
  O(m^{d-2} R^{-d}).
\ee

\subsubsection{Case (c): $d=2$} \label{AppA.2.3}

Let us now consider the case $d=2$. Also in this case the integral is infrared
divergent for $m^2\to 0$. However we cannot proceed as in the case $d<2$,
since the subtracted integral in Eq. \reff{I1R-dlt2} is ultraviolet divergent. 
First we set $q = v/R$ and expand $\overline{\Pi}_{R}(v/R)$ obtaining
\be
I_{1,R}(m^2) \approx \, 
   {1\over R^2} \int {d^2v\over (2\pi)^2} 
   {1 - \Pi(v) \over \Pi(v) + m^2} 
   - {1+m^2\over R^4} \int {d^2v\over (2\pi)^2}
     {\Pi_1(v)\over (\Pi(v) + m^2)^2} + O(R^{-6}).
\label{I1R-2d-eq1}
\ee
Since $\Pi_1(v)\sim v^4$, we can simply take the limit $m\to 0$ in the last
term. Let us now deal with the first one. Because of the lattice symmetry
and of property (v) we have, for $v\to 0$,
\be
\Pi(v) = v^2 + \alpha_1 (v^2)^2 + \alpha_2 v^4 + o(v^4),
\label{Pi-smallv-expansion}
\ee
where $v^4 = v_1^4 + v_2^4$.
Then we rewrite 
\bea
&& \int {d^2v\over (2\pi)^2}
   {1 - \Pi(v) \over \Pi(v) + m^2} =\, 
 \int {d^2v\over (2\pi)^2} 
 \left\{ {1 - \Pi(v) \over \Pi(v) + m^2} - 
         {e^{-v^2} \over v^2 + m^2} + 
         {e^{-v^2} [\alpha_1 (v^2)^2 + \alpha_2 v^4]\over (v^2 + m^2)^2}
 \right\} 
\nonumber \\
&& + \int {d^2v\over (2\pi)^2} \left\{
     {e^{-v^2} \over v^2 + m^2} - 
     {e^{-v^2} [\alpha_1 (v^2)^2 + \alpha_2 v^4]\over (v^2 + m^2)^2}
 \right\}.
\eea
The first integral can be expanded in powers of $m^2$ neglecting terms 
of order $m^4 \log m^2$, while the second one 
can be computed exactly. We obtain finally
\bea
&& \hskip -1truecm \int {d^2v\over (2\pi)^2}
   {1 - \Pi(v) \over \Pi(v) + m^2} =\, 
   - {1\over 4\pi} (\log m^2 + \gamma_E) - 
     {m^2\over 4\pi} (\log m^2 + \gamma_E - 1)  
\nonumber \\
&& \hskip -0.5truecm
- {m^2\over 16\pi} (4\alpha_1 + 3 \alpha_2) (2 \log m^2 + 2 \gamma_E + 1)
+ \int {d^2v\over (2\pi)^2}
    \left[{1 - \Pi(v) \over \Pi(v)} - {e^{-v^2}\over v^2}\right]
\nonumber \\
&& \hskip -0.5truecm
   - m^2 \int {d^2v\over (2\pi)^2}
   \left\{ {1 - \Pi(v) \over \Pi(v)^2} - 
   {e^{-v^2}\over (v^2)^2} + 
   {2 e^{-v^2} [\alpha_1 (v^2)^2 + \alpha_2 v^4] \over (v^2)^3}\right\}
   + O(R^{-4} \log R^2)
\nonumber \\ [-2mm]
{}
\label{I1R-2d-eq2}
\eea
We obtain finally, up to terms $o(R^{-4})$,
\be
I_{1,R}(m^2) \approx\, - {1\over 4\pi R^2} \log m^2 + {C_2\over R^2} -
   {m^2\over 8\pi R^2} (4\alpha_1 + 3\alpha_2 + 2) \log m^2 + 
   + {m^2 G_1\over R^2} + {G_2\over R^4},
\label{I1R-2d}
\ee
where
\be
C_2 \equiv\, - {\gamma_E\over 4\pi} + 
   \int {d^2v\over (2\pi)^2}
   \left[ {1 - \Pi(v)\over \Pi(v)} - {e^{-v^2}\over v^2}\right],
\ee
and $G_1$ and $G_2$ can be computed from 
Eqs. \reff{I1R-2d-eq1} and \reff{I1R-2d-eq2}.
For the interaction defined in Eq. \reff{Jrhoflat} and for the 
domains $D^{(1)}$, $D^{(2)}$, $D^{(3)}$ we have respectively
\bea
C_2 &=&   -0.04578786  \qquad \hbox{\rm for $D^{(1)}$ and $D^{(2)}$ }, 
\nonumber \\
C_2 &=&   -0.05045\hphantom{786} \qquad \hbox{\rm for $D^{(3)}$}. 
\eea
The equality of $C_2$ for the domains $D^{(1)}$ and $D^{(2)}$ follows 
from the identity \reff{relation-Omega0-D1-D2}.
Analogously
\be
J_{1,R}(m^2) = \, 
 {1\over 4\pi m^2 R^2} + 
 {1\over 8 \pi R^2} (4\alpha_1 + 3 \alpha_2 + 2) (\log m^2 + 1)
 - {G_1\over R^2} + o(R^{-2}).
\ee

\subsection{The two-loop integral} \label{AppA.3}

We wish now to discuss the two-loop integral  
\be
I_{2,R}(m^2) \, \equiv\, \int {d^dq\over (2\pi)^d} {d^dk\over (2\pi)^d}\, 
{(1 - \overline{\Pi}_R(q))
 (1 - \overline{\Pi}_R(k))
 (1 - \overline{\Pi}_R(q+k))\over 
     (\overline{\Pi}_R(q) + m^2) (\overline{\Pi}_R(k) + m^2) 
     (\overline{\Pi}_R(q+k) + m^2)},
\ee
in the crossover limit $R\to\infty$, $m\to 0$ with 
$m^2 R^{2d/(4-d)}$ fixed.

Since we wish to compute the integral in the large-$R$ limit, we can
rescale the internal momenta and use properties (iv), (vi) in order to rewrite 
$I_{2,R}(m^2)$ in the form
\bea
\hskip -1.5truecm
I_{2,R}(m^2) &\approx& {1\over R^{2d}}
    \int {d^dq\over (2\pi)^d} {d^dk\over (2\pi)^d}\, 
{(1 - \Pi(q))(1 - \Pi(k))(1 - \Pi(q+k))\over 
     (\Pi(q) + m^2) (\Pi(k) + m^2) (\Pi(q+k) + m^2)}
\nonumber \\
&& \hskip -2truecm 
   - {3\over R^{2d+2}} 
    \int {d^dq\over (2\pi)^d} {d^dk\over (2\pi)^d}\,
{\Pi_1(q)(1 - \Pi(k))(1 - \Pi(q+k))\over
     (\Pi(q) + m^2)^2 (\Pi(k) + m^2) (\Pi(q+k) + m^2)}\; ,
\eea
where the integrals are extended over $\R^{2d}$. The neglected terms 
are of order $O(R^{2d+4})$ and $O(R^{2d+4} m^{2d-4})$.
$I_{2,R}(m^2)$ is infrared divergent for $d\le 3$ and therefore we will
distinguish three cases: (a) $d>3$; (b) $d<3$; (c) $d=3$.

\subsubsection{Case (a): $3<d<4$} \label{AppA.3.1}

For $d>3$ the integral is finite for $m^2\to 0$. In analogy with the 
discussion of Sec. \ref{AppA.2.1} we have
\bea
\hskip -2truecm
&& I_{2,R}(m^2) \approx \overline{I}_{2,R} + 
\nonumber \\ 
\hskip -1.5truecm
&& + {m^{2d-6}\over R^{2d}}
    \int {d^dq\over (2\pi)^d} {d^dk\over (2\pi)^d}\,
    \left[{1\over (q^2 + 1) (k^2 + 1) ((q+k)^2 + 1)} - 
          {1\over q^2 k^2 (q+k)^2}\right] + o(R^{-2d}),
\nonumber \\ [-2mm]
{}
\eea
where
\bea
\hskip -1truecm
\overline{I}_{2,R} &=& \int {d^dq\over (2\pi)^d} {d^dk\over (2\pi)^d}\,
{(1 - \overline{\Pi}_R(q))(1 - \overline{\Pi}_R(k))
 (1 - \overline{\Pi}_R(q+k))\over
     \overline{\Pi}_R(q)\overline{\Pi}_R(k)
     \overline{\Pi}_R(q+k)} 
\nonumber \\
\hskip -2.2truecm
&\approx &
{1\over R^{2 d}}
    \int {d^dq\over (2\pi)^d} {d^dk\over (2\pi)^d}\,
{(1 - \Pi(q))(1 - \Pi(k))(1 - \Pi(q+k))\over
     \Pi(q)\Pi(k)\Pi(q+k)} + O(R^{-2d-2}).
\eea

\subsubsection{Case (b): $d<3$} \label{AppA.3.2}

For $d<3$ the integral is infrared divergent for $m^2\to 0$.
As we did in Sec. \ref{AppA.2.2} we have
\bea
I_{2,R}(m^2) & \approx & {m^{2d-6}\over R^{2d}}
    \int {d^dq\over (2\pi)^d} {d^dk\over (2\pi)^d}\,
    {1\over (q^2 + 1) (k^2 + 1) ((q+k)^2 + 1)}
\nonumber \\
&&  \hskip -3.0truecm 
  + {1\over R^{2d}} 
    \int {d^dq\over (2\pi)^d} {d^dk\over (2\pi)^d}\,
   \left[{(1 - \Pi(q))(1 - \Pi(k))(1 - \Pi(q+k))\over
     \Pi(q)\Pi(k)\Pi(q+k)} - 
     {1\over q^2 k^2 (q+k)^2}\right] + o(R^{-2d}).
\nonumber \\ [-2mm]
{}
\label{I2R-lt3}
\eea
In two dimensions we have explicitly
\be
I_{2,R}(m^2) = {1\over m^2 R^4} \left({1\over 24\pi^2} \psi'(1/3) - 
       {1\over 36}\right) + O(R^{-4}).
\ee

\subsubsection{Case (c): $d=3$} \label{AppA.3.3}

For $d=3$ the integral is infrared divergent for $m^2\to 0$. However 
we cannot proceed as in the case $d<3$ because the subtracted integral in
Eq. \reff{I2R-lt3} is ultraviolet divergent. We will obtain the 
asymptotic behaviour using the same method we used in two dimensions
to deal with $I_{1,R}(m^2)$. We write
\bea
&& \hskip -1.8truecm 
  I_{2,R} (m^2) =\, {1\over R^6}  I_2^{\rm cont} (m^2)+ 
    {F_3\over R^8} + 
\nonumber \\
&& \hskip -1.2truecm {1\over R^6}
   \int {d^3q\over (2\pi)^3} {d^3k\over (2\pi)^3}\, 
  \left[
  {(1 - \Pi(q))(1 - \Pi(k))(1 - \Pi(q+k))\over \Pi(q)\Pi(k)\Pi(q+k)} - 
  {e^{-q^2 - k^2 - (q+k)^2}\over q^2 k^2 (q+k)^2}\right],
\nonumber \\ [-2mm]
{} 
\end{eqnarray}
where
\bea
I_2^{\rm cont} (m^2) &\equiv& 
   \int {d^3q\over (2\pi)^3} {d^3k\over (2\pi)^3}\,
   {e^{-q^2 - k^2 - (q+k)^2}\over  
      (q^2+m^2) (k^2+m^2) ( (q+k)^2+m^2)},
\\
F_3 &\equiv& - 3
    \int {d^3q\over (2\pi)^3} {d^3k\over (2\pi)^3}\,
{\Pi_1(q)(1 - \Pi(k))(1 - \Pi(q+k))\over
     \Pi(q)^2 \Pi(k)\Pi(q+k)}.
\eea
where all integrals are extended over $\R^6$ and terms of order 
$O(R^{-10})$ have been neglected. In order to compute $I_2^{\rm cont} (m^2)$,
let us define
\begin{eqnarray}
&& \hskip -2.0truecm 
P(x,m)  \equiv \int {d^3p\over (2\pi)^3}
      {e^{-p^2 + ip\cdot x}\over p^2 + m^2} 
\nonumber \\
&& \hskip -1truecm 
=  - {e^{m^2}\over 8\pi |x|}
   \left[ 2 \sinh m|x| + e^{-m|x|} {\rm erf}\left(m - {|x|\over2}\right) -
                         e^{m |x|} {\rm erf}\left(m + {|x|\over2}\right)
   \right],
\end{eqnarray}
where
\be
{\rm erf}(x) \, =\, {2\over \sqrt{\pi}} \int_0^x dt\, e^{-t^2}.
\ee
Then, for $m\to 0$,
\be
I_2^{\rm cont} (m^2) = \,
  - {1\over 32 \pi^2} \left[\log (9 m^2) + 2\gamma_E\right]
  +\, 4 \pi \int_0^\infty dx\, 
   \left[x^2 P^3(x,0) - {1\over 64\pi^3} {1\over x+1}\right] + O(m).
\ee
Collecting all terms we have
\be
I_{2,R}(m^2) =\, 
 - {1\over 32 \pi^2 R^6} \log  m^2 + {C_3\over R^6} + {F_3\over R^8} + 
   O(R^{-9}),
\label{I2R-d3}
\ee
where $C_3$ is defined as
\bea
\hskip -1truecm
C_3 &\equiv& - {1\over 16 \pi^2} \left(\log 3 + \gamma_E\right)
   +\, 4 \pi \int_0^\infty dx\,
   \left[x^2 P^3(x,0) - {1\over 64\pi^3} {1\over x+1}\right]
\nonumber \\
&& \hskip -2.0truecm + 
\int {d^3q\over (2\pi)^3} {d^3k\over (2\pi)^3}\,
  \left[
  {(1 - \Pi(q))(1 - \Pi(k))(1 - \Pi(q+k))\over \Pi(q)\Pi(k)\Pi(q+k)} -
  {e^{-q^2 - k^2 - (q+k)^2}\over q^2 k^2 (q+k)^2}\right].
\label{defC3}
\eea
We have  computed the constant $C_3$ for the three different domains introduced
at the beginning.
We obtain
\bea
C_3 &=&   -0.0127 \qquad \hbox{\rm for $D^{(1)}$}, 
\nonumber \\
C_3 &=&   -0.0127 \qquad \hbox{\rm for $D^{(2)}$}, 
\nonumber \\
C_3 &=&   -0.0129 \qquad \hbox{\rm for $D^{(3)}$}. 
\eea

\section{Large-$N$ limit} \label{AppB}

In this Appendix we will compute the crossover functions in the large-$N$ limit.
In App. \ref{AppB.2} we will discuss our general model for $d> 2$,
while in App. \ref{AppB.1} we will consider the two dimensional case
for the $N$-vector model.

\subsection{Crossover limit for $2<d<4$} \label{AppB.2}

In this Appendix we will study the model introduced in Sect. 3 
in the large-$N$ limit following the strategy of Refs. 
\cite{Sarbach-Fisher_78,Emery,Sarbach-Schneider_76_77}. 
We write $V(\varphi) = NW(\varphi^2)$ and study the limit
$N\to\infty$ with $\beta$ and $W(x)$ fixed. The basic trick consists
in rewriting 
\be
e^{-NW(\varphi^2)} \sim \int d\rho\, d\lambda\, 
   \exp\left[ -{N\over2} \lambda (\varphi^2 - \rho) - NW(\rho)\right].
\ee
The saddle point is given by the equations
\be
W'(\rho) = {1\over2}\beta V_R (1 + m^2), \qquad\qquad
\rho \beta V_R =\, {1\over 1 + m^2} [1 + I_{1,R} (m^2)],
\label{gapeqV}
\ee
where
$I_{1,R} (m^2)$ is defined in Eq. \reff{I1R-def}, while the two-point
function is given by
\be
\< \varphi_0\, \varphi_x \> =\, {1\over \beta}
   \int {d^d p\over \overline{\Pi}_R(p) + m^2}.
\ee
The critical point corresponds to $m^2 = 0$ and therefore 
the critical values of $\rho$ and $\beta$ satisfy the equations
\be
W'(\rho_c) = {1\over2}\beta_c V_R , \qquad\qquad
\rho_c \beta_c V_R =\,  1 + \overline{I}_{1,R}. 
\ee
The critical value $\rho_c$ is the solution of the equation
\be 
   1 + \overline{I}_{1,R} =\, 2 \rho_c W'(\rho_c).
\label{eq:B5}
\ee
We will not need to solve Eq. \reff{eq:B5} explicitly,
we will only assume $W(x)$ to be such that a positive solution exists.
Since $\overline{I}_{1,R}\sim R^{-d}$,
for $R\to\infty$ we can expand 
\be
\rho_c = \sum_{n=0}^\infty {\rho_{cn} \overline{I}^n_{1,R}},
\ee
where
\be
  2 \rho_{c0} W'(\rho_{c0}) = 1,
\label{rho0c-equation}
\ee
and $\rho_{c1}$, $\rho_{c2}$, $\ldots$ can be computed iteratively in terms 
of the derivatives of $W(\rho)$ computed at $\rho = \rho_{c0}$. Correspondingly
we obtain the expansion of $\beta_c$ for $R\to\infty$. We have
\be
\beta_c V_R =\, 
  {1\over \rho_{c0}} +\, 
  {2\rho_{c0}\, W_2\over 1 + 2 \rho_{c0}^2\, W_2}\, \overline{I}_{1,R} +
  {\rho^2_{c0} (W_3 - 4 \rho_{c0}\, W_2^2)\over (1 + 2 \rho_{c0}^2\, W_2)^3}
  \, \overline{I}_{1,R}^2 + O(R^{-3 d}).
\label{eq:B8}
\ee
where $W_n = W^{(n)}(\rho_{c0})$.
Let us now discuss the scaling behaviour. Using Eqs. 
\reff{gapeqV} and \reff{I1R-dgt2}, we have
\be
(\rho \beta - \rho_c\beta_c)V_R = \, 
  A_d {m^{d-2}\over R^d} - m^2 + 
  (E_d - \sigma) {m^2\over R^d} + o(m^2 R^{-d})
\label{gapeqV-1}
\ee
where $A_d = (4\pi)^{-d/2} \Gamma(1 - d/2)$ and 
$\sigma$ is defined in Eq. \reff{sigma-def}. 
Now let us introduce\footnote{
The normalization of $t$ is chosen so that the results can be directly
compared with those of Sect. \ref{sec5}. One could have defined 
$t\equiv (\beta_c - \beta)/\beta_c$, as in Sec. \ref{sec3}. This choice 
does not change the leading crossover curve, but changes the corrections.}
\be
\delta \equiv \rho - \rho_c, \qquad\qquad
t\equiv {\beta_c - \beta\over \beta_{c,\rm MF} }= 
 \rho_{c0} V_R(\beta_c - \beta).
\ee
Using the first gap equation and Eq. \reff{eq:B8} we obtain
\be
\delta=\, {1\over 2 \rho_{c0}\, W''(\rho_c)}
   \left[ m^2 - t + {\sigma\over s_\infty} {m^2\over R^d} \right] +\,
   O(t^2,m^4,t m^2, t R^{-2d}).
\ee
where 
\be
s_\infty = {1 + 2\rho^2_{c0}\, W_2\over 2 \rho_{c0}^2 W_2},
\label{sinfty_def}
\ee
and we have assumed $W_2\not = 0$. Going back to Eq. \reff{gapeqV-1}
we obtain finally
\be
\left(s_\infty + {s_{1,\infty}\over R^d}\right)
 (m^2 - t) - A_d {m^{d-2}\over R^d} - (E_d-\sigma) {m^2\over R^d} =
 O(t m^2, t R^{-2d}, m^2 R^{-2d}),
\label{gapeqV-2}
\ee
where 
\be
s_{1,\infty}=\, {\sigma\over \rho^2_{c0} W_2 s_\infty}
\left(1 - {W_3\over 4 \rho_{c0} W_2^2}\right).
\ee
Notice that this equation is also valid in the $N$-vector model
with $s_\infty=1$, $s_{1,\infty}=0$.
Let us now consider the critical crossover limit. We introduce 
\be
\widetilde{m}^2\equiv m^2 R^{2d/(4-d)}, \qquad\qquad
\widetilde{t} \equiv  tR^{2d/(4-d)},
\ee
and expand
\be
\widetilde{m}^2 = f_m(\widetilde{t}) + {1\over R^d} g_m(\widetilde{t}) +
    o(R^{-d}).
\ee
The universal crossover function $f_m(\widetilde{t})$ is given by
\be
s_\infty [f_m(\widetilde{t}) - \widetilde{t}\,] - A_d
          f_m(\widetilde{t})^{(d-2)/2}=\,0.
\label{eq-crossover-largeN}
\ee
This results shows clearly the universality of $f_m(\widetilde{t})$. 
Indeed all the model dependence is included in the constant $s_\infty$
that can be eliminated with a proper rescaling of $f_m(\widetilde{t})$
and $\widetilde{t}$. 

It is important to notice that Eq. \reff{eq-crossover-largeN} has a 
solution for $\widetilde{t}\to 0$ only if $s_\infty>0$. Indeed we can
rewrite the equation as 
\be
\widetilde{t} = - {A_d\over s_\infty}f_m(\widetilde{t})^{(d-2)/2}\, 
   +\, f_m(\widetilde{t}).
\ee
For $\widetilde{t}\to 0$, $f_m(\widetilde{t})\to 0$, and thus we can neglect
the last term in the right-hand side. Since the left-hand side is positive
--- we are considering the high-temperature phase --- we should have
$A_d/s_\infty < 0$. Since $A_d < 0$, we obtain $s_\infty > 0$.
Notice that, requiring the stability of the 
free energy in the limit $R\to\infty$ one obtains the condition $W_2>0$,
which, using Eq. \reff{sinfty_def}, gives again $s_\infty > 0$. 
It is important to remark, as we shall show below, that this condition is
equivalent to the requirement $\overline{a}_4<0$ that was
introduced in Sec. \ref{sec4.1}.

In three dimensions Eq. \reff{eq-crossover-largeN} can be solved explicitly 
finding
\be
f_m(\widetilde{t}) = 
   {1\over 64 \pi^2 s^2_\infty}
   \left[\sqrt{1 + 64 \pi^2 s^2_\infty \widetilde{t}} - 1\right]^2.
\ee
It is easy to compute the correction function $g_m(\widetilde{t})$, 
obtaining
\be
g_m(\widetilde{t}) =\, f_m(\widetilde{t})\,
   \left[(E_d-\sigma) - \lambda f_m(\widetilde{t})^{(d-4)/2}
   \right]
   \left[s_\infty - {d-2\over2} A_d f_m(\widetilde{t})^{(d-4)/2}\right]^{-1},
\ee
where $\lambda = {s}_{1,\infty} A_d/s_\infty$.

Let us now compute the asymptotic behaviour of $f_m(\widetilde{t})$ and
$g_m(\widetilde{t})$. For $\widetilde{t}\to\infty$ we have
\bea
f_m(\widetilde{t}) &=& 
  \widetilde{t}\left[ 1 + {A_d\over s_\infty} \widetilde{t}^{(d-4)/2} + 
        O(\widetilde{t}^{(d-4)})\right],
\\
g_m(\widetilde{t}) &=&
  {E_d - \sigma\over s_\infty} \widetilde{t} 
  \left[1 + O(\widetilde{t}^{(d-4)/2})\right].
\eea
For $\widetilde{t}\to 0$ we have
\bea
f_m(\widetilde{t}) &=& \mu \widetilde{t}^\gamma 
   \left[1 - {2\mu \widetilde{t}^\Delta\over d-2} +
    O(\widetilde{t}^{2\Delta}) \right],
\\
g_m(\widetilde{t}) &=&
    {s_{1,\infty}\over s_\infty} {2\mu \widetilde{t}^\gamma\over d-2}
    \left(1 + O(\widetilde{t}^\Delta)\right),
\eea
where $\gamma = 2/(d-2)$, $\Delta = (4-d)/(d-2)$, 
$\mu = (s_\infty/|A_d|)^\gamma$.

We can also compute the crossover function for the susceptibility. Since
\be
\chi = {1\over \beta V_R m^2},
\label{chi-largeN}
\ee
we have
\bea
f_\chi(\widetilde{t}) &=& {\rho_{c0}\over f_m(\widetilde{t})},\\
g_\chi(\widetilde{t}) &=& - {1\over\rho_{c0}} g_m(\widetilde{t})
   f_\chi(\widetilde{t})^2 - {\sigma\over s_\infty} f_\chi(\widetilde{t}).
\eea
While $f_\chi(\widetilde{t})$ is universal apart from normalization factors,
the function $g_\chi(\widetilde{t})$ is model-dependent. For this class 
of models, we have a three-parameter 
family of functions $g_\chi(\widetilde{t})$ parametrized, for instance, 
by $s_{1,\infty}$, $s_\infty$, and $E_d - \sigma$.
The fact that the $g_\chi(\widetilde{t})$ depends only on 
three parameters should be due to the large-$N$ limit. For general 
values of $N$ we expect $g_\chi(\widetilde{t})$ to be 
a non-trivial functional of the Hamiltonian.
In Fig. \ref{correzioni_largeN} we report 
$g_\chi(\widetilde{t})/f_\chi(\widetilde{t})$
for three different models: the $N$-vector model and the theories 
corresponding to the potentials
\bea
W(\varphi^2) &=& \varphi^2 + (\varphi^2 - 1)^2,
\\
W(\varphi^2) &=& \varphi^2 + (\varphi^2 - 1)^2 + (\varphi^2)^3,
\eea
using in all cases the coupling \reff{Jrhoflat} and the domain 
\reff{domainD1}. Notice that although the curves are quantitatively
different, the qualitative behaviour is similar. 

We wish now to compare the results presented above with the explicit 
calculations of Sec. \ref{sec4.3}. 
The basic ingredient is the large-$N$ expansion of the integral
\be
\int_0^\infty dx\, x^{N+k-1} e^{-NW(x^2)} =\, 
{1\over2} \int_0^\infty dy\, y^{k/2-1} 
    e^{N(\log y/2 - W(y))}.
\label{eq:B23}
\ee
The saddle point equation is 
\be
{1\over 2y} = W'(y),
\ee
which is exactly the equation defining $\rho_{c0}$, cf. 
Eq. \reff{rho0c-equation}. The large-$N$ expression of Eq. \reff{eq:B23}
is obtained defining
$y = \rho_{c0} + z/\sqrt{N}$ and expanding in powers of $1/\sqrt{N}$. 
In this way we obtain
\bea
\overline{a}_2 &=& \rho_{c0}, \\
\overline{a}_4 &=& - {6\rho_{c0}^2\over s_\infty}, 
\label{a4bar_largeN}\\
\overline{a}_6 &=& {15\over s_\infty^3 W_2^3}
   (12 \rho_{c0} W_2^2 + 16 \rho_{c0}^3 W_2^3 - W_3).
\eea
These expressions are obtained assuming $W_2>0$. As it can be seen
from Eq. \reff{a4bar_largeN}, this condition is equivalent to 
requiring $\overline{a}_4<0$.
Using the previous expressions, it is easy to verify all the formulae reported 
in Sect. \ref{sec4.3}. Notice that in the $N$-vector case 
$s_\infty= \rho_{c0} = 1$. 

All the considerations we have presented above apply if $W_2 > 0$. 
If $W_2 = 0$, but $W_3\not = 0$, the leading behaviour of $\delta$ changes
and, for $R\to\infty$, we have 
\be
\delta\to {R^d\over 2 \rho_{c0}^2 \sigma W_3}(m^2 - t) \equiv q R^d (m^2 - t).
\ee
Thus, for $R\to\infty$, keeping only the leading terms, we have
\be
(q + \rho_{c0}) (m^2 - t) - \rho_{c0} A_d {m^{d-2}\over R^{2d}} \approx 0.
\ee
If we scale
\be
\widetilde{m}^2 = R^{4d/(4-d)} m^2\qquad\qquad
\widetilde{t}   = R^{4d/(4-d)} t,
\ee
we obtain again the same crossover scaling function. 
To interpret these results we should notice that 
$W''(\rho_c) = 0$ corresponds to the tricritical point. 
Therefore, here we are considering theories that for any finite 
$R$ have a standard critical point, while for $R\to\infty$ converge to
the mean-field tricritical point which is also a Gaussian point. 
We find that the scaling crossover functions
are unchanged, although the scaling variables are different. 
In the framework of the introduction, these theories correspond to models
in which the bare coupling $u$ scales also with $R$ as $1/R^d$, so that 
the Ginzburg parameter $G$ becomes $G=(R^{-2d})^{2/(d-4)} = R^{-4d/(d-4)}$.

In order to have different crossover functions we must consider a family
a potentials such that $W''(\rho_c) = 0$ for all $R$, i.e. consider 
theories at the tricritical point for any value of $R$. With our
field normalization it is impossible to realize such a case, unless 
$W(\varphi^2)$ depends explicitly on $R$, 
i.e. $W(\varphi^2) = W(\varphi^2,R)$. Assuming therefore
$\partial^2 W(\rho_c,R)/\partial\rho^2 = 0$,
a simple computation gives 
\be
\delta^2 \approx {1\over \rho_{c0} W_3} (m^2 - t),
\ee
where $W_n = \partial^n W(\rho,\infty)/\partial\rho^n$ 
evaluated at $\rho = \rho_{c0}$.
Then we have 
\be
\left({m^2 - t\over \rho_{c0} W_3}\right)^{1/2} + 
   \rho_{c0} A_d{m^{d-2}\over R^d} = 0.
\ee
Rescaling\footnote{These rescalings can be derived in the formalism 
presented in the introduction starting from a Hamiltonian with 
a $\phi^6$ coupling.}
\be
\widetilde{m}^2 = R^{2d/(3-d)} m^2,\qquad\qquad
\widetilde{t}   = R^{2d/(3-d)} t,
\ee
we obtain the equation for the crossover function
\be
s^{\rm tr} \left(\widetilde{m}^2 - \widetilde{t}\right)^{1/2} - 
   A_d \widetilde{m}^{d-2}=\, 0.
\ee
valid, of course, for $d<3$. 
One can go further and define multicritical crossover functions. 
If $W''(\rho_c,R) = \ldots W^{(n)}(\rho_c,R) =0$, then, by a rescaling
\be
\widetilde{m}^2 = R^{2nd/(2 + 2n - nd)} m^2,\qquad\qquad
\widetilde{t}   = R^{2nd/(2 + 2n - nd)} t,
\ee
we obtain for $R\to\infty$
\be
s^{(n)} \left(\widetilde{m}^2 - \widetilde{t}\right)^{1/n} -
   A_d \widetilde{m}^{d-2} =0,
\ee
for a suitable constant $s^{(n)}$.

\subsection{Crossover limit in two dimensions} \label{AppB.1}

In this Appendix we will discuss the critical crossover limit in two
dimensions but we will consider only the $N$-vector model, since it 
already exhibits all the general features.

The gap equation is given by
\be
\beta V_R =\,  \int {d^2q\over (2\pi)^2}\, 
   {1\over \overline{\Pi}_R(q) + m^2}  = 
   {1\over 1 + m^2} (1 + I_{1,R}(m^2)).
\label{gapequation2d}
\ee
The asymptotic behaviour of $I_{1,R}(m^2)$ is reported in App. \ref{AppA.2.3}.
Now define $\widehat{t}$ from
\be
\beta V_R = 1 + {1\over 4 \pi R^2} \log R^2 + {a_0\over R^4} \log R^2 + 
  {a_1\over R^4} - {\widehat{t}\over R^2}.
\ee
Here we have introduced two free parameters $a_0$ and $a_1$ that 
represent the possible ambiguity in the definition of 
$\beta^{\rm (exp)}_{c,R}$. Then assume for $R\to\infty$ at $\widehat{t}$ fixed
that 
\be
m^2 = {1\over R^2} f_m(\widehat{t}) + {1\over R^4} g_m(\widehat{t},\log R)
   + O(R^{-6} \log R^2).
\ee
Using the gap equation we obtain for the leading term
\be
f_m(\widehat{t}) + {1\over 4\pi} \log f_m(\widehat{t}) = \widehat{t} + C_2.
\label{crossoverlimitmtilde2d}
\ee
Eq. \reff{crossoverlimitmtilde2d} defines implicitly the crossover curve for 
the correlation length $f_{\xi}(\widehat{t}) = 1/f_m(\widehat{t})$ 
and for
the susceptibility $f_\chi(\widehat{t})=f_{\xi}(\widehat{t})$. It is easy
to check that this equation has the correct limits. 
For $\widehat{t}\to -\infty$ we have the standard asymptotic-scaling behaviour
\be
f_m(\widehat{t}) =\, e^{-4\pi |\widehat{t}| + 4\pi C_2}
    \left[1 + O(e^{-4\pi |\widehat{t}|})\right],
\ee
while, for $\widehat{t}\to+\infty$, we have
\be
f_m(\widehat{t}) =\, {\widehat{t}}\, 
    \left[1 - {1\over 4\pi \widehat{t}} \log \widehat{t} +
  {C_2\over \widehat{t}} + 
   O\left(\widehat{t}^{-2}\log \widehat{t}\right)\right].
\ee
This expansion agrees with the results presented in Sec. \ref{sec4.2}.

For the correction term we obtain
\bea
\hskip -0.8truecm
g_m(\widehat{t},\log R) &=& 
   {4 \pi f_m(\widehat{t})\over 1 + 4 \pi f_m(\widehat{t})} 
   \left[ {1\over 8\pi} (4 \alpha_1 + 3\alpha_2) f_m(\widehat{t}) - a_0
   \right] \log R^2 
\nonumber \\
&& \hskip -3.2truecm
  - {4 \pi f_m(\widehat{t})\over 1 + 4 \pi f_m(\widehat{t})}
    \left[ {1\over 8\pi} (4 \alpha_1 + 3\alpha_2) f_m(\widehat{t})
    \log f_m(\widehat{t}) + a_1 - G_2 - (G_1 - C_2)f_m(\widehat{t})
    \right].
\eea
Notice that in general there is no choice of $a_0$ which allow
to cancel the logarithmic term. On the other hand, if one chooses
Hamiltonians such that $\alpha_1=\alpha_2 = 0$, the logarithmic term
cancels. This class of Hamiltonians are called 
Symanzik tree-level improved \cite{Symanzik_81}.

For $\widehat{t}\to-\infty$ we have
\be
g_m(\widehat{t},\log R) \approx 4\pi f_m(\widehat{t})
  \left[- a_0 \log R^2 + a_1 - G_2 + O(e^{-4\pi|\widehat{t}|})\right],
\ee
while for $\widehat{t}\to+\infty$ we have
\be
g_m(\widehat{t},\log R) \approx {1\over 8\pi} (4 \alpha_1 + 3\alpha_2)\,
    \widehat{t}\, \log(R^2 \widehat{t})- (G_1 - C_2)\, \widehat{t} + 
    O(\log \widehat{t}).
\ee
Notice that if one takes $a_0=0$ and $a_1 = G_2$, the corrections 
are strongly reduced in the limit $\widehat{t}\to-\infty$. 
On the other hand, in the mean-field limit, the corrections do not depend 
on the scaling of $\beta$.

Finally we wish to compute $g_\chi(\widehat{t},\log R)$. 
Using Eq. \reff{chi-largeN} we have
\be
g_\chi(\widehat{t},\log R) = \,
- {g_m(\widehat{t},\log R)\over f_m(\widehat{t})^2} + 
  {4 \pi \widehat{t} - \log R^2\over 4\pi f_m(\widehat{t})}.
\ee
For $\widehat{t}\to\infty$ we have
\be
g_\chi(\widehat{t},\log R) = \,
  - {1\over 8\pi\widehat{t}}(4 \alpha_1 + 3\alpha_2+2)\,\log R^2 + 1
  - {1\over 8\pi\widehat{t}} (4 \alpha_1 + 3\alpha_2)\, \log \widehat{t} +\,
    {G_1 - C_2\over \widehat{t}},
\ee
which agrees with the perturbative result \reff{eq:6.6}.

\clearpage 


\begin{figure}
\vspace*{-1cm} \hspace*{-0cm}
\begin{center}
\epsfxsize = 0.9\textwidth
\leavevmode\epsffile{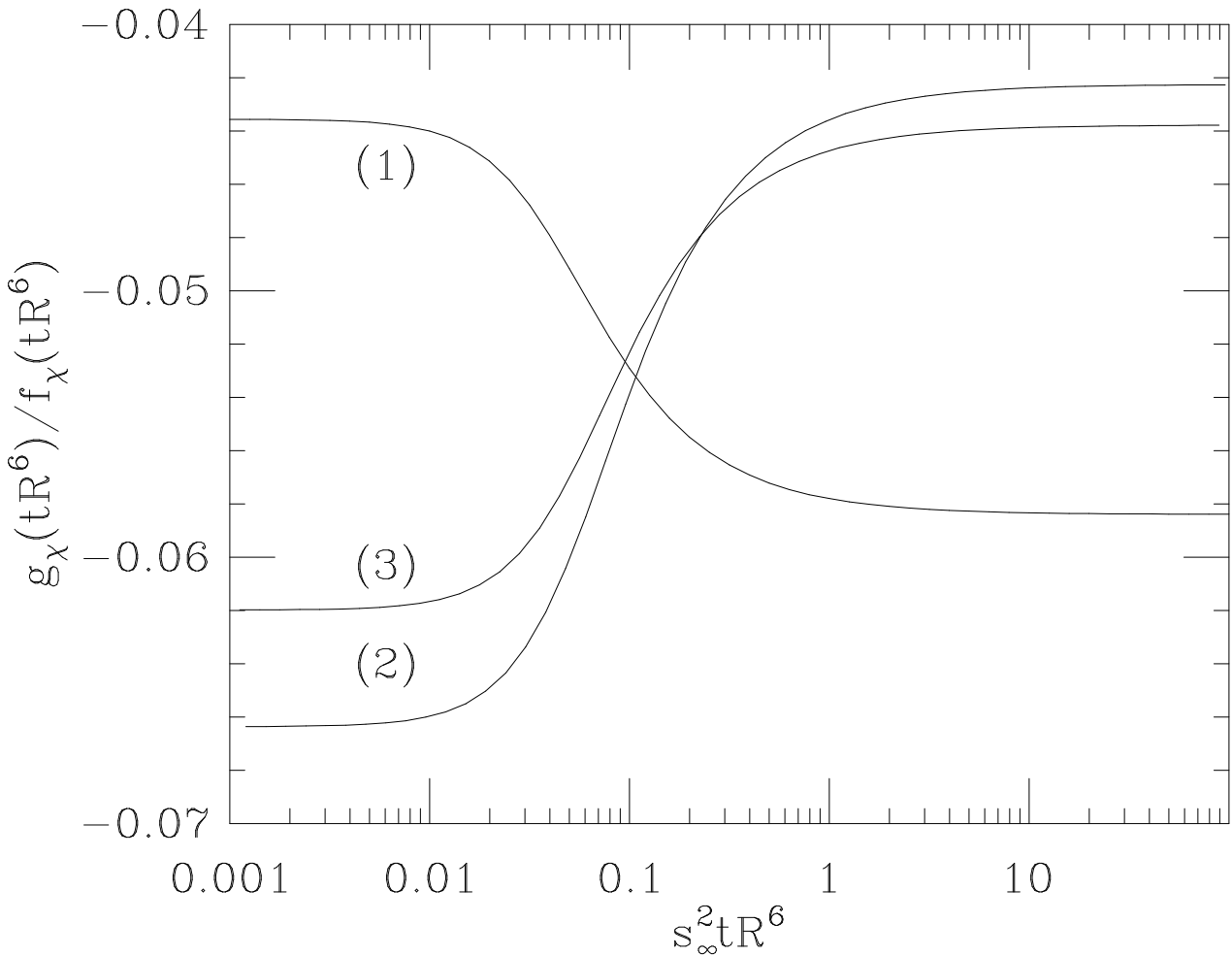}
\end{center}
\vspace*{-1cm}
\caption{Ratio $g_\chi(\widetilde{t})/f_\chi(\widetilde{t})$
in the large-$N$ limit
for three different models: (1) $N$-vector model; (2) 
potential $W(\varphi^2) = \varphi^2 + (\varphi^2 - 1)^2$; 
(3) potential $W(\varphi^2) = \varphi^2 + (\varphi^2 - 1)^2 + (\varphi^2)^3$.
In all cases $J(x)$ is given in Eq. \protect\reff{Jrhoflat} 
with domain \protect\reff{domainD1}. $s_\infty$ is a constant defined in 
Eq. \protect\reff{sinfty_def}.
}
\label{correzioni_largeN}
\end{figure}

\begin{figure}
\vspace*{-1cm} \hspace*{-0cm}
\begin{center}
\epsfxsize = 0.9\textwidth
\leavevmode\epsffile{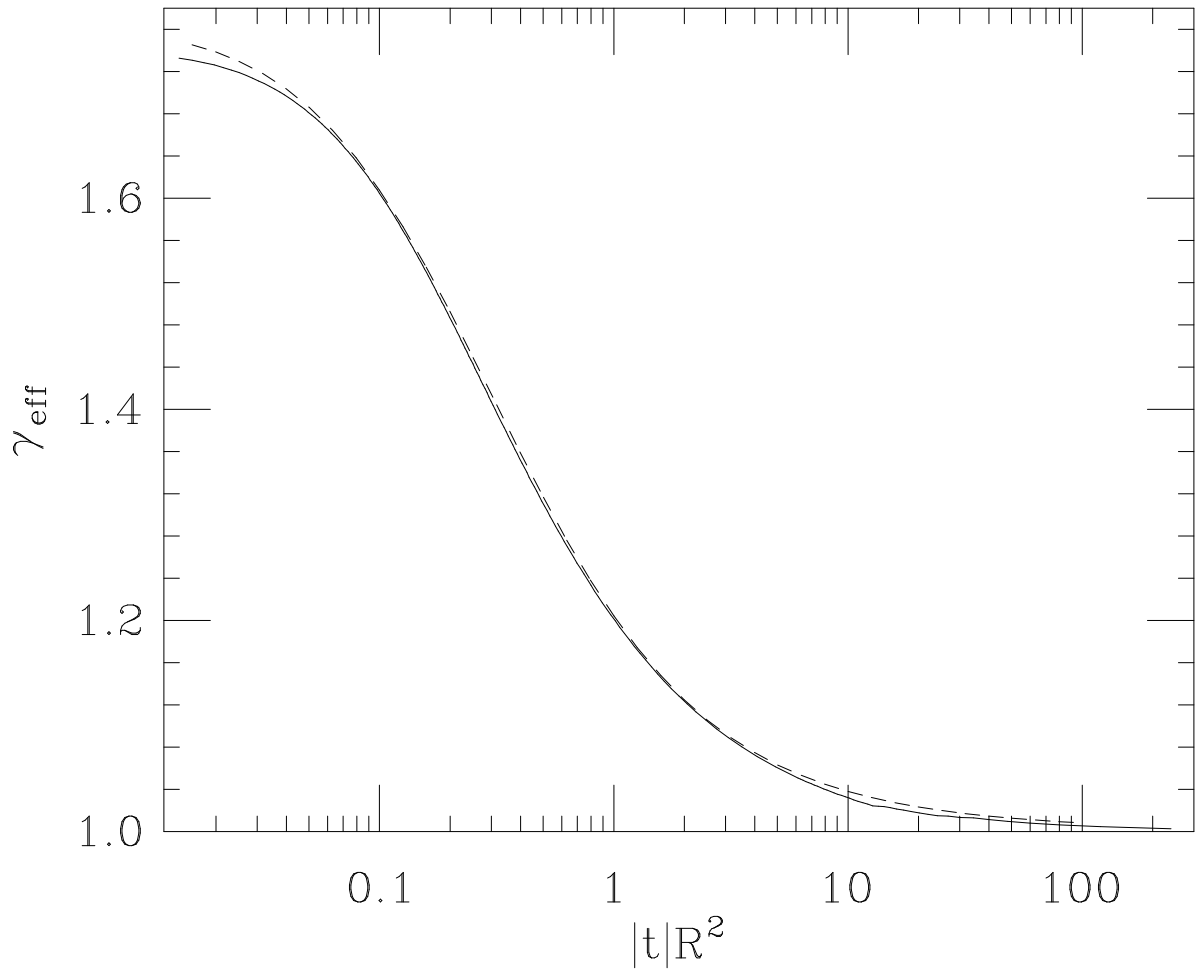}
\end{center}
\vspace*{-1cm}
\caption{
Effective susceptibility exponent as a function of
$\widetilde{t}$ in the high-temperature phase of the two-dimensional
Ising model. The dashed line represents our interpolation of the 
numerical results of Ref. \protect\cite{L-B-B-prl}.
In the mean-field limit $\gamma_{\rm eff}=1$, while for $\widetilde{t}\to 0$,
$\gamma_{\rm eff} = 7/4$.
}
\label{gammaeff2d}
\end{figure}

\begin{figure}
\vspace*{-1cm} \hspace*{-0cm}
\begin{center}
\epsfxsize = 0.9\textwidth
\leavevmode\epsffile{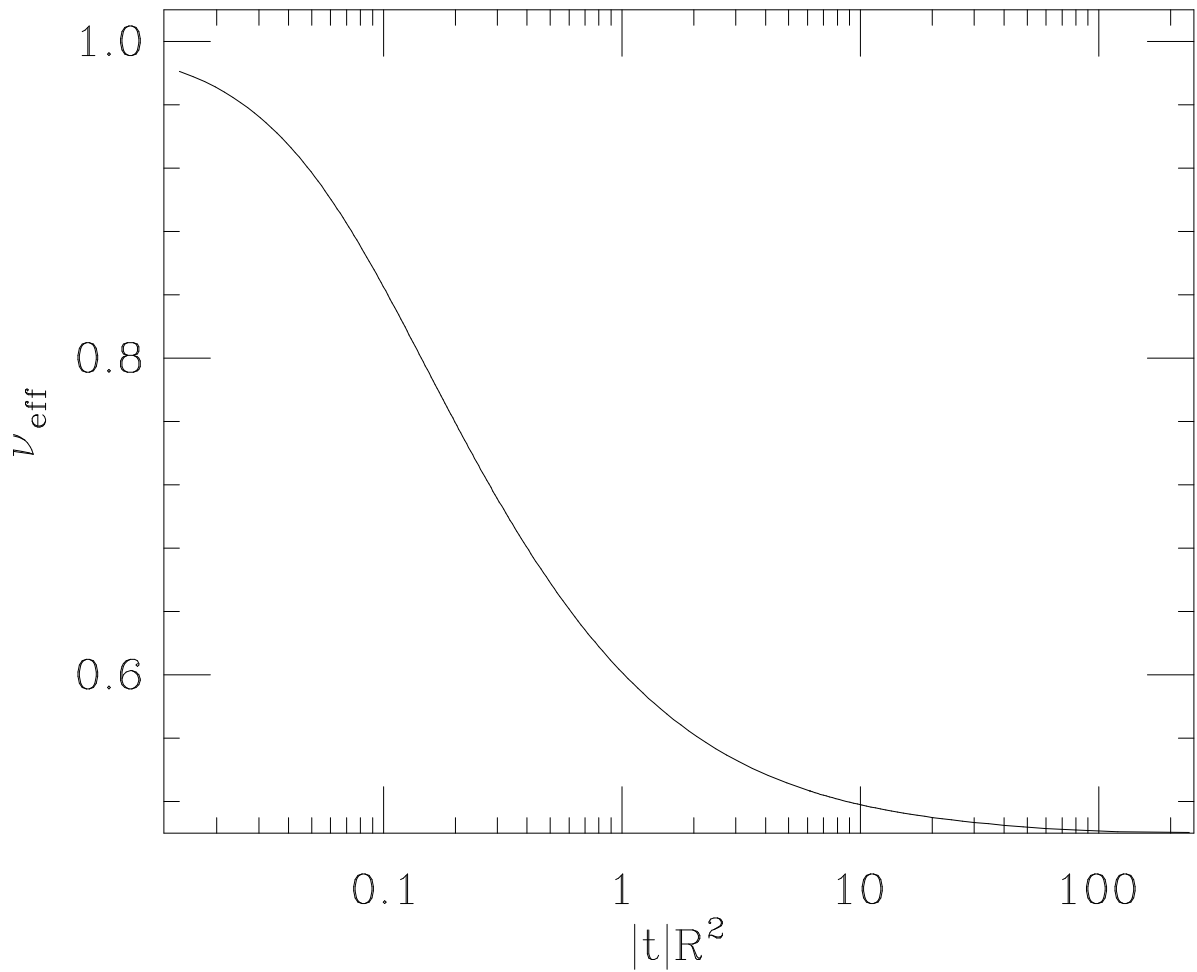}
\end{center}
\vspace*{-1cm}
\caption{
Effective correlation-length exponent as a function of
$\widetilde{t}$ for the high-temperature phase of the two-dimensional
Ising model. In the mean-field limit $\nu_{\rm eff}=1/2$, while for 
$\widetilde{t}\to0$, $\nu_{\rm eff} = 1$.
}
\label{nueff2d}
\end{figure}

\begin{figure}
\vspace*{-1cm} \hspace*{-0cm}
\begin{center}
\epsfxsize = 0.9\textwidth
\leavevmode\epsffile{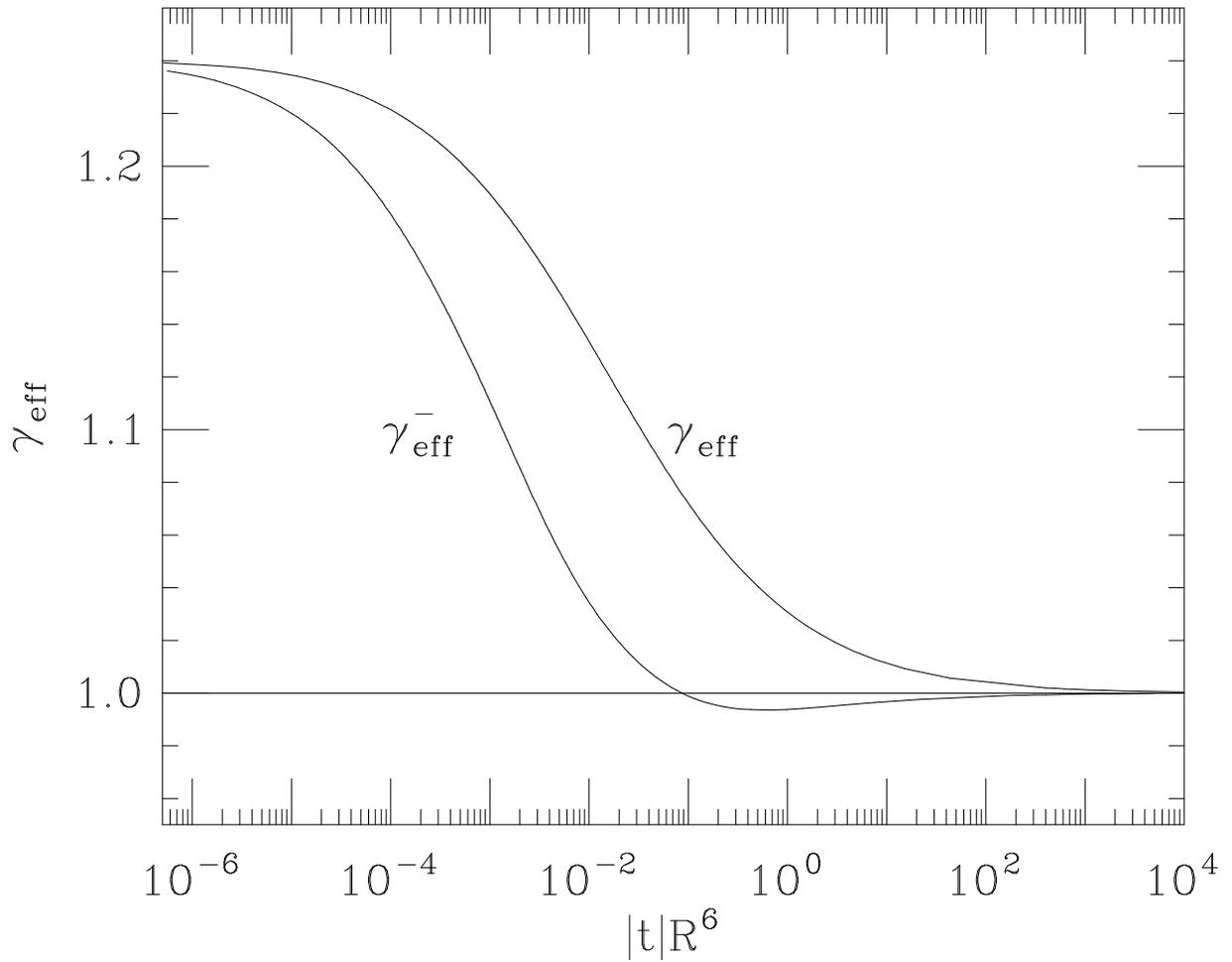}
\end{center}
\vspace*{-1cm}
\caption{
Effective susceptibility exponent as a function of
$\widetilde{t}$ for the high- ($\gamma_{\rm eff} $) and
low- ($\gamma_{\rm eff}^- $) temperature phase of the three-dimensional
Ising model. In the mean-field limit $\gamma_{\rm eff} = 1$, while for 
$|\widetilde{t}|\to 0$, $\gamma_{\rm eff}\approx 1.237$.
}
\label{gammaeff}
\end{figure}

\begin{figure}
\vspace*{-1cm} \hspace*{-0cm}
\begin{center}
\epsfxsize = 0.9\textwidth
\leavevmode\epsffile{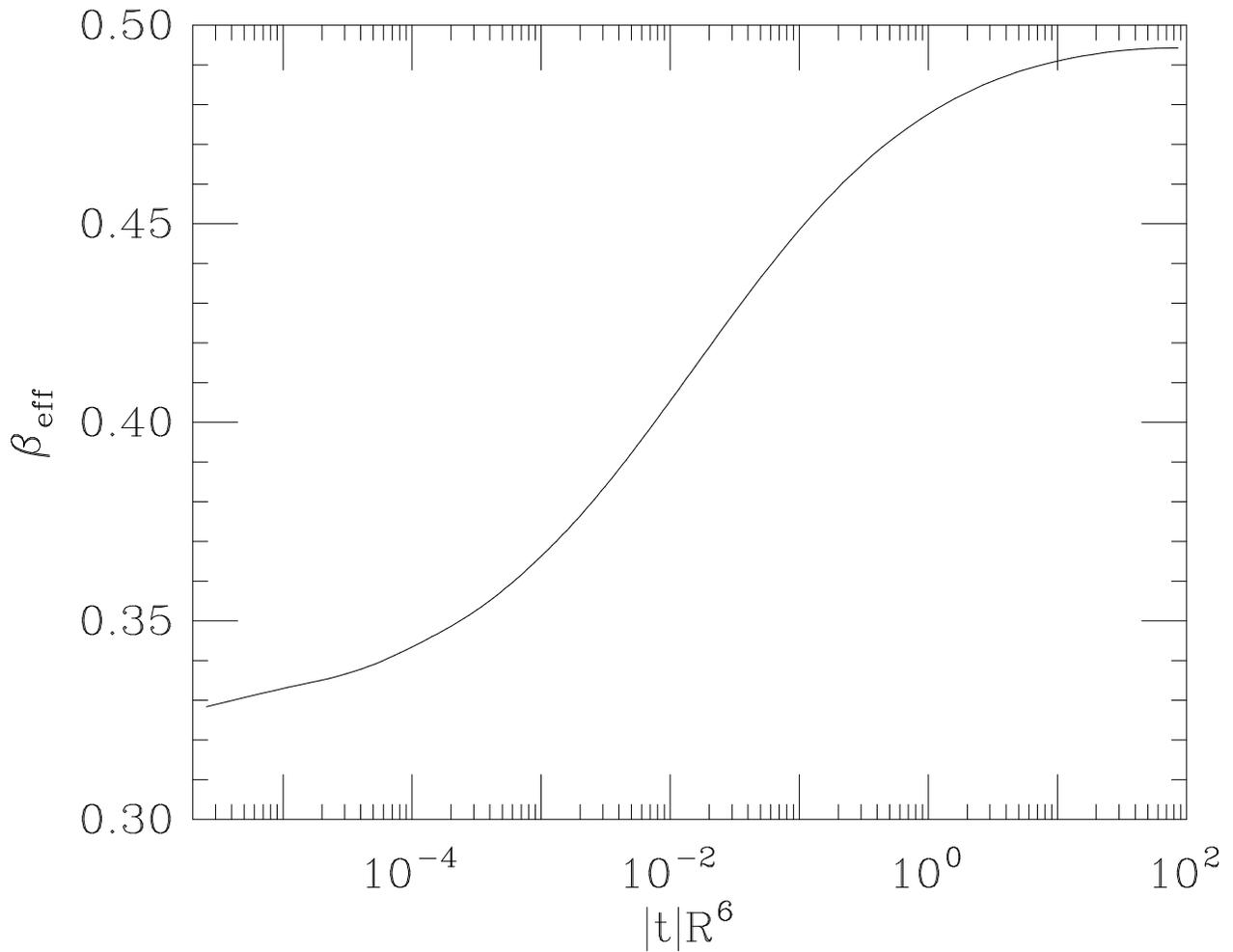}
\end{center}
\vspace*{-1cm}
\caption{
Effective magnetization exponent as a function of
$\widetilde{t}$ in the 
low-temperature phase of the three-dimensional
Ising model. In the mean-field limit $\beta_{\rm eff} = 1/2$,
while for $|\widetilde{t}|\to 0$, $\beta_{\rm eff} \approx 0.327$.
}
\label{betaeff}
\end{figure}

\begin{figure}
\vspace*{-1cm} \hspace*{-0cm}
\begin{center}
\epsfxsize = 0.9\textwidth
\leavevmode\epsffile{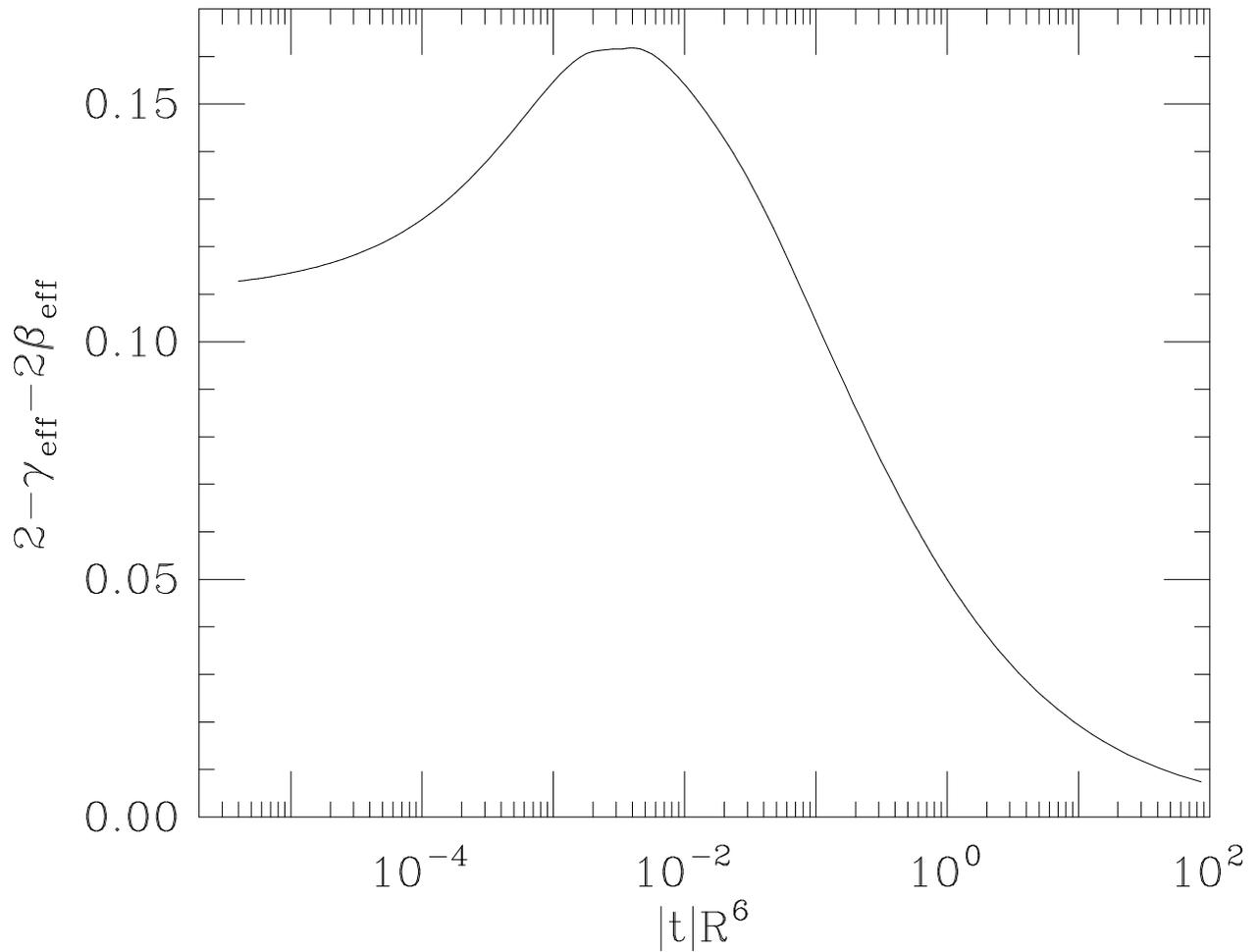}
\end{center}
\vspace*{-1cm}
\caption{
Combination $2 - \gamma_{\rm eff}^- - 2\beta_{\rm eff}$
as a function of $\widetilde{t}$ in the
low-temperature phase of the three-dimensional
Ising model. In the mean-field limit this combination vanishes,
while for $\widetilde{t}\to 0$ it is equal to the 
specific-heat exponent $\alpha \approx 0.109$.
}
\label{scalingrel}
\end{figure}

\begin{figure}
\vspace*{-1cm} \hspace*{-0cm}
\begin{center}
\epsfxsize = 0.9\textwidth
\leavevmode\epsffile{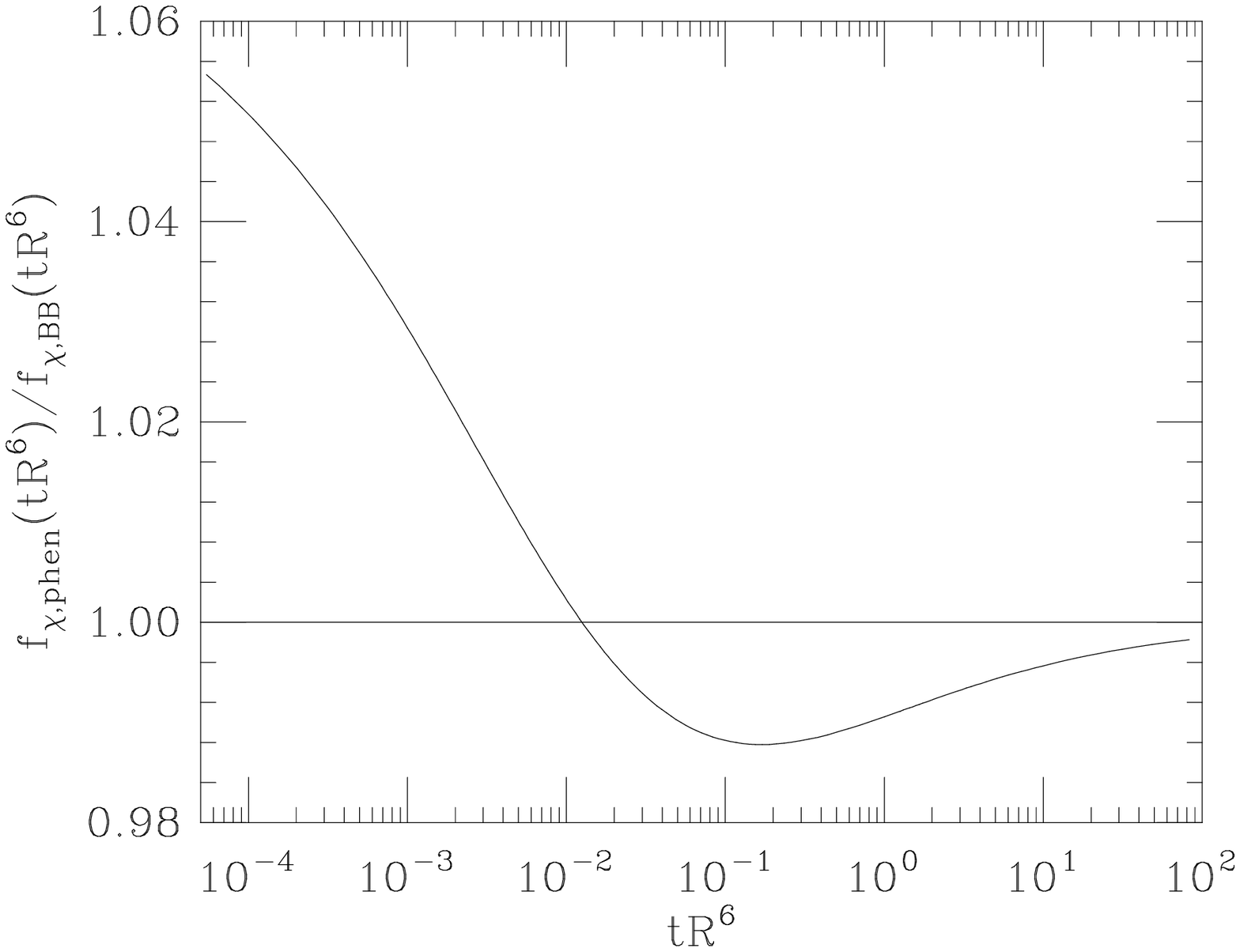}
\end{center}
\vspace*{-1cm}
\caption{Ratio 
$f_{\chi,\rm phen}(\widetilde{t})/f_{\chi,\rm BB}(\widetilde{t})$
as a function of $\widetilde{t}$.
}
\label{confrontoSengersBB}
\end{figure}

\begin{figure}
\vspace*{-1cm} \hspace*{-0cm}
\begin{center}
\epsfxsize = 0.9\textwidth
\leavevmode\epsffile{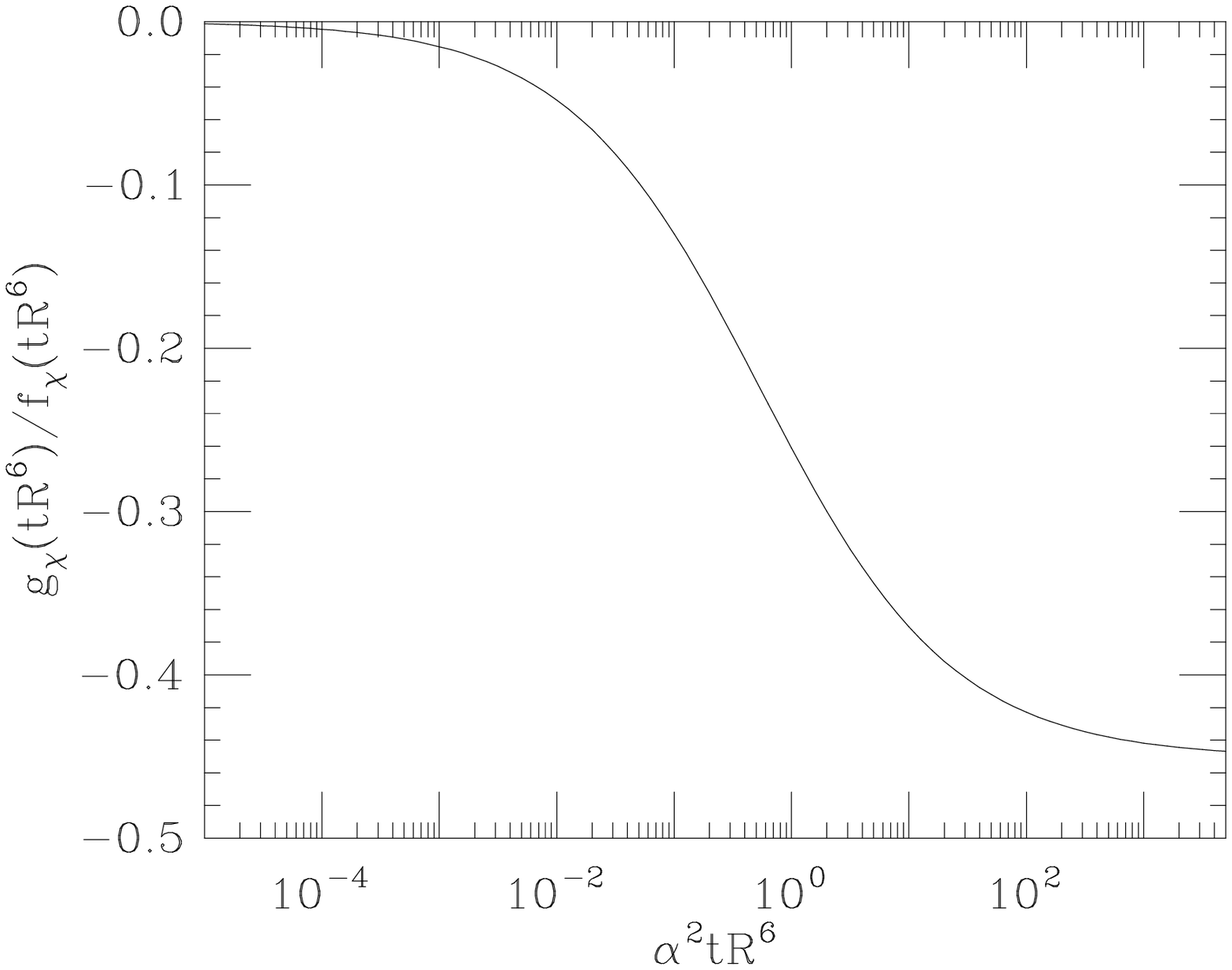}
\end{center}
\vspace*{-1cm}
\caption{$g_\chi(\widetilde{t})/f_\chi(\widetilde{t})$ in the model of 
Ref. \protect\cite{Luijten-Maryland}
as a function of $\alpha^2\widetilde{t}$.
}
\label{ratiofg_Sengers}
\end{figure}


\begin{thebibliography}{199}

\bibitem{Corti-Degiorgio_85}
M.~Corti and  V.~Degiorgio,
Phys.\ Rev.\ Lett.\ {\bf 55} (1985) 2005.

\bibitem{Dietler-Cannel_88}
G. Dietler and D. S. Cannel,
Phys.\ Rev.\ Lett.\ {\bf 60} (1988) 1852.

\bibitem{Anisimov-etal_95}
M.~A.~Anisimov, A.~A.~Povodyrev,
V.~D.~Kulikov, and J.~V.~Sengers,
Phys.\ Rev.\ Lett.\ {\bf 75} (1995) 3146.

\bibitem{Anisimov-etal_96}
M.~A.~Anisimov, A.~A.~Povodyrev,
V.~D.~Kulikov, and  J.~V.~Sengers,
Phys.\ Rev.\ Lett.\ {\bf 76} (1996) 4095.

\bibitem{Jacob-etal_98}
J. Jacob, A. Kumar, M. A. Anisimov, A. A. Povodyrev, and 
J. V. Sengers, Phys. Rev. {\bf E 58} (1998) 2188.

\bibitem{Bagnuls-Bervillier_84} C.~Bagnuls and  C.~Bervillier,
J. Phys. Lett. (Paris) {\bf 45} (1984) L-95.

\bibitem{Bagnuls-Bervillier_85} C.~Bagnuls and  C.~Bervillier,
Phys.\ Rev.\ {\bf B 32} (1985) 7209.

\bibitem{Bagnuls-etal_87}
C.~Bagnuls, C.~Bervillier, D. I. Meiron, and B. G. Nickel,
Phys.\ Rev.\ {\bf B 35} (1987) 3585.

\bibitem{Fisher_86}
M.~E.~Fisher, Phys.\ Rev.\ Lett.\ {\bf 57} (1986) 1911.

\bibitem{Bagnuls-Bervillier_87}
C.~Bagnuls and C.~Bervillier,
Phys.\ Rev.\ Lett.\ {\bf 58} (1987) 435.

\bibitem{Schloms-Dohm_89}
R. Schloms and V. Dohm, Nucl. Phys. {\bf B328} (1989) 639.

\bibitem{Chen-etal_90}
Z.~Y.~Chen, P.~C.~Albright,
and J.~V.~Sengers, Phys.\ Rev.\ {\bf A 41} (1990) 3161.

\bibitem{Anisimov-etal_92}
M.~A.~Anisimov, S.~B.~Kiselev, J.~V.~Sengers, and S.~Tang,
Physica {\bf A 188} (1992) 487.

\bibitem{Belyakov-Kiselev_92}
M.~Y.~Belyakov and  S.~B.~Kiselev,
Physica {\bf A 190} (1992) 75.

\bibitem{Thouless_69}
D. J. Thouless, Phys. Rev. {\bf 181} (1969) 954.

\bibitem{M-B} K.~K.~Mon and K.~Binder,
Phys.\ Rev.\ {\bf E 48} (1993) 2498.

\bibitem{L-B-B-pre}
E.~Luijten, H.~W.~J.~Bl\"ote, and K.~Binder,
Phys.\ Rev.\ {\bf E 54} (1996) 4626.

\bibitem{L-B-B-prl}
E.~Luijten, H.~W.~J.~Bl\"ote, and K.~Binder,
Phys.\ Rev.\ Lett.\ {\bf 79} (1997) 561;
Phys.\ Rev.\ {\bf E 56} (1997) 6540.

\bibitem{Luijten-Binder_98}
E.~Luijten and K.~Binder,
Phys.\ Rev.\ {\bf E 58} (1998) R4060.

\bibitem{PRV_longrange} 
A. Pelissetto, P. Rossi, and E. Vicari, 
Phys. Rev. {\bf E 58} (1998) 7146.

\bibitem{Ginzburg_60} V.~L.~Ginzburg,
Fiz.\ Tverd.\ Tela {\bf 2} (1960) 2031 [Sov. Phys. Solid State,
{\bf 2} (1960) 1824].

\bibitem{Luijten-FSS}
E.~Luijten, {\em Critical properties of the three-dimensional
equivalent-neighbor model and crossover scaling in finite systems},
{\tt cond-mat/9811332}.

\bibitem{Luijten-Maryland}
M. A. Anisimov, E. Luijten, V. A. Agayan, J. V. Sengers, 
and K. Binder, {\em Shape of crossover between mean-field and asymptotic
critical behavior in a three-dimensional Ising lattice}, 
{\tt cond-mat/9810252}.

\bibitem{nosotros-preparation}
S. Caracciolo, M. S. Causo, A. Pelissetto, P. Rossi, and 
E. Vicari, in preparation.

\bibitem{CCPRV_Lattice98}
S. Caracciolo, M. S. Causo, A. Pelissetto, P. Rossi, and
E. Vicari, {\em Crossover scaling from classical to non-classical
critical behaviour}, Nucl. Phys. B (Proc. Suppl.) (1999).
{\tt hep-lat/9809101}.

\bibitem{Parisi_Cargese}
G.~Parisi, Carg\`{e}se Lectures (1973),
J.\ Stat.\ Phys.\ {\bf 23} (1980) 49.

\bibitem{Baker-etal_77_78} G.~A.~Baker, Jr.,  B.~G.~Nickel,
M.~S.~Green, and D.~I.~Meiron,
Phys.\ Rev.\ Lett.\ {\bf 36} (1977) 1351;
G.~A.~Baker, Jr., B.~G.~Nickel, and D.~I.~Meiron,
Phys.\ Rev.\ {\bf B 17} (1978) 1365.

\bibitem{Murray-Nickel_91}
D. B. Murray and B.~G.~Nickel, {\em
Revised estimates for critical exponents for the continuum
$n$-vector model in 3 dimensions}, unpublished
Guelph University report (1991).

\bibitem{Dohm} V.~Dohm,
Z. Phys. {\bf B 60} (1985) 61;
{\bf B 61} (1985) 193.

\bibitem{K-S-D} H.J.~Krause, R.~Schloms, and V.~Dohm,
Z. Phys. {\bf B 79} (1990) 287.

\bibitem{C-G-L-T} K.~G.~Chetyrkin, S.~G.~Gorishny,
S.~A.~Larin, and F.~V.~Tkachov,
Phys.\ Lett.\ {\bf B 132} (1983) 351.

\bibitem{K-N-S-C-L} H.~Kleinert, J.~Neu,
V.~Schulte-Frohlinde,
K.~G.~Chetyrkin,  and S.~A.~Larin,
Phys.\ Lett.\ {\bf B 272} (1991) 39;
Erratum {\bf B 319} (1993) 545.

\bibitem{LeGuillou-ZinnJustin_80}
J.~C.~Le Guillou and  J.~Zinn-Justin,
Phys.\ Rev.\ {\bf B 21} (1980) 3976.

\bibitem{ZinnJustin_book}
J. Zinn-Justin, {\em ``Quantum field theory and critical phenomena"},
third edition,
(Clarendon, Oxford, 1996).

\bibitem{PV_gstar}
A. Pelissetto and E. Vicari, 
Nucl. Phys. {\bf B519} [FS] (1998) 605; {\em Non-analyticity of the 
Callan-Symanzik $\beta$-function of $O(N)$ models}, 
Nucl. Phys. B (Proc. Suppl.) (1999), {\tt hep-lat/9809041}.

\bibitem{Butera-Comi_98}
P. Butera and M. Comi, Phys. Rev. {\bf B 58} (1998) 11552.

\bibitem{Guida-ZinnJustin_98}
R. Guida and J. Zinn-Justin, J. Phys. {\bf A 31} (1998) 8103.

\bibitem{Symanzik_73}
K. Symanzik, Lett. Nuovo Cimento {\bf 8} (1973) 771.

\bibitem{Parisi_79}
G. Parisi, Nucl. Phys. {\bf B150} (1979) 163.

\bibitem{Bergere-David_82}
M. C. Berg\`ere and F. David,
Ann. Phys. (N.Y.) {\bf 142} (1982) 416.

\bibitem{Lipatov} L.~N.~Lipatov, Zh. Eksp. Teor. Fiz. 
{\bf 72} (1977) 411 [Sov.\ Phys. --- JETP {\bf 45}
(1977) 216].

\bibitem{B-L-Z}
E.~Br\'ezin, J.~C.~Le Guillou, and J. Zinn-Justin,
Phys. Rev. {\bf D 15} (1977) 1544.


\bibitem{Brezin-etal_DG6} 
E. Br\'ezin, J. C. Le Guillou, and J. Zinn-Justin,
in ``Phase Transitions and Critical Phenomena", Vol. 6,
C. Domb and M. S. Green eds. (Academic, London--New York--San Francisco, 1976).

\bibitem{Baker_62} G. A. Baker, Phys. Rev. {\bf 126} (1962) 2071.

\bibitem{Hubbard_72} J. Hubbard, Phys. Lett. {\bf 39A} (1972) 365.

\bibitem{Brout_60}
R. Brout, Phys. Rev. {\bf 118} (1960) 1009.

\bibitem{Dalton-Domb_66}
N. W. Dalton and C. Domb, Proc. Phys. Soc. (London)
{\bf 89} (1966) 873.

\bibitem{Vaks-Larkin-Pikin_66}
V. G. Vaks, A. I. Larkin, and S. A. Pikin,
Zh. Eksp. Teor. Fiz. {\bf 51} (1966) 361 
[Sov. Phys. --- JETP {\bf 24} (1967) 240].

\bibitem{Caracciolo-Pelissetto_98}
S. Caracciolo and A. Pelissetto,
Phys. Rev. {\bf D 58} (1998) 105007; Nucl. Phys. B (Proc. Suppl.)
{\bf B53} (1997) 693.

\bibitem{Caracciolo-etal_98}
S. Caracciolo, A. Montanari, and A. Pelissetto,
{\em Testing the Efficiency of Different Improvement Programs},
{\tt hep-lat/9812014}.

\bibitem{Luijten-private}
E. Luijten, private communication.

\bibitem{Chen-etal_90b}
Z. Y. Chen, A. Abbaci, S. Tang, and J. V. Sengers, 
Phys. Rev. {\bf A 42} (1990) 4470.

\bibitem{Nicoll-Bhattacharjee_81}
J. F. Nicoll and J. K. Bhattacharjee, Phys. Rev. {\bf B 23} (1981) 389.

\bibitem{Nicoll-Albright_85}
J. F. Nicoll and P. C. Albright,
Phys. Rev. {\bf B 31} (1985) 4576.

\bibitem{Melnichenko-etal_97}
Y. B. Melnichenko, M. A. Anisimov, A. A. Povodyrev, G. D. Wignall,
J. V. Sengers, and W. A. Van Hook, Phys. Rev. Lett. {\bf 79} (1997) 5266.

\bibitem{Sarbach-Fisher_78}
S. Sarbach and M. E. Fisher, Phys. Rev {\bf B 18} (1978) 2350.

\bibitem{Emery}
V. J. Emery, Phys. Rev. {\bf B 11} (1975) 239; {\bf B 11} (1975) 3397.

\bibitem{Sarbach-Schneider_76_77}
S. Sarbach and T. Schneider, 
Phys. Rev. {\bf B 13} (1976) 464; {\bf B 16} (1977) 347.

\bibitem{Symanzik_81}
K. Symanzik, in {\em ``Mathematical problems in
theoretical physics"}, R. Schrader et al. eds., (Springer, Berlin, 1982);
Nucl. Phys. {\bf B226} (1983) 187, 205.

\end{thebibliography}
\end{document}